\documentclass[11pt]{article}
\pdfoutput=1
\usepackage[letterpaper, margin=1in]{geometry}
\usepackage{amsmath, amssymb}
\usepackage{cite}
\usepackage{graphicx}
\usepackage[font=small,labelfont=bf]{caption}

\usepackage{xcolor}
\usepackage{slashed}
\usepackage{cancel}
\usepackage{bm}
\usepackage{verbatim}
\usepackage{tabularx}

\definecolor{Mahogany}{rgb}{0.62,0.24,0.15}
\definecolor{DarkRed}{rgb}{0.6,0,0}
\definecolor{DarkGreen}{rgb}{0,0.6,0}
\definecolor{DarkBlue}{rgb}{0,0,0.6}
\usepackage[colorlinks=true,linktocpage=true,linkcolor=DarkBlue,citecolor=DarkGreen,urlcolor=DarkGreen]{hyperref}

\usepackage{setspace}
\usepackage{authblk}
\usepackage{physics}
\usepackage{caption}
\usepackage{subcaption}
\usepackage{tablefootnote}

\usepackage{dsfont}

\usepackage{etoolbox}
\appto\appendix{\addtocontents{toc}{\protect\setcounter{tocdepth}{1}}}

\usepackage[greek,english]{babel}

\newcommand{\be}{\begin{equation}}
\newcommand{\ee}{\end{equation}}
\def\bga{\begin{aligned}}
\def\eda{\end{aligned}}
\def\bgp{\begin{pmatrix}}
\def\edp{\end{pmatrix}}

\newcommand{\Emin}{E_{\textrm{min}}}
\newcommand{\Emax}{E_{\textrm{max}}}
\newcommand{\Eminhat}{\hat{E}_{\textrm{min}}}
\newcommand{\Emaxhat}{\hat{E}_{\textrm{max}}}

\newcommand{\BR}{\textrm{BR}}

\newcommand{\Sec}[1]{Section~\ref{#1}}

\newcommand{\App}[1]{Appendix~\ref{#1}}

\newcommand{\Fig}[1]{Fig.~\ref{#1}}
\newcommand{\Eq}[1]{Eq.~(\ref{#1})}

\definecolor{goodgreen}{rgb}{0,.6,0.4}

\begin{document}
 
\title{\bf 
\LARGE Complementary Signals of Lepton Flavor Violation at a High-Energy Muon Collider
}
\author{Samuel Homiller\footnote{shomiller@g.harvard.edu},~ Qianshu Lu\footnote{qianshu$\_$lu@g.harvard.edu},~ and Matthew Reece\footnote{mreece@g.harvard.edu}\\
\small \textit{Department of Physics, Harvard University, Cambridge, MA 02138}
}

\date{\today}

\maketitle

\begin{abstract}
\begin{spacing}{1.05}\noindent
\normalsize
A muon collider would be a powerful probe of flavor violation in new physics. There is a strong complementary case for collider measurements and precision low-energy probes of lepton flavor violation (as well as CP violation). We illustrate this by studying the collider reach in a supersymmetric scenario with flavor-violating slepton mixing. We find that the collider could discover sleptons and measure the slepton and neutralino masses with high precision, enabling event reconstruction that could cleanly separate flavor-violating new physics signals from Standard Model backgrounds. The discovery reach of a high-energy muon collider would cover a comparably large, and overlapping, range of parameter space to future $\mu \to e$ conversion and electron EDM experiments, and unlike precision experiments could immediately shed light on the nature of new physics responsible for flavor violation. This complementarity strengthens the case that a muon collider could be an ideal energy-frontier laboratory in the search for physics beyond the Standard Model.
\end{spacing}
\end{abstract}

\begin{spacing}{1.15}

\clearpage

\tableofcontents

\vskip 1cm


\section{Introduction}

Since discovering the Higgs boson, the Large Hadron Collider has continued to push the energy frontier, and in the absence of any clear signs of new physics, has set ever-stronger constraints on potential extensions to the Standard Model (SM). At the same time, a great deal of what we know about the SM and what lies beyond it comes from precision tests of Standard Model particles. In fact, depending on the assumptions about physics beyond the Standard Model (BSM), our strongest constraints on new states come from searches for the electron and neutron electric dipole moments, constraints on flavor-violating muon decays, or measurements of Kaon mixing. If one assumes that new physics responsible for flavor-violating neutral currents arises with anarchic, $\mathcal{O}(1)$ coefficients at a common scale, the scale of new physics must be $\gtrsim 10^6\,\textrm{TeV}$~\cite{Isidori:2010kg}: far beyond the reach of colliders.

However, given that the SM itself has a very non-generic flavor structure---the fermion masses span five orders of magnitude, and the quark mixing matrix has a clear, hierarchical structure---it is natural to think that the structure of physics beyond the Standard Model will be non-generic as well. Coupling this reasoning with the Higgs hierarchy problem, the sensitivity of the electroweak scale to quantum corrections, there is good reason to expect that we might find flavorful new physics near the TeV scale. This accentuates the need for new colliders to probe yet shorter distances and higher energies.

In this light, a high-energy muon collider is a uniquely well-suited option.
A muon collider would have immense capability to extend our reach for new physics to much higher masses by colliding fundamental objects at multi-TeV energies~\cite{Delahaye:2019omf, AlAli:2021let, Aime:2022flm, DeBlas:2022wxr}. In doing so, such a collider can probe the same physics that we hope to see signals of in low-energy experiments. This raises the tantalizing prospect of not only discovering new physics at the TeV scale, but learning qualitatively and quantitatively about its structure, so that we can make headway on the fundamental puzzles of the Standard Model.

Among the numerous processes that can be tested reciprocally at both low- and high-energies, we will focus in this paper on those with charged lepton flavor violation (LFV). As many of the most sensitive searches for LFV involve muons, these are the most naturally linked with a muon collider. Indeed, we will see in an explicit example that searches for mixed-flavor final states at a collider and processes such as $\mu \to e\gamma$, $\mu \to 3 e$ and $\mu$-to-$e$ transitions are probing the same UV physics. An investigation of complementarity with LFV probes is also timely, as a number of experiments searching for the aforementioned processes are expected to come online with orders of magnitude greater sensitivity than their predecessors in the coming decade~\cite{Baldini:2018uhj, Middleton:2022fvu}. 

Throughout this work, we will focus on the minimal supersymmetric standard model (MSSM) as a canonical example of how lepton flavor violation can arise in models of beyond the Standard Model (BSM) physics. Absent some underlying symmetry structure, the spectrum of the MSSM naturally yields an assortment of new contributions to flavor-changing processes and sources of CP violation. However, while the MSSM contains a large number of new parameters, we will see that only a few of them are necessary to describe the lepton-flavor violating signatures at a high-energy muon collider and the complementary signals in low-energy probes.

While our main goal is to understand the complementarity of collider and low-energy probes, we will also revisit the expectations for lepton flavor violation rates in the MSSM.  We will mostly work within a simple, minimal framework that leads to both collider and low-energy signals. Within this framework, we will highlight the regions of parameter space that are particularly well motivated in concrete models, with an emphasis on how flavor violation in the MSSM is intimately connected to how supersymmetry is broken at higher scales. This framework serves as a useful benchmark for not only the collider reach, but also for prospective measurements of $\mu \to e\gamma$, $\mu \to 3e$ and $\mu$-to-$e$ conversion in nuclei. 

Realistically, of course, nature is unlikely to subscribe to a simple construction that is most convenient for studying particular processes. With this in mind, we will also explore how moving beyond this minimal framework changes the expected rates for LFV processes at low energies. As we will see, the next-to-minimal ingredients also give rise to a nonzero electric dipole moment (EDM) for the electron. This further reinforces the potential of a high-energy muon collider in being particularly well-suited to explore the same fundamental physics being tested at low energies. 

The rest of this paper is structured as follows: in \Sec{sec:mssm_clfv}, we review the minimal scenario in which charged lepton flavor violation arises in the MSSM and discuss the contributions to LFV processes at low energies.
\Sec{sec:mssm_at_muc} is devoted to a discussion of signatures of the MSSM at a muon collider, including an estimate of how precisely the slepton and neutralino masses can be measured. 
In \Sec{sec:reach}, we present the muon collider reach for the flavor-violating processes at high-energies and compare to future constraints Mu3e, Mu2e, and PRISM/PRIME.
Finally, in \Sec{sec:nonminimal} we discuss how the LFV signatures can change when we include states beyond the minimal ingredients, and the complementarity with searches for electric dipole moments.
We conclude in \Sec{sec:conclusion}. 
The appendices contain more discussion on how flavor-violating terms arise in realistic models of supersymmetry breaking and provide more details on the slepton and neutralino mass measurement and the reconstruction procedure used to compute the collider reach.

\section{Charged Lepton Flavor Violation in the MSSM}
\label{sec:mssm_clfv}

Charged lepton flavor violation arises in the MSSM when there are contributions to the slepton masses that are not diagonal in the same basis as the SM leptons.
When this occurs, the physical sleptons are mixtures of different flavors, and their interactions with the SM leptons and  neutralinos/charginos will be flavor-violating.

As the Yukawa interactions dictated by supersymmetry are necessarily aligned with the lepton interactions, the flavor-violating terms must arise from soft supersymmetry-breaking terms, which depend on the underlying model of supersymmetry breaking. In this sense, the flavor structure of supersymmetry is inextricably tied with understanding the dynamics of supersymmetry breaking. We will say more about this in \App{app:flavor}.

For our purposes, it will suffice to consider only the right-handed sleptons and the lightest neutralino of the full MSSM spectrum. The left-handed and color-charged scalars, as well as the gluino, are assumed to acquire soft masses with parametrically larger values that lie outside the reach of the high-energy collider we consider. We will further assume that the lightest neutralino is pure Bino, $\tilde{B}$, with mass $M_1$. Aside from being phenomenologically convenient, this assumption is also quite natural, as the left-handed and color-charged particles typically receive additional contributions to their masses from $SU(2)_L$ and $SU(3)_c$ gauge interactions that the right-handed sleptons and Bino do not. This spectrum bears some resemblance to those studied from a UV perspective in ref.~\cite{Agashe:2022uih} in the context of the hierarchy problem and the muon $(g-2)$. The LHC bounds were compared to the flavor constraints for a similar spectrum in ref.~\cite{Calibbi:2015kja} as well.

For our collider studies, we will be interested in the case that $M_1$ is less than the slepton masses, so that the sleptons decay via $\tilde{\ell}_i \to \ell_j \tilde{B}$, with the neutralino appearing as missing momentum.
Finally, for simplicity, we will largely ignore the effects of the stau, and assume that its mass is reasonably well separated from the selectron and smuon. We will revisit the effects of stau mixing with the other sleptons again at the end of the section and in much more detail in \Sec{sec:nonminimal}.

The result of this suite of motivated assumptions is that the flavor-violating slepton spectrum is reduced to a $2\times 2$ mixing problem. We write the mass matrix of the right-handed selectron and smuon as
\be
\label{eq:slepton_mass_matrix}
\bgp \tilde{e}_R^{\dagger} & \tilde{\mu}_R^{\dagger} \edp
\bgp \mathcal{M}_R^2 \edp
\bgp \tilde{e}_R \\ \tilde{\mu}_R \edp
\equiv
\bgp \tilde{e}_R^{\dagger} & \tilde{\mu}_R^{\dagger} \edp
\bgp
m_R^2 + \Delta^{RR}_{ee} & \Delta^{RR}_{e\mu} \\ 
\big(\Delta^{RR}_{e\mu}\big)^* & m_R^2 + \Delta^{RR}_{\mu\mu}
\edp
\bgp \tilde{e}_R \\ \tilde{\mu}_R \edp
\ee
where $m_R^2$ is the flavor universal part of the mass matrix, and the $\Delta^{RR}_{ij}$ represent flavor non-universal entries. The matrix $\mathcal{M}_R^2$ in \Eq{eq:slepton_mass_matrix} is diagonalized by a unitary matrix $U_R$ parametrized by a mixing angle $\theta_R$ and a phase $\varphi_R$,
\be
U_R^{\dagger} \mathcal{M}_R^2 U_R = \textrm{diag}(m_{\tilde{e}_1}^2, m_{\tilde{e}_2}^2),
\quad
\text{where}
\quad
U_R = \bgp \cos\theta_R & -\mathrm{e}^{i\varphi_R} \sin\theta_R \\ \mathrm{e}^{-i\varphi_R}\sin\theta_R & \cos\theta_R \edp .
\ee
From \Eq{eq:slepton_mass_matrix}, we can compute the average and difference of the mass-squared eigenvalues, the mixing angle $\theta_R$ and the phase $\varphi_R$:
\begin{align}\label{eq:mixing_defs}
\overline{m^2} & \equiv \frac{1}{2}\big(m_{\tilde{\ell}_1}^2 + m_{\tilde{\ell}_2}^2\big) 
	= m_R^2 + \frac{1}{2}\big( \Delta^{RR}_{ee} + \Delta^{RR}_{\mu\mu}\big) \nonumber\\
\Delta m^2 & \equiv m_{\tilde{\ell}_2}^2 - m_{\tilde{\ell}_1}^2 
	= \sqrt{ \big|\Delta^{RR}_{ee} - \Delta^{RR}_{\mu\mu}\big|^2 + 4 \big|\Delta^{RR}_{e\mu}\big|^2 } \nonumber\\
\sin 2\theta_R & = \frac{2 |\Delta^{RR}_{e\mu}|}{\sqrt{|\Delta^{RR}_{ee} - \Delta^{RR}_{\mu\mu}|^2 + 4 |\Delta_{e\mu}|^2}} \nonumber \\
\varphi_R & = \arg(\Delta^{RR}_{e\mu}) 
\end{align}
where we've defined $\overline{m^2}$ and $\Delta m^2$ for later convenience.

The phase $\varphi_R$ can be removed from the slepton mass matrix by phase rotations on $\tilde{e}_R$, $e_R$ and $e_L$ such that all (s)electron interaction vertices and the electron mass stay real. Without the $\varphi_R$ phase, the slepton mixing problem can thus be completely described in our scenario by the mean slepton mass squared, $\overline{m^2}$, the mass splitting, $\Delta m^2 / \overline{m^2}$, and the mixing angle, $\sin(2\theta_R)$. To make the parameters responsible for flavor violation more manifest, we will rewrite the slepton mass matrix as
\be
\mathcal{M}_R^2 = \overline{m^2}\mathds{1} + \frac{\Delta m^2}{2}U_R \begin{pmatrix}1 &0 \\ 0&-1\end{pmatrix} U^{\dagger}_R.
\ee
The off-diagonal element of the second term breaks the $U(2)$ rotation that mixes the sleptons, and is proportional to $\Delta m^2 \sin(2\theta_R)$. This is why the usual normalized parameter describing the amount of flavor violation, $\delta^{RR}_{\mu e}$, is~\cite{Gabbiani:1996hi}
\be \label{eq:deltaRReq}
\delta^{RR}_{\mu e} \equiv \frac{\big(\mathcal{M}_R^2\big)_{\mu e}}{\sqrt{\big(\mathcal{M}_R^2\big)_{\mu\mu} \big(\mathcal{M}_R^2\big)_{ee}}} 
= \frac{\Delta m^2}{2 \overline{m^2}} \sin(2\theta_R) .
\ee

As we will discuss more in \Sec{sec:reach}, a high-energy muon collider can study the flavor violation in the MSSM via slepton pair production with searches for the flavor-violating decays~\cite{Arkani-Hamed:1996bxi, Krasnikov:1995qq}.
As we will see, the three parameters above---along with the Bino mass, $M_1$---suffice to completely characterize the reach.
Before turning to the collider reach, however, we first review the low-energy signatures of LFV. 
We will see that these same four parameters largely govern the lepton flavor violating signatures in muon decays and $\mu$-to-$e$ transitions as well.

\subsection{Low-Energy Signatures of Lepton Flavor Violation}
\label{subsec:precision}

In this section, we discuss the constraints from low-energy precision measurements of muons. 
In particular, we'll focus on the constraints from $\mu \to e\gamma$, $\mu \to 3e$ and $\mu$-to-$e$ conversion in atomic nuclei, since these arise without any field content beyond the mixed right-handed sleptons and the Bino that we will also consider for the collider reach.
Beyond this minimal scenario, additional constraints from flavor-violating mixings with the stau and searches for the electron electric dipole moment can arise, but these depend on additional parameters, so we delay a discussion of them until \Sec{sec:nonminimal}.

There is a large literature on the rate of charged lepton flavor violation processes in supersymmetric theories, too large to review here; key papers and useful entry points to this literature include~\cite{Hisano:1995cp, Kitano:2002mt, Masina:2002mv, Arganda:2005ji, Arganda:2007jw, Altmannshofer:2013lfa, Moroi:2013sfa, Ellis:2016yje, Calibbi:2017uvl, Crivellin:2018mqz}. Here, we will only extract some results in particular limits and regions of parameter space that are of interest for our work.

\begin{figure}[h]
\centering
\includegraphics[width = 0.55\textwidth]{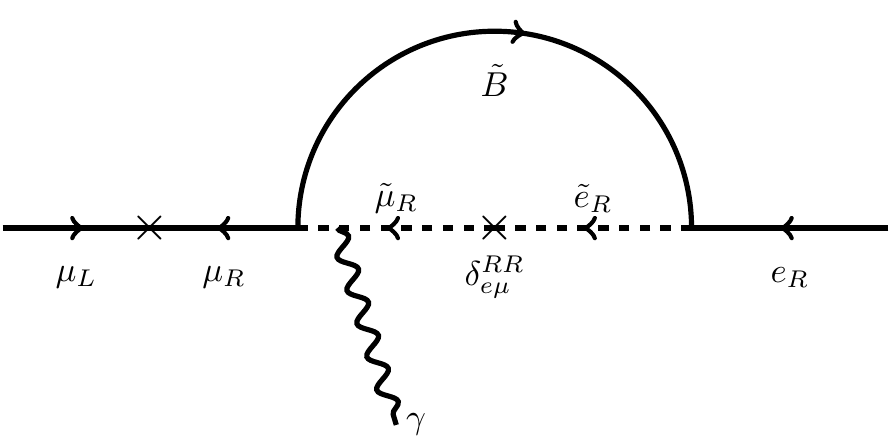}
\caption{The minimal one-loop contribution to the $\mu \to e\gamma$ dipole associated with the flavor-violating slepton collider signal of interest. The right-handed selectron and smuon are both relatively light, and contribute to $\mu \to e\gamma$ through a loop involving a bino and the off-diagonal slepton mass matrix contribution $\delta_{e\mu}^{RR}$.}
\label{fig:lfv_dipole}
\end{figure}

In the situations of interest to us, the dominant contributions to CLFV processes are from contributions to dipole operators.
Given a general dipole operator below the electroweak scale,
\be
{\cal L} \supset \mathcal{A}^R_{ij}\, {\bar \ell}_i \sigma^{\mu \nu} P_R \ell_j F_{\mu \nu} + \mathrm{h.c.},
\ee
the $\mu \to e\gamma$ decay rate is given by
\be
\mathrm{Br}(\mu \to e\gamma) = \frac{48 \pi^3 \alpha}{m_\mu^2 G_F^2} \left(|\mathcal{A}^R_{e\mu}|^2 +|\mathcal{A}^R_{\mu  e}|^2\right).
\ee
The minimal set of ingredients we consider already leads to a contribution to the dipole operator above, as illustrated in Fig.~\ref{fig:lfv_dipole}.
This diagram has a mass insertion on the external muon line. Because the diagram is not 1LPI, one way to interpret this is that integrating out the sleptons and bino generates an operator of the form ${\bar e}_{j}^\dagger {\overline \sigma}^\mu \overset{\leftrightarrow}{D^\nu} {\bar e}_{i} B_{\mu \nu}$, which becomes the more familiar dipole operator $L_j {\overline\sigma}^{\mu \nu}{\bar e}_i B_{\mu \nu}$ only after applying the equation of motion for the lepton fields.

The mixing of ${\tilde e}_R$ and ${\tilde \mu}_R$ leads to two slepton mass eigenstates ${\tilde \ell}_{1,2}$. Then the dipole amplitude in this  case is
\be 
\mathcal{A}^R_{\mu e} = m_\mu \frac{\alpha_Y}{8\pi} \sin(2\theta_R)\left[\frac{1}{m^2_{{\tilde \ell}_1}} {\bar A}\left(\frac{|M_1|^2}{m^2_{{\tilde \ell}_1}}\right)-\frac{1}{m^2_{{\tilde \ell}_2}} {\bar A}\left(\frac{|M_1|^2}{m^2_{{\tilde \ell}_2}}\right) \right],
\ee
where the loop function is defined as (in notation following~\cite{Aloni:2021wzk}; see also~\cite{Hisano:1995cp,Crivellin:2018qmi})
\be
{\bar A}(r) = \frac{2r^3 + 3 r^2- 6 r + 1 - 6 r^2 \log(r)}{6(1-r)^4}.
\ee
In the case that the sleptons are very nearly degenerate, the mixing angle $\theta_R$ can be large. Nonetheless, the $\mu \to e\gamma$ rate is suppressed by the approximate degeneracy, because the two terms inside the brackets nearly cancel. Indeed, at the beginning of the section we discussed that flavor violation is controlled by the product $\Delta m^2 \sin(2\theta_R)$. Expanding in small $\Delta m^2$, we find
\be
\label{eq:dipole_minimal_mia}
\mathcal{A}^R_{\mu e} \approx m_\mu \frac{\alpha_Y}{\pi} \frac{\delta^{RR}_{\mu e}}{\overline{m^2}} f_{1n}\left(\frac{|M_1|^2}{\overline{m^2}}\right),
\ee
where 
\be
f_{1n}(r) = \frac{-1 + 9 r + 9 r^2 - 17 r^3 + 6  r^2(r+3) \log(r)}{24 (r-1)^5}.
\ee
This latter result appears in the literature via the mass insertion  approximation; see, e.g.,~\cite{Calibbi:2017uvl}. These results are related via $(r {\bar A}(r))' = 4 f_{1n}(r)$. 

Thus, even when the right-handed selectron and smuon are heavily mixed, so that each slepton mass eigenstate has an $O(1)$ probability of decaying to either an electron or a muon, the $\mu \to e\gamma$ rate is still suppressed due to the mass degeneracy between the states. The mass insertion approximation continues to be valid,  as reflected in the smallness of $\delta^{RR}_{\mu e}$, despite the large mixing angle. This is a consequence of a super-GIM mechanism: in the limit of degenerate masses, the total rate is proportional to $\sum_i U_{1i} U_{2i}^*$, where $U_{ji}$ denotes the matrix element between flavor eigenstate $j$ and mass eigenstate $i$. By unitarity, this sum is zero. As a result, when flavor mixing is very large due to a near-degeneracy in masses, constraints from dedicated CLFV experiments become weaker, and relatively light new states are allowed.

The current limit $\BR(\mu\to e\gamma) < 4.2 \times 10^{-13}$ ($90\%$ C.L.) was set by the MEG experiment~\cite{TheMEG:2016wtm}. The upgrade MEG-II will improve this bound to $< 6\times 10^{-14}$~\cite{Baldini:2018nnn}. The constraints on slepton mixing in our scenario from this constraint will be shown in \Sec{sec:reach}, along with constraints from other lepton flavor violating processes to which we now turn.

\medskip

The minimal dipole contribution in \Eq{eq:dipole_minimal_mia} also contributes to the $\mu \to 3e$ decay rate.
In the situations we are interested here, the dipole contribution dominates over box and penguin diagrams, and the branching ratio can be reliably estimated as~\cite{Hisano:1995cp, Arganda:2005ji}
\be
\BR(\mu \to 3e) = \frac{\alpha}{3\pi} \Big( \log\frac{m_{\mu}^2}{m_e^2} - \frac{11}{4} \Big)\, \BR(\mu \to e\gamma)
 \simeq \frac{1}{162}\, \BR(\mu \to e\gamma) .
\ee
The current bound on $\mu \to 3e$ decays comes from the SINDRUM experiment~\cite{Bellgardt:1987du}, which set a limit $\BR(\mu \to 3e) < 1.09 \times 10^{-12}$.
This constraint will significantly improve, to $5\times 10^{-15}$ with the first phase of the Mu3e experiment~\cite{Mu3e:2020gyw}, with further upgrades expected to push the sensitivity to $< 10^{-16}$ in the future.

The dipole operator can also dominate the contributions to the $\mu$-to-$e$ conversion in nuclei, particularly if the squarks are much heavier than the sleptons.
Assuming dipole dominance, the conversion rate in a nucleus $N$ takes the simple form~\cite{Kitano:2002mt, Ellis:2016yje}
\be
\bga
\textrm{CR}(\mu \to e)_{N} = \BR(\mu\to e\gamma) \times 
 \begin{cases}
\frac{\alpha}{3} & N = \textrm{Al}, \\
\frac{\alpha}{2} & N = \textrm{Au}.
\end{cases}
\eda
\ee
The best constraints available now are from SINDRUM~II, 
which set a limit on the conversion rate in $\textrm{Au}$ at $< 7.0 \times 10^{-13}$~\cite{Bertl:2006up}.
There are a number of upcoming experiments that are expected to improve these bounds significantly using aluminum nuclei: COMET is projected to set a limit of $7 \times 10^{-15}$ ($2.6 \times 10^{-17}$) in Phase-I (Phase-II)~\cite{Adamov:2018vin, Angelique:2018svf}, while Mu2e projects an eventual Phase-II sensitivity of $2.5 \times 10^{-18}$~\cite{Bartoszek:2014mya, Abusalma:2018xem}. A future experiment, PRISM/PRIME, aims to eventually push this sensitivity to the $\sim 10^{-19}$ level~\cite{Kuno:2005mm, Pasternak:2010zz, Baldini:2018uhj}.

Thus far, we have completely ignored the effects of mixing with the stau. 
As emphasized in~\cite{Altmannshofer:2009ne} and further in~\cite{Moroi:2013sfa, McKeen:2013dma, Altmannshofer:2013lfa}, the SUSY contributions to dipole operators in theories with flavor violation can be dominated by mixing with the third generation, due to the much larger Yukawa couplings that are then accessible. We will discuss these effects, along with analogous contributions to the electron electric dipole moment, in \Sec{sec:nonminimal}.

\section{The MSSM at a High-Energy Muon Collider}
\label{sec:mssm_at_muc}

In \Sec{sec:reach}, we will discuss how the physics leading to the low-energy LFV signals described in the previous section can be studied directly at a high-energy muon collider. First, it is worth reviewing the prospects of a high-energy muon collider in more generality---in particular the capabilities in searching for superpartners.

A high-energy muon collider combines the advantages of proton and electron-positron machines.
Because of their larger mass, muons produce less synchrotron radiation than electrons and positrons, so they can be readily accelerated to TeV-scale energies in relatively compact circular accelerators. 
At the same time, the muons are not composite objects like protons, so they can utilize the full center of mass energy in their collisions to produce heavy new states. 
Because the colliding objects are second-generation fermions themselves, muon colliders have a particular aptitude for testing physics with generation-specific couplings~\cite{Capdevilla:2020qel, Buttazzo:2020ibd, Yin:2020afe, Capdevilla:2021rwo, Chen:2021rnl, Huang:2021biu, Li:2021lnz, Asadi:2021gah, Haghighat:2021djz, Bandyopadhyay:2021pld, Dermisek:2021mhi, Capdevilla:2021kcf, Medina:2021ram}. 
Finally, high-energy muons have a large probability to radiate collinear gauge bosons in their collisions~\cite{Chen:2016wkt, Han:2020uid, Han:2021kes, AlAli:2021let, Ruiz:2021tdt}, and can thus act efficiently as ``vector boson colliders'', with large rates for VBF processes that can be used both to search for new physics~\cite{Eichten:2013ckl, Buttazzo:2018qqp, Chakrabarty:2014pja, Bandyopadhyay:2020otm, Liu:2021jyc, Han:2021udl, Liu:2021akf, Chen:2022msz, Bao:2022onq} and to make precision measurements of the electroweak sector~\cite{DiLuzio:2018jwd, Buttazzo:2020uzc, Chiesa:2020awd, Costantini:2020stv, Han:2020pif, Chiesa:2021qpr, Cepeda:2021rql, Chen:2021pqi, Spor:2022mxl, Buonincontri:2022ylv}. That the muons are color-neutral, and produce far fewer hadronic backgrounds, makes them particularly well-suited for this purpose.

These advantages come at the cost of working with unstable particles, which are more challenging to produce with high-intensities while maintaining a high-quality, low-emittance beam~\cite{Neuffer:1983jr}. 
The decays of the muons also lead to a large ``beam-induced background'' (BIB), which must be effectively mitigated by a combination of shielding and precision timing in the detectors~\cite{Ally:2022rgk}.
Recent advances have begun demonstrating that these challenges can be overcome~\cite{Antonelli:2015nla, MICE:2019jkl, Jindariani:2022gxj, Bartosik:2022ctn}, however, coinciding with a surge of interest in the physics potential of a high-energy muon collider facility~\cite{Long:2020wfp, Franceschini:2021aqd, Aime:2022flm, DeBlas:2022wxr, Cesarotti:2022ttv}.

In what follows, we will consider muon colliders with center of mass energies of $3$, $10$ and $30\,\textrm{TeV}$, with corresponding integrated luminosities of $1$, $10$, and $10\,\textrm{ab}^{-1}$. As for the detectors, we will only assume coverage up to $|\eta| < 2.5$, and require that charged particles have $p_T \geq 25\,\textrm{GeV}$, and otherwise ignore any reconstruction efficiencies or secondary effects of the BIB. As we are primarily interested in signals involving high-energy charged leptons, we do not expect details of the detector design to significantly affect any of our results.

The direct reach for superpartners at high-energy muon colliders was explored in detail in ref.~\cite{AlAli:2021let}. 
There, the rates for a number of simplified model production processes were estimated and some of the leading candidates for discovery signatures were discussed, particularly in relation to the hierarchy problem.
The prospects for discovering superpartners were also studied in the context of WIMP candidates in refs.~\cite{Han:2020uak,Capdevilla:2021fmj, Bottaro:2021snn}. In particular, \cite{Capdevilla:2021fmj} provides a proof-of-concept that a disappearing track signature can be searched for even in the presence of heavy contamination due to BIB.

In general, any superpartner with electroweak quantum numbers, such as the sleptons of interest to us here, will be pair-produced in the $s$-channel with a rate that falls like $1/s$, but is still at the $O(\text{few tens})~\textrm{ab}$ level, even at a $30\,\textrm{TeV}$ collider.
As the superpartner mass approaches $\sqrt{s} / 2$, the $s$-channel cross section scales like the velocity $\beta$ for fermionic superpartners and like $\beta^3$ for scalar ones. This leads to roughly an order of magnitude drop in the cross section for scalars with masses $\sim 0.9 \times \sqrt{s} / 2$, but with sufficient luminosity the reach for electroweak particles can extend all the way up to half the center of mass energy.

In summary, a high-energy muon collider would have the reach to discover any new charged superpartners if they have masses less than approximately half the center of mass energy. In the scenarios we have discussed in \Sec{sec:mssm_clfv}, this means that the first signals of supersymmetry could be seen in the pair production of sleptons, perhaps in tandem with signatures from low-energy experiments. The results in the next subsection, and in \Sec{sec:reach}, will demonstrate that beyond discovering the new states, a muon collider can measure a number of properties of the spectrum and deliver interesting lessons about the nature of the underlying supersymmetric theory.

\subsection{Measurement of the Slepton and Neutralino Masses}
\label{subsec:mass_measurement}

As we will discuss in detail in Section~\ref{sec:reach}, knowledge of the slepton and neutralino masses can be used to suppress the backgrounds to flavor-violating processes, allowing for more precise studies of the SUSY spectrum.
It is thus worthwhile to understand how well these masses can be measured at a muon collider, assuming that sleptons and neutralinos are within the collider reach, and will be discovered via flavor-conserving channels. 
The measurement of superpartner masses at a lepton collider is a well-studied problem, and a number of detailed studies have been performed for $e^+e^-$ colliders with energies in the $250\,\textrm{GeV}$ -- $3\,\textrm{TeV}$ range~\cite{AguilarSaavedra:2001rg, Freitas:2004re, Conley:2010jk, Blaising:2012vd, Battaglia:2013bha}.

However, there are several important differences in performing this measurement at a high-energy muon collider.
Synchrotron radiation is highly suppressed relative to $e^+e^-$ beams of comparable energies, and as a result, beamsstrahlung effects can essentially be ignored. While initial state radiation exists, the collision energy is expected to be much more sharply defined than at, e.g., CLIC, and designs typically project a spread on the beam energy of $\delta E / E \sim 0.1\%$~\cite{Delahaye:2019omf, Neuffer:2018yof, Zimmermann:2018wfu}.
On the other hand, SM backgrounds at higher energy colliders are expected to be quite large, and the rates for vector boson fusion (VBF) processes especially are significantly larger than at lower energies.
This is compounded by the fact that collisions at a muon collider suffer from a large beam-induced background.
Methods of effectively mitigating the BIB are the subject of ongoing research, but a common proposal is to make use of ``shielding nozzles'' that cover the far-forward regions of the detector, suppressing the BIB at the cost of detector coverage~\cite{Ally:2022rgk}.
Such a scheme makes it difficult to veto events with forward-going leptons, which could otherwise greatly reduce the VBF background.
A threshold scan to determine the masses may be possible, but likely only for a small range of energies, making a shape-based determination more important.

In light of these differences, an updated investigation of the attainable precision is warranted. 
A comprehensive analysis requires not only a precise understanding of the detector performance, but also a careful treatment of the systematic uncertainties, both of which are beyond the scope of this work.
However, the problem at hand is primarily kinematic, and depends only on measuring the energies of charged leptons, so a reasonable estimate of the precision can be obtained with simple parton-level simulations of the relevant processes. These estimates will still be enlightening in understanding how the unique muon collider environment can be dealt with, and suffice for mitigating the backgrounds to the flavor-violating processes we are interested in.\footnote{A similar study has been performed in ref.~\cite{Freitas:2011ti}. Here, we consider a different set of cuts to mitigate the $\gamma\gamma$ induced background, and a different set of muon collider energies.}\\

As in the $e^+e^-$ collider case, the slepton and neutralino masses can be measured at a muon collider via the flavor-conserving pair production processes:
\begin{align}\label{eq:fc_processes}
\mu^+ \, \mu^- & \to \tilde{\mu}_R^+ \, \tilde{\mu}_R^- \to \mu^+ \, \mu^-\, \tilde{B} \, \tilde{B}, \nonumber\\
\mu^+ \, \mu^- & \to \tilde{e}_R^+ \, \tilde{e}_R^- \to e^+ \, e^- \, \tilde{B} \, \tilde{B},
\end{align}
by measurements of the final state leptons.\footnote{For the purposes of estimating the precision attainable on the masses, we will assume that the sleptons are flavor eigenstates, i.e., that it is the selectron mass we are measuring in the final states with $e^+ e^-$ and missing momentum, and the smuon mass in final states with $\mu^+\mu^-$. This is only strictly true in the limit $\sin2\theta_R \to 0$ (or $\Delta m^2 / \overline{m^2} \to 0$, when the masses are identical), but we evaluate the precision for all $\sin(2\theta_R)$, taking the effect on the signal cross section into account.}
Energy and momentum conservation dictates that the slepton energy equal the beam energy, and the lab-frame energy of the charged lepton decay product is therefore fully determined by the (rest frame) decay angle, $\theta_0$. 
The lepton energy is then found to be,
\begin{equation}
\label{eq:lepton_energy}
E_{\ell} = \frac{\sqrt{s}}{4} \bigg( 1 - \frac{ M_1^2 }{m_{\tilde{\ell}}^2} \bigg) \big( 1 + \beta \cos\theta_0 \big) .
\end{equation}
In the above, $\beta = \sqrt{1 - 4 m_{\tilde{\ell}^2} / s}$ is the velocity of the slepton in the lab frame. The lepton energies are distributed uniformly between the two limits of Eq.~\eqref{eq:lepton_energy} obtained when $\cos\theta_0 = 1$ ($\Emax$) and $\cos\theta_0 = -1$ ($\Emin$).
These endpoints, $\Emin$, $\Emax$, are therefore related to the slepton and neutralino masses via:
\begin{equation}\label{eq:mass_endpoint_relations}
m_{\tilde{\ell}}^2 = s\, \frac{\Emin \Emax}{(\Emin + \Emax)^2},
\quad
M_1^2 = m_{\tilde{\ell}}^2 \bigg( 1 - \frac{2(\Emin + \Emax)}{\sqrt{s}}\bigg) .
\end{equation}
The problem of measuring the slepton and neutralino masses then becomes simply the problem of measuring the endpoints of the slepton energy distribution in the presence of backgrounds and detector effects.

In practice, the endpoints of the spectrum (and the corresponding uncertainty) can be determined via an unbinned likelihood fit to the data. This requires a functional description of the expected shapes of the distribution, both for signal and background events. To determine the expected precision on the masses, we simulate a large sample of the signal and background events. Toy datasets with the expected number of events (after applying a simple set of cuts to mitigate the background) can then be generated, and the likelihood fit performed. Repeating this procedure for a large number of toys gives an estimate of the expected precision on the slepton and neutralino masses.
More details of this procedure are left to Appendix~\ref{app:mass_meas}, where we describe the simulation procedure, the selection cuts used to mitigate the backgrounds, the fitted shapes of the signal and background distributions, and the details of the likelihood analysis.

\begin{figure}[h]
\centering
\includegraphics[width=8cm]{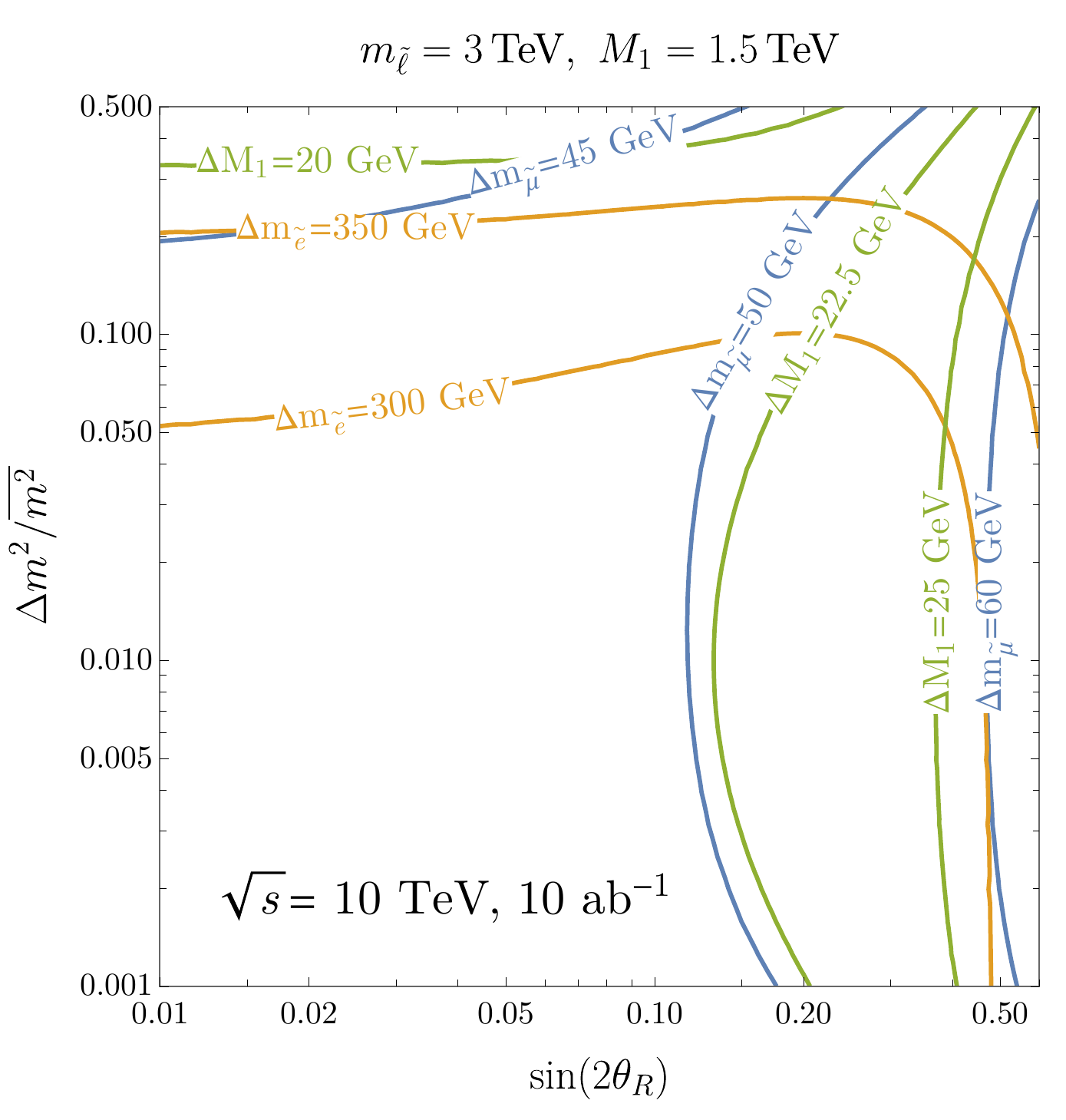}
~
\includegraphics[width=8cm]{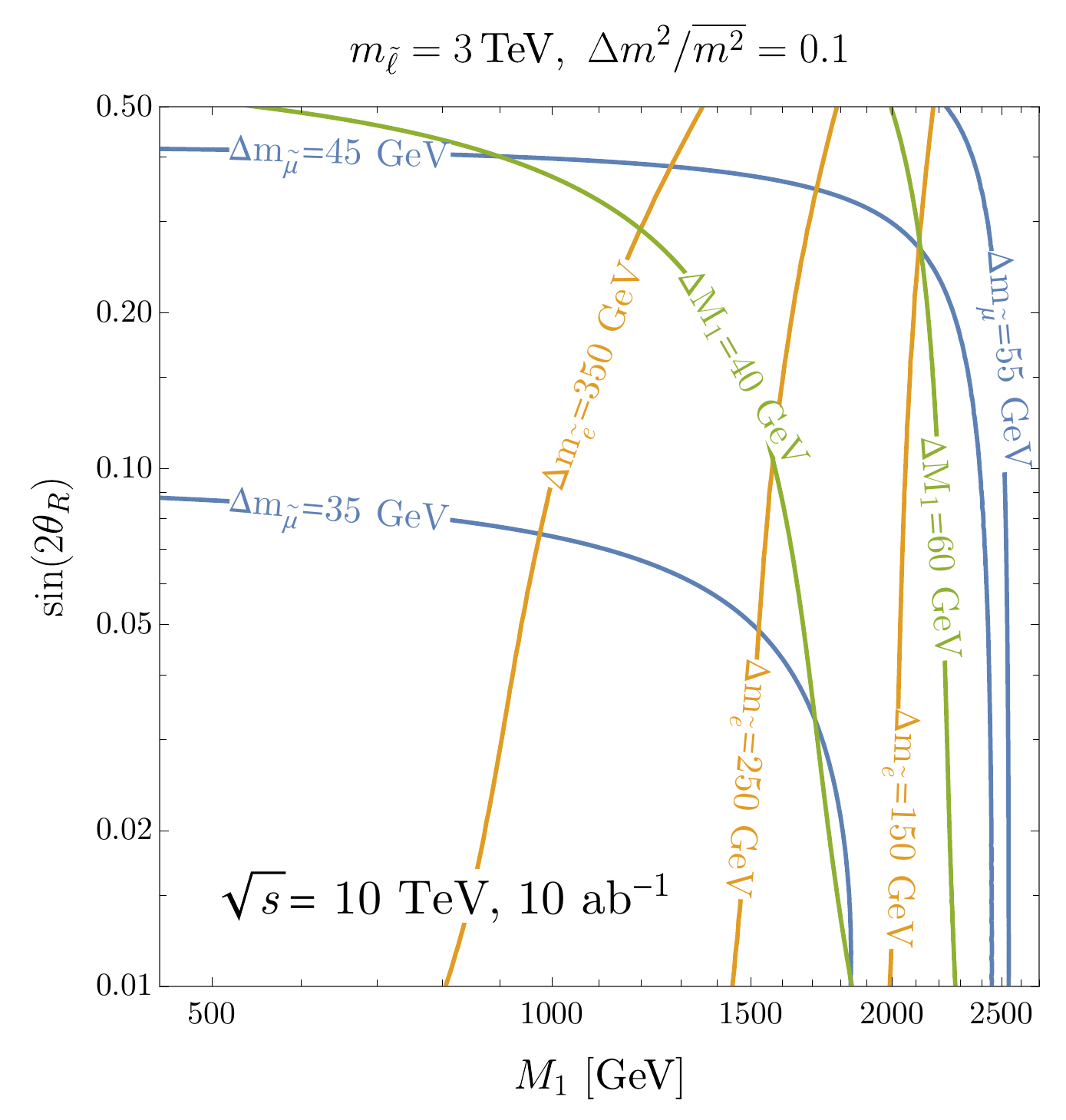}
\caption{
Contours of the expected precision on the selectron, smuon and neutralino masses as a function of $\sin2\theta_R$ and $\Delta m^2/\overline{m^2}$ for fixed $m_{\tilde{\ell}}$ and $M_1$ (left) and as a function of $M_1$ and $\sin2\theta_R$ for fixed $m_{\tilde{\ell}}$ and $\Delta m^2 / \overline{m^2}$ (right), all at $\sqrt{s} = 10\,\textrm{TeV}$, assuming $10\,\textrm{ab}^{-1}$ of data.}
\label{fig:mass_meas_summary}
\end{figure}

The results of the likelihood analysis for $\sqrt{s} = 10\,\textrm{TeV}$ are summarized in Fig.~\ref{fig:mass_meas_summary}.
We show contours of the expected precision (at 95\% C.L.) on the selectron and smuon masses ($m_{\tilde{\mu}}$ and $m_{\tilde{e}}$) and on the neutralino mass, $M_1$, in orange, blue, and green, respectively.
In the left panel, the precision is shown as a function of the slepton mixing angle, $\sin2\theta_R$ and mass splitting $\Delta m^2 / \overline{m^2}$, as defined in Eq.~\eqref{eq:mixing_defs}, with the average slepton mass fixed to $3\,\textrm{TeV}$ and $M_1 = 1.5\,\textrm{TeV}$.
The uncertainties on the smuon and neutralino masses vary from $\sim 45$ to $60\,\textrm{GeV}$ and $\sim 20$ to $25\,\textrm{GeV}$,
increasing slightly towards larger mixing angles where the $\mu^+\mu^- \tilde{B}\tilde{B}$ cross section decreases. Correspondingly, the rate for the $e^+e^-$ final state process increases, and the precision on the selectron varies from $\sim 350$ to $300\,\textrm{GeV}$.

The right panel shows the same contours but as a function of $\sin 2\theta_R$ and $M_1$, with the mean slepton mass $m_{\tilde{\ell}} = 3\,\textrm{TeV}$ and $\Delta m^2 / \overline{m^2} = 0.1$.
In this plane, the same behavior for large mixing angles is observed, while the precision on all the masses improves for small $M_1$ but gets rapidly worse as $M_1$ approaches $m_{\tilde{\ell}}$, where the cross section drops significantly due to the small phase space available for the leptons and the endpoints approach the kinematic limits of the detector, where they are harder to measure accurately.

The results for $\sqrt{s} = 3$ and $30\,\textrm{TeV}$ show very similar behavior, with the precision on the selectron, smuon, and neutralino varying from 25 to 50, 10 to 50 and 20 to $60\,\textrm{GeV}$, respectively at $3\,\textrm{TeV}$ and from 600 to 750, 200 to 350 and 100 to $200\,\textrm{GeV}$ at $30\,\textrm{TeV}$. We show the results here for the benchmark luminosities assumed above, but the precision on each of the masses scales like the square root of the integrated luminosity, absent any systematic effects that we have neglected.

\section{Direct Reach for Flavor-Violating Signatures}
\label{sec:reach}

\begin{figure}[h]
\centering
\includegraphics[width=0.7\textwidth]{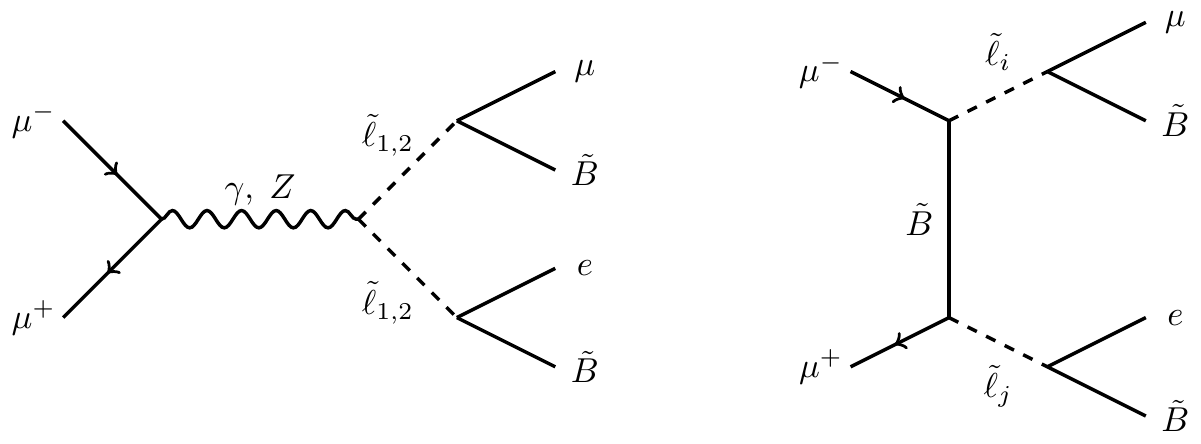}
\caption{Main production mechanism for the flavor-violating signal at the muon collider. For the $t$-channel process, $\tilde{\ell}_i$ and $\tilde{\ell}_j$ can be any of the two slepton mass eigenstates, $\tilde{\ell}_1$ and $\tilde{\ell}_2$.}\label{fig:sig_Feyn}
\end{figure}

Charged lepton flavor violation can be searched for in a high-energy muon collider through the direct production process $\mu^+ \mu^- \rightarrow \mu e + X$. In the simplified MSSM model of interest described in \Sec{sec:mssm_clfv}, the signal process is 
\be
\mu^ + \mu^- \rightarrow \mu e \tilde{B}\tilde{B},
\ee
where $\tilde{B}$ is the lightest neutralino that we have assumed to be a pure Bino which would appear as missing momentum. The dominant contribution to the signal process is pair production,
\be
\mu^ + \mu^- \rightarrow \tilde{\ell}_{i} \tilde{\ell}_{j} \rightarrow \mu e \tilde{B}\tilde{B},
\ee
as shown in \Fig{fig:sig_Feyn}, where $\tilde{\ell}_{i,j}$ are any of $\tilde{\ell}_1$ and $\tilde{\ell}_2$, the slepton mass eigenstates that are mixtures of the flavor eigenstates $\tilde{\mu}_R$ and $\tilde{e}_R$. The signal cross section is completely determined by the four parameters in the model, 
\be
\overline{m^2}, \; \; \frac{\Delta m^2}{\overline{m^2}}, \;\; \sin(2\theta_R),\;\; M_1,
\ee
as are the low-energy observables discussed in \Sec{subsec:precision}. This allows a direct comparison of the muon collider constraint with bounds coming from low-energy observables. The strong complementarity between the high-energy and low-energy experiments will be demonstrated in detail in the rest of this section.

There are two backgrounds to the signal process. The first is $\tilde{\tau}^+\tilde{\tau}^-$ pair production decaying into $\tau^+\tau^-\tilde{B}\tilde{B}$, with the pair of taus decaying leptonically to $\mu$-$e$ final states (the staus cannot directly decay into $\mu$ or $e$ because our minimal MSSM model outlined in \Sec{sec:mssm_clfv} assumes that flavor violation is confined in the $\mu$-$e$ sector).\footnote{Of course, in general all mixing parameters will be present, and one would need a more complicated global fit. Nonetheless, we believe our results are a useful starting point for calibrating expectations about the muon collider's capabilities.} But this background has a negligible cross section compared to the more dominant Standard Model background discussed in the next paragraph, even when the $\tilde{\tau}$ is light, because of the branching ratio suppression coming from demanding specific leptonic decay of the $\tau$s.

The second, larger background for the signal process is Standard Model pair production of $W^+ W^-$ that decay leptonically to $\mu$-$e$ final states. Even though the background process has a significantly larger cross section in the range of collider energies and model parameters we consider, the background can be rejected efficiently by reconstructing the neutralino kinematics, using the measurement of the slepton and neutralino masses as discussed in \Sec{subsec:mass_measurement}. The kinematic reconstruction procedure is discussed in detail in \App{app:reconstruction}. The key quantities from the reconstruction analysis are the ``reconstructed energies'' for the two neutralinos, which are the neutralino energies calculated without the $y$-component of their three-momenta (see \App{app:reconstruction} for the definition of the frame of reference and the formula in terms of visible particle momenta),
\be
\mathcal{E}_{\tilde{B}_a} \equiv \sqrt{p^2_{\tilde{B}_{a,x}}+p^2_{\tilde{B}_{a,z}}+m^2_{\tilde{B}}},\quad \mathcal{E}_{\tilde{B}_b} \equiv \sqrt{p^2_{\tilde{B}_{a,x}}+(p_{{\rm inv}}-p_{\tilde{B}_{a,z}})^2+m^2_{\tilde{B}}}.
\ee
Here $\{a,b\}$ labels the two neutralinos and their corresponding sleptons and leptons, i.e. $\tilde{B}_a$ and $\ell_a$ are produced from $\tilde{\ell}_a$ decay, and  $\tilde{B}_b$ and $\ell_b$ are produced from $\tilde{\ell}_b$ decay. The reconstructed energies put a lower bound on the true total energy carried by the neutralinos
\be
E_{\tilde{B}_a}+E_{\tilde{B}_b} \geq \mathcal{E}_{\tilde{B}_a}+ \mathcal{E}_{\tilde{B}_b}.
\ee
By energy conservation, $\mathcal{R} \equiv E_{\rm CM}-E_{\ell_a}-E_{\ell_b}-\mathcal{E}_{\tilde{B}_a}- \mathcal{E}_{\tilde{B}_b}$ must be positive. Since we do not know which of the two slepton mass eigenstates is involved in the particular signal event that we are reconstructing the kinematics for, we calculate $\mathcal{R}$ for all four possibilities of $\tilde{\ell}_a = \tilde{\ell}_{1,2}$ and $\tilde{\ell}_b = \tilde{\ell}_{1,2}$, take the largest value of $\mathcal{R}$ among the four cases as our final result, and demand that $\max\mathcal{R}>0$. While for signal events $\max\mathcal{R}>0$ is almost always true, background events rarely satisfy this criterion since the events are being reconstructed using the wrong particle masses. The precise efficiency varies with collider energy and particle masses, but in general signal events pass $\sim 98\%$ of the time, while background event efficiencies are around $\sim 0.2\%$. 

The reconstruction efficiency for the signal events becomes worse once we take into account the uncertainty in the measurements of slepton and neutralino masses found in \Sec{subsec:mass_measurement}. We will use the mass measurement uncertainty $\Delta m$ as an estimate of how far the measured mass is from the true mass,
\be
m_{{\rm measured}} = m_{{\rm true}}\pm \Delta m.
\ee
Specifically for the purpose of kinematic reconstruction, the sign of $\Delta m$ matters. For example, if we do kinematic reconstruction analysis with a slepton mass that is lower than the true value, what the analysis deems the ``correct'' phase space would be smaller than the true phase space spanned by the signal events, thus rejecting more signal events. Among the eight different combinations of $\Delta m$ signs for the three particle masses, $m_{\tilde{\ell}_1}$, $m_{\tilde{\ell}_2}$, and $M_1$, the following ``low-low-high'' combination lowers the signal reconstruction efficiency the most,
\be
m_{\tilde{\ell}_1{\rm measured}} = m_{\tilde{\ell}_1{\rm true}} - \Delta m_{\tilde{\ell}_1},\;\; m_{\tilde{\ell}_2{\rm measured}} = m_{\tilde{\ell}_2{\rm true}} - \Delta m_{\tilde{\ell}_2},\;\; m_{\tilde{B}{\rm measured}} = m_{\tilde{B}{\rm true}}+\Delta m_{\tilde{B}},
\ee
because the phase space that the reconstruction analysis considers correct is the smallest. The signal efficiency of the $\mathcal{R}>0$ cut in the ``low-low-high'' configuration strongly depends on the value of the true masses and the mass errors. For example, in the extreme case where the $m_{\tilde{B}{\rm measured}}$ becomes higher than $m_{\tilde{\ell}_{1,2}{\rm measured}}$, the signal efficiency would approach zero. In our study, the muon collider reach for lepton flavor-violating signal will be computed in two ways, first with the true particle masses in the reconstruction analysis, second with the ``low-low-high'' particle masses which will yield a weaker collider reach, capturing the uncertainty in collider reach due to the error in the mass measurements. Now, because the background distribution of $\mathcal{R}$ is much wider than the signal distribution, it is in principle possible to place the reconstruction cut at a negative value, $\max\mathcal{R} > -C$, to improve signal efficiency for the ``low-low-high'' configuration without passing many background events. We do not consider this optimization of the kinematic reconstruction cut in our study, and simply cut at $\max \mathcal{R} > 0$ for all events and mass configurations.

In the rest of the section, we will discuss in detail the direct reach of flavor-violating processes at muon colliders with $\sqrt{s} = 3$, 10, and 30 TeV. All signal and background events are simulated using \textsc{MadGraph5}~\cite{Alwall:2014hca}. For the signal events we use an extended MSSM UFO model that allows general flavor violation in the lepton sector, created using the \textsc{FeynRules} packages from \cite{Duhr:2011se}.

\begin{figure}[h]
\centering
\begin{subfigure}[b]{0.49\textwidth}
 \centering
 \includegraphics[width=\textwidth]{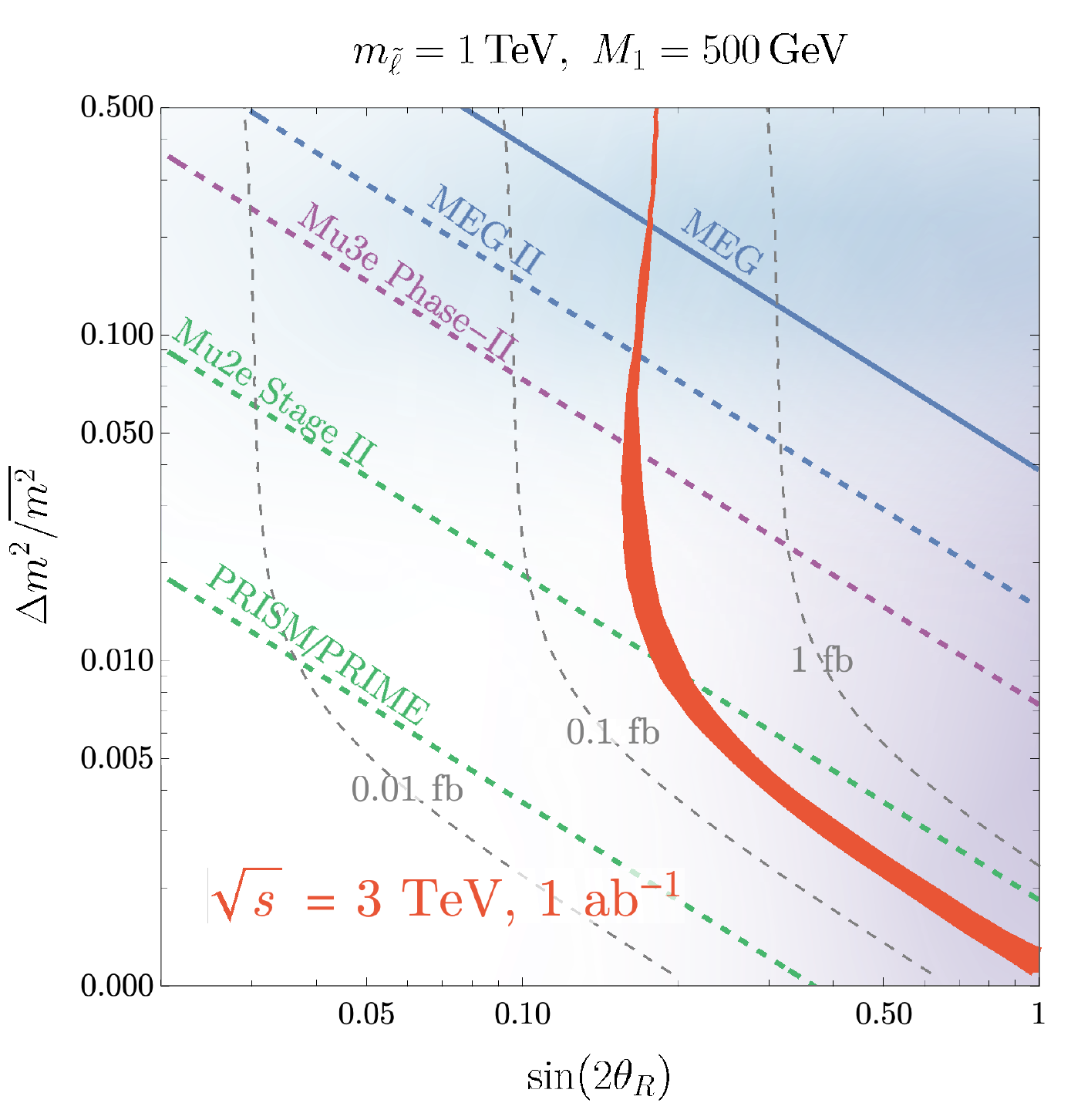}
\end{subfigure}
\hfill
\begin{subfigure}[b]{0.49\textwidth}
 \centering
 \includegraphics[width=\textwidth]{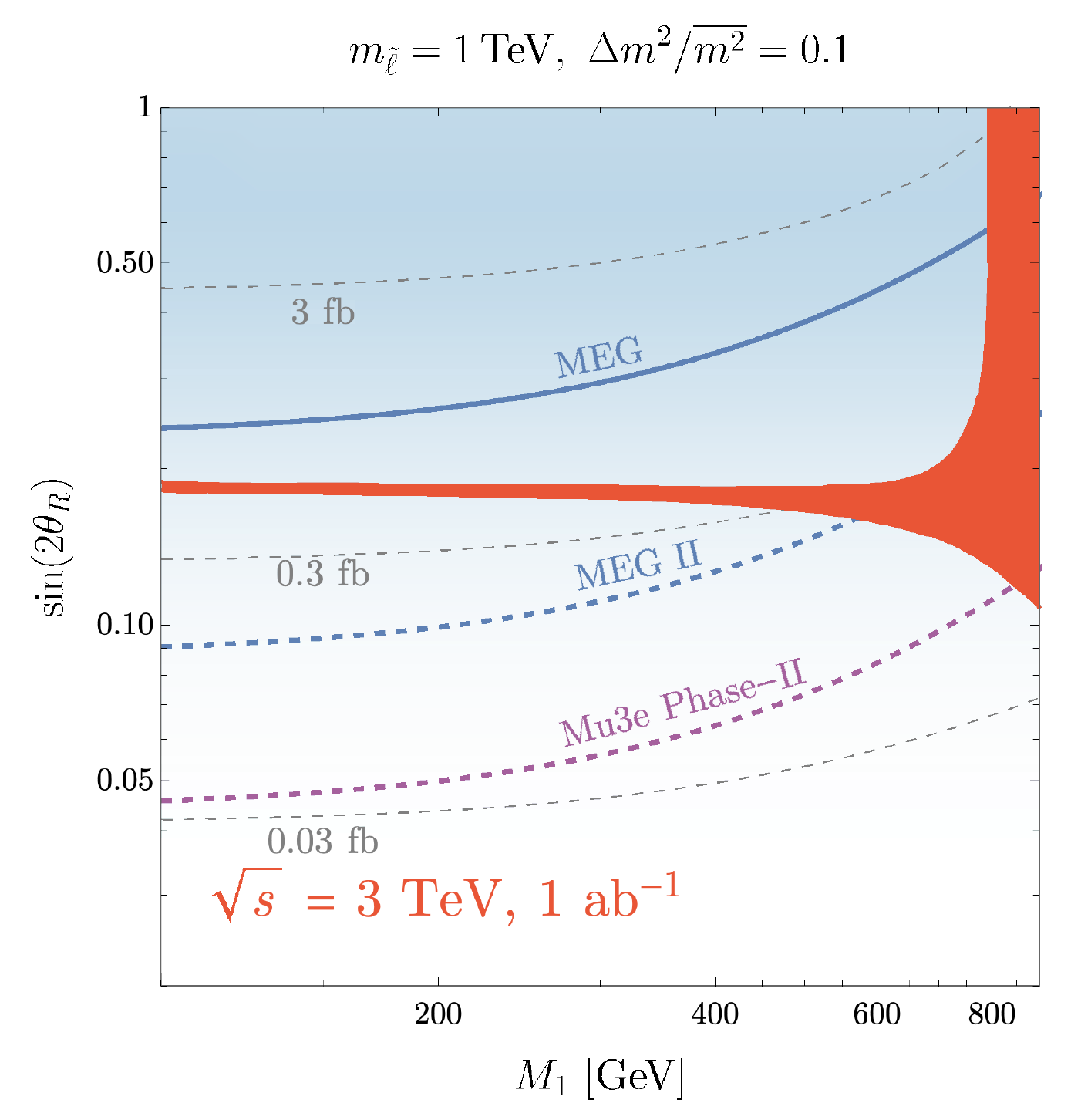}
\end{subfigure}
\hfill
\caption{5$\sigma$ discovery reach of the lepton-flavor violating MSSM by a 3 TeV muon collider with 1 ${\rm ab}^{-1}$ of data in the $\Delta m^2 /\overline{m^2}$ vs $\sin 2\theta_R$ plane and the $\sin 2\theta_R$ vs $M_1$ plane (red band). The width of the collider reach band shows the uncertainty due to inexact slepton and neutralino mass measurements. The gray contours are cross sections of the signal process. Both plots assume a mean slepton mass of 1 TeV. In the left plot we fix the neutralino mass $M_1 = 500$ GeV, while in the right figure $\Delta m^2/\overline{m^2}$ is fixed to 0.1. The current (solid) and expected (dashed) limits from low-energy lepton flavor violation experiments are indicated by the blue, purple and green lines. The purple and blue shaded lightly shaded regions indicate parameters preferred in Gauge-Mediated Supersymmetry Breaking scenarios and flavor-dependent mediator scenarios, respectively (as described in Appendix~\ref{app:flavor}).}
\label{fig:collider_3TeV}
\end{figure}

\begin{figure}[h]
\centering
\includegraphics[width=0.8\textwidth]{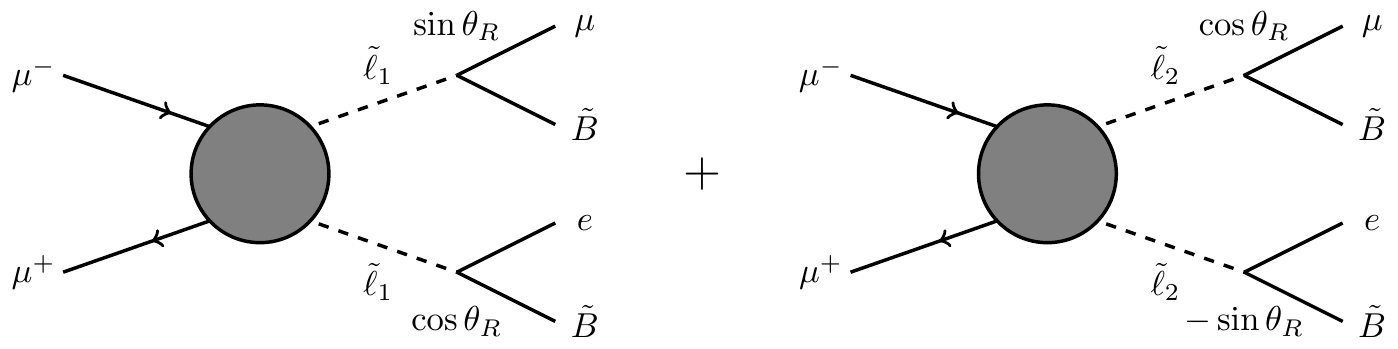}
\caption{When mass splitting of the two mass eigenstates are small, interference between the two Feynman diagrams suppresses signal cross section due to the opposite coupling.}\label{fig:Feyn_interference}	
\end{figure}

Fig.~\ref{fig:collider_3TeV} shows the $5\sigma$ reach of a 3 TeV muon collider with 1 ${\rm ab}^{-1}$ of data, assuming a mean slepton mass of $\overline{m^2}$ = (1 TeV$)^2$. The remaining three-dimensional parameter space of the model is sliced in two different ways: in the $\Delta m^2 /\overline{m^2}$ vs $\sin 2\theta_R$ plane where we assume $M_1$ = 500 GeV and in the $\sin 2\theta_R$ vs $M_1$ plane where we assume $\Delta m^2 /\overline{m^2}$ = 0.1. In both planes, the reach in $\sin 2\theta_R$ with fixed $\Delta m^2 /\overline{m^2}$ or fixed $M_1$ will improve by a facor of $10^{1/4}$ if collider luminosity increases by a factor of 10, since the signal cross section is proportional to $(\sin 2 \theta_R)^2$. In the $\Delta m^2 /\overline{m^2}$ vs $\sin 2\theta_R$ plane, we see that the collider reach deteriorates quickly in $\sin2\theta_R$ as the mass splitting becomes small. This suppression was pointed out in ref.~\cite{Arkani-Hamed:1996bxi} as a consequence of a quantum interference effect between the two mass eigenstates contained in the flavor eigenstates. In particular, the probability of the gauge eigenstate selectron, $\tilde{e}$, decaying into a final state containing a muon, $f_{\mu}$, is calculated as the product of the time-integrated probability of the selectron oscillating into a smuon gauge eigenstate, $\tilde{\mu}$, and the probability of a smuon decaying into a final state containing a muon. The probability $P(\tilde{e}\rightarrow f_{\mu})$ is found to be\footnote{The $\bar{m}^2$ in this formula is the square of the average, $\bar{m}^2 = (\frac{1}{2}(m_{\tilde{\ell}_1}+m_{\tilde{\ell}_2}))^2$, which naturally appears since the probability is due to interference between two mass eigenstates that evolve with different energies, $E = m_{\tilde{\ell}_{1,2}}$. Elsewhere in the paper we refer to the average of the square, $\overline{m^2} = \frac{1}{2}(m_{\tilde{\ell}_1}^2+m_{\tilde{\ell}_2}^2)$. In the limit of small mass splitting, the two values approach each other, $\overline{m}^2 - \bar{m}^2 = \frac{1}{4}\Delta m^2$.}
\be
\begin{split}
P(\tilde{e}\rightarrow f_{\mu}) &= P(\tilde{e}\rightarrow \tilde{\mu}){\rm Br}(\tilde{\mu}\rightarrow f_{\mu})\\
&= \frac{\int_0^{\infty}\dd t |\braket{\tilde{\mu}}{\tilde{e}(t)}|^2}{\int_0^{\infty}\dd t \braket{\tilde{e}(t)}{\tilde{e}(t)}}{\rm Br}(\tilde{\mu}\rightarrow f_{\mu})\\
&= \frac{1}{2}\sin 2\theta_R \frac{\left(\Delta m^2\right)^2}{4 \bar{m}^2\Gamma^2 + \left(\Delta m^2\right)^2}{\rm Br}(\tilde{\mu}\rightarrow f_{\mu}),
\end{split}
\ee
where $\Gamma$ is the widths of the two mass eigenstates which are assumed to be equal, and $\tilde{e}(t)$ is the time-evolved wavefunction of a selectron eigenstate prepared at $t = 0$. $\tilde{e}(t)$ is a mixture of selectron and smuon eigenstates due to the oscillation effect, and the oscillation effect vanishes as the mass splitting between the two eigenstates approaches zero. This interference effect can also be understood from a Feynman diagram perspective without using the language of flavor oscillations. When the two mass eigenstates have similar masses, the two pair production diagrams shown in \Fig{fig:Feyn_interference}, which are the dominant contributions to the flavor-violating process $\mu^+\mu^-\rightarrow \mu e \tilde{B}\tilde{B}$, cancel due to the opposite sign of the flavor-violating coupling, $\pm \sin \theta_R$. The narrow-width approximation is breaking down here because the mass splitting between the two intermediate states becomes smaller than their individual widths, and the interference effect cannot be neglected.

The width of the collider reach contour indicates the uncertainty coming from error in slepton and neutralino mass measurements. In the $\Delta m^2 /\overline{m^2}$ vs $\sin 2\theta_R$ plane, the uncertainty varies mildly since the mass measurement errors are relatively small compared to the available phase space. In the $\sin 2\theta_R$ vs $M_1$ plane, as $M_1$ becomes closer to $m_{\tilde{\ell}_1}$ and $m_{\tilde{\ell}_2}$, the available phase space for $e$/$\mu$ final states becomes smaller. When the measured masses are the same as the true masses, the reconstruction efficiency of the signal events is roughly constant as $M_1$ increases, but background events have smaller reconstruction efficiency since it is more difficult for the event kinematics to land within the small phase space. Thus the best possible collider reach (lower edge of the red band) is stronger as $M_1$ increases. On the other hand, the reach in the ``low-low-high'' configuration (upper edge of the red band) defined earlier in the section deteriorates quickly as $M_1$ increases, because the smuon and bino mass measurement uncertainties increase quickly with $M_1$, as shown in Fig.~\ref{fig:mass_meas_summary}. At around $M_1 \approx 750$ GeV, the mass measurement errors are so large such that
\be
m_{\tilde{\ell}_{1,2}{\rm measured}} \approx  M_{1{\rm measured}},
\ee
causing the kinematic reconstruction analysis efficiency to approach zero for the signal events. Therefore the width of the collider reach diverges at $M_1 \gtrsim 750$ GeV, as shown in the right panel of Fig.~\ref{fig:collider_3TeV}.

\begin{figure}
\centering
\begin{subfigure}[b]{0.49\textwidth}
 \centering
 \includegraphics[width=\textwidth]{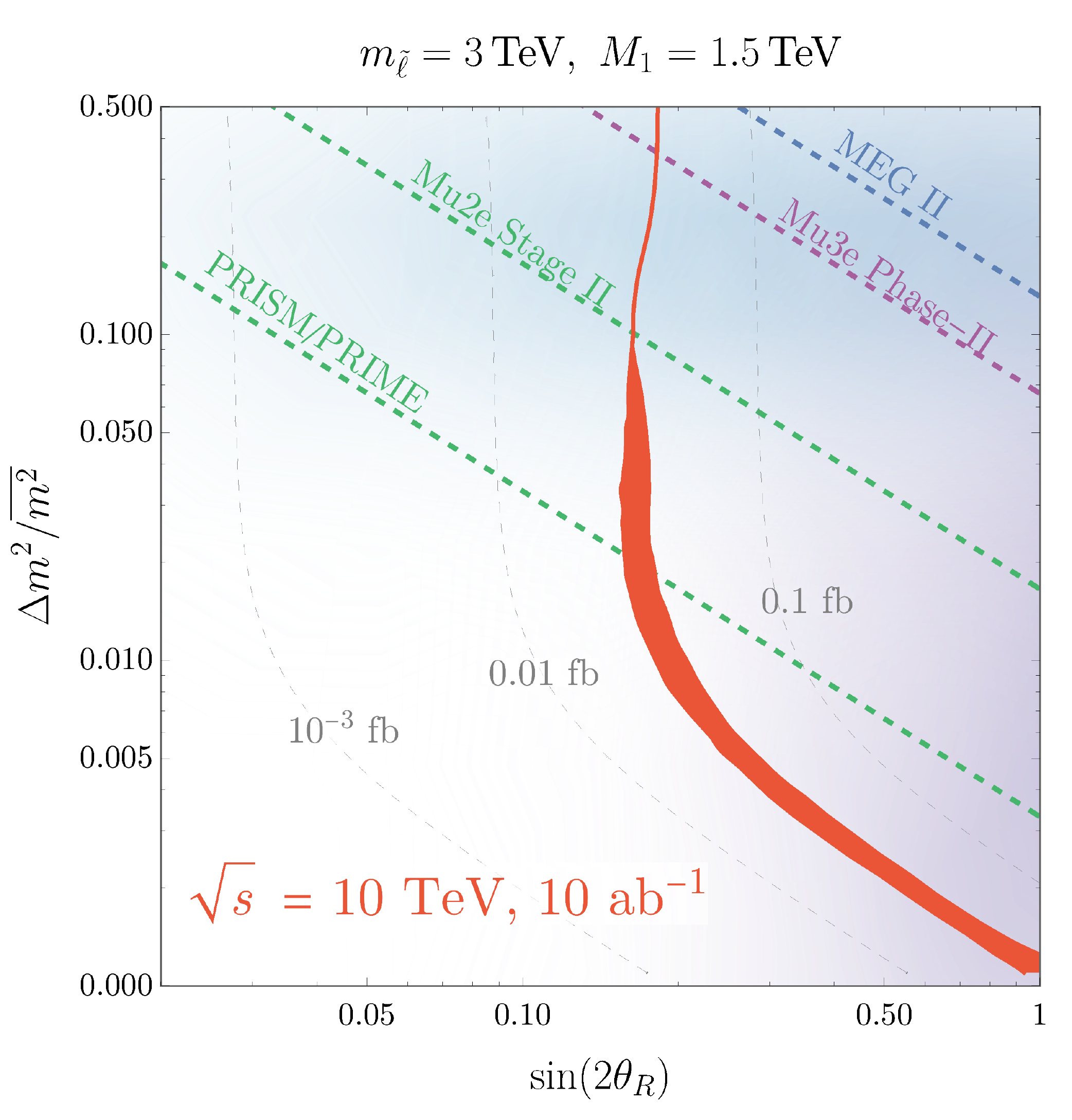}
\end{subfigure}
\hfill
\begin{subfigure}[b]{0.49\textwidth}
 \centering
 \includegraphics[width=\textwidth]{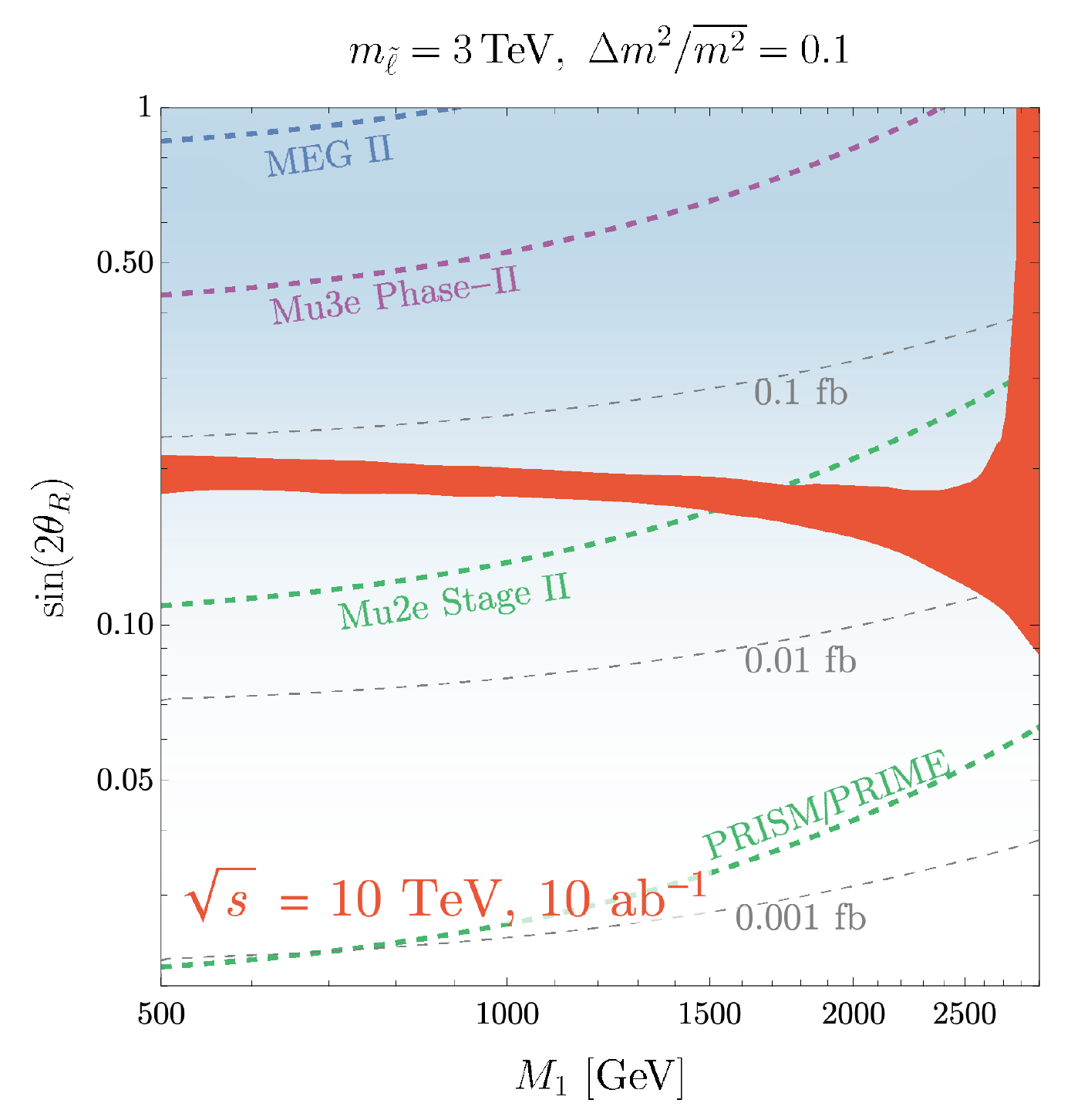}
\end{subfigure}
\begin{subfigure}[b]{0.49\textwidth}
 \centering
 \includegraphics[width=\textwidth]{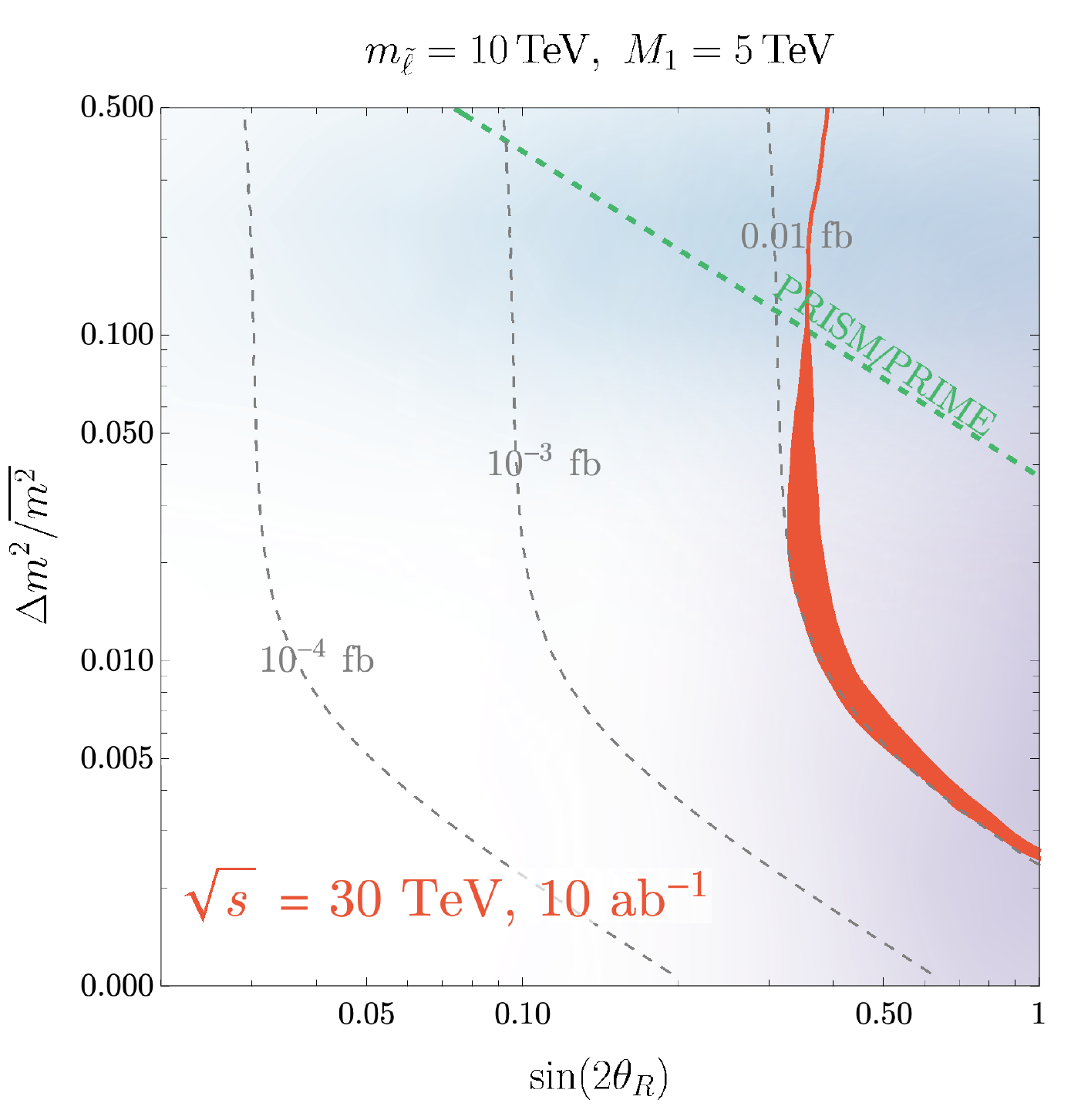}
\end{subfigure}
\hfill
\begin{subfigure}[b]{0.49\textwidth}
 \centering
 \includegraphics[width=\textwidth]{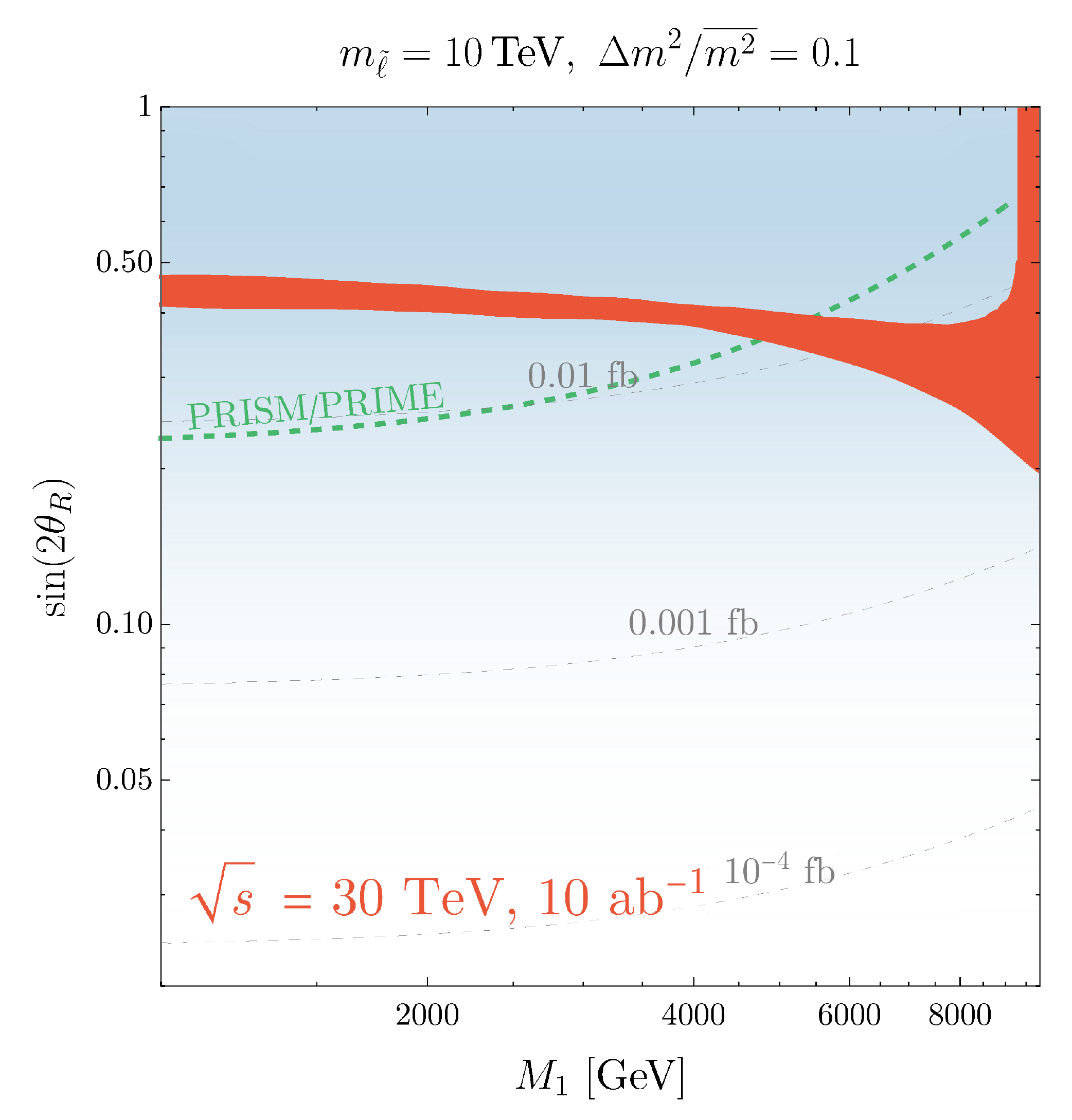}
\end{subfigure}
\hfill
\caption{5$\sigma$ discovery reach of lepton-flavor violating MSSM from a 10 TeV (top) or 30 TeV (bottom) muon collider, both with 1 ${\rm ab}^{-1}$ of data, in $\Delta m^2 /\overline{m^2}$ vs $\sin 2\theta_R$ plane and $\sin 2\theta_R$ vs $M_1$ plane (red band). The width of the collider reach band shows the uncertainty due to inexact slepton and neutralino mass measurements. The gray contours are cross sections of the signal process. The expected limits from low-energy lepton flavor violation experiments are indicated by the blue, purple and green dashed lines. The purple and blue shaded lightly shaded regions indicate parameters preferred in Gauge-Mediated Supersymmetry Breaking scenarios and flavor-dependent mediator scenarios, respectively.}
\label{fig:collider_higherTeV}
\end{figure}

The straight lines in blue, purple, and green in Fig.~\ref{fig:collider_3TeV} shows current (solid) and projected (dashed) constraints coming from low-energy experiments. The contours show a great complementarity between the high energy muon collider and low-energy experiments. A 3 TeV muon collider would produce competitive bounds with the projected Mu2e Stage II experiment. Furthermore, the complementary slicing of the parameter space by the muon collider and the low-energy experiments helps elucidate the underlying mechanism of lepton flavor violation when a flavor-violating signal is discovered at one of the experiments. 

Lastly, the lightly shaded purple and blue regions in Fig.~\ref{fig:collider_3TeV} indicate motivated parameter regions from different UV origins of the flavor mixing. Large mixing angles are motivated in models involving gauge-mediated supersymmetry breaking (GMSB), indicated by the purple region, while larger mass splittings are motivated in scenarios where the messengers carry flavor-dependent charges, such as models where $L_{\mu}-L_{\tau}$ is gauged, indicated by the blue regions. We see that a 3 TeV muon collider will constrain a significant fraction of the motivated parameter ranges.

Fig.~\ref{fig:collider_higherTeV} shows the $5\sigma$ reach of a 10 TeV and 30 TeV muon collider with 10  ${\rm ab}^{-1}$ of data, assuming a mean slepton mass of $\overline{m^2}$ = (3 TeV$)^2$ and (10 TeV$)^2$, respectively. Many qualitative features of the collider reach contours are similar to the 3 TeV case. We again observe the interference effect which deteriorates the collider reach at small mass splitting in the $\Delta m^2 /\overline{m^2}$ vs $\sin 2\theta_R$ plane, and the divergence of the uncertainty width with $M_1$ in the $\sin 2\theta_R$ vs $M_1$ plane. The strength of the high energy muon collider compared to low-energy experiments is more evident at higher energies. We see that a 10 TeV muon collider will probe parameter space unreached by the final stage PRISM/PRIME experiment.

\section{Beyond the Minimal Spectrum: LFV, EDM and Collider Complementarity}
\label{sec:nonminimal}

\subsection{Complementarity with the electron EDM: minimal flavor violation}

One of the most stringent precision constraints on sleptons and neutralinos arises from electron EDM experiments, among which the best current limit is from the ACME II experiment~\cite{ACME:2018yjb}. The electron EDM can arise at one loop in diagrams involving a CP-violating phase (see, e.g.,~\cite{Ibrahim:2007fb}). There is no direct analogue of the minimal $\mu \to e\gamma$ diagram in Fig.~\ref{fig:lfv_dipole}, because only the off-diagonal $\delta^{RR}$ entries admit a complex phase. As a result, any comparison  of collider reach with the electron EDM constraint necessarily involves additional assumptions about the SUSY spectrum and interactions. For realizations of SUSY with large CP-violating phases, the EDM constraint can already rule out slepton masses of order 10 TeV~\cite{Cesarotti:2018huy}. However, many well-motivated scenarios for SUSY breaking predict smaller phases. In this section, we will discuss some plausible  assumptions about the dominant sources of the EDM, and compare  the reach of precision experiments with collider experiments. Further discussion of the underlying assumptions about SUSY breaking may be found in Appendix~\ref{app:flavor}.

\begin{figure}[!h]
\centering
\includegraphics [width = \textwidth]{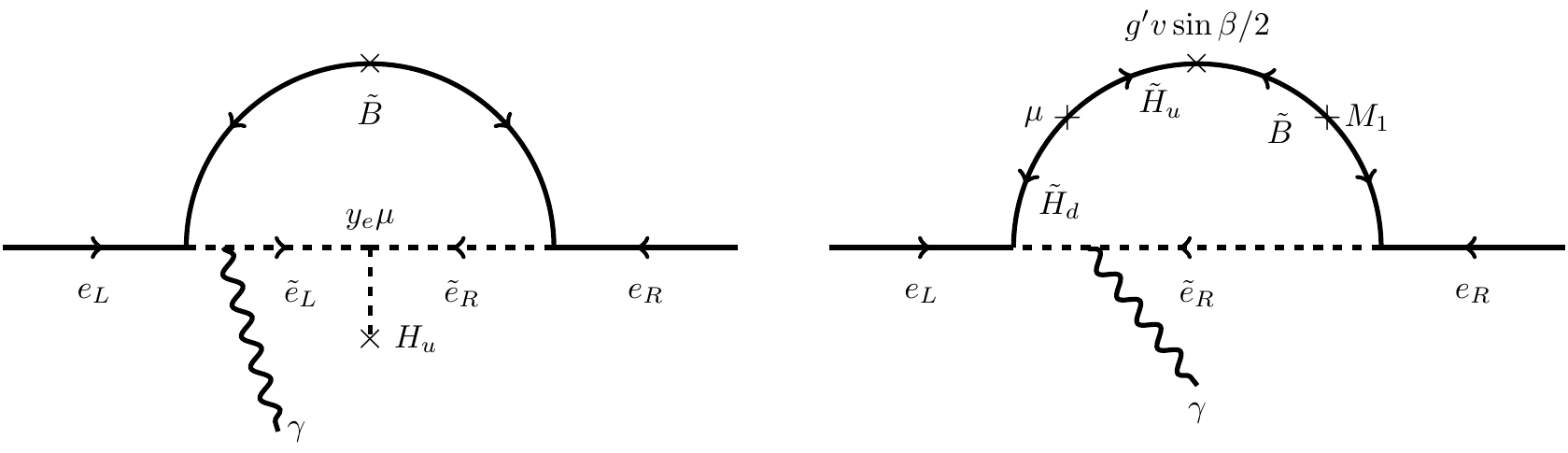}
\caption{
One-loop diagrams contributing to the electron EDM involving higgsinos or right-handed selectrons, but without flavor violation. Both diagrams may be sensitive to a subleading, CP-violating phase $\arg(\mu M_1)$ arising from gravity-mediated contributions. At left: a diagram that is sensitive to $\mu$ through the mixing of ${\tilde e}_L$ and ${\tilde e}_R$. At right: a diagram involving only ${\tilde e}_R$ together with bino--higgsino mixing.
}
\label{fig:eEDMdiagrams}
\end{figure}

In models with minimal flavor violation, the electron  EDM will be proportional to $m_e$ itself.  In this case, the leading contribution to the electron EDM could arise from the diagrams in Fig.~\ref{fig:eEDMdiagrams}. Both diagrams are sensitive to the phase $\arg(\mu M_1)$, where the higgsino mass parameter $\mu$ enters in one diagram through the mixing of left- and right-handed selectrons, and in the other diagram through bino-higgsino mixing. In the limit where the left-handed selectron and higgsino are much heavier than the right-handed selectron and bino, the sum of these contributions can be estimated as
\begin{equation}
d_e \approx -e \frac{\alpha_Y}{4\pi} \frac{m_e \, \tan \beta}{m_{{\tilde \ell}_R}^2} \mathrm{Im}\left[\frac{\mu M_1}{m_{{\tilde \ell}_L}^2 } + \frac{M_1}{\mu}\right] B\left(\frac{|M_1|^2}{m_{{\tilde \ell}_R}^2}\right),
  \label{eq:eEDMselectrononly}
\end{equation}
where the loop function $B(r)$ (following the notation of~\cite{Ellis:2008zy}) is defined as
\begin{equation}
B(r) = \frac{1 - r^2 + 2r \log r}{2(1-r)^3}.
  \label{eq:loopfunctionB}
\end{equation}
This formula captures the full diagram up to precision $O(m_{{\tilde \ell}_R}^2/m_{{\tilde \ell}_L}^2)$ and $O(|M_1/\mu|^2)$. In our numerical results below, we will include the full one-loop contribution of the mixed left- and  right-handed selectrons and the mixed bino and higgsinos.

\begin{figure}[!ht]
\centering
\begin{subfigure}[b]{0.49\textwidth}
 \centering
 \includegraphics[width=\textwidth]{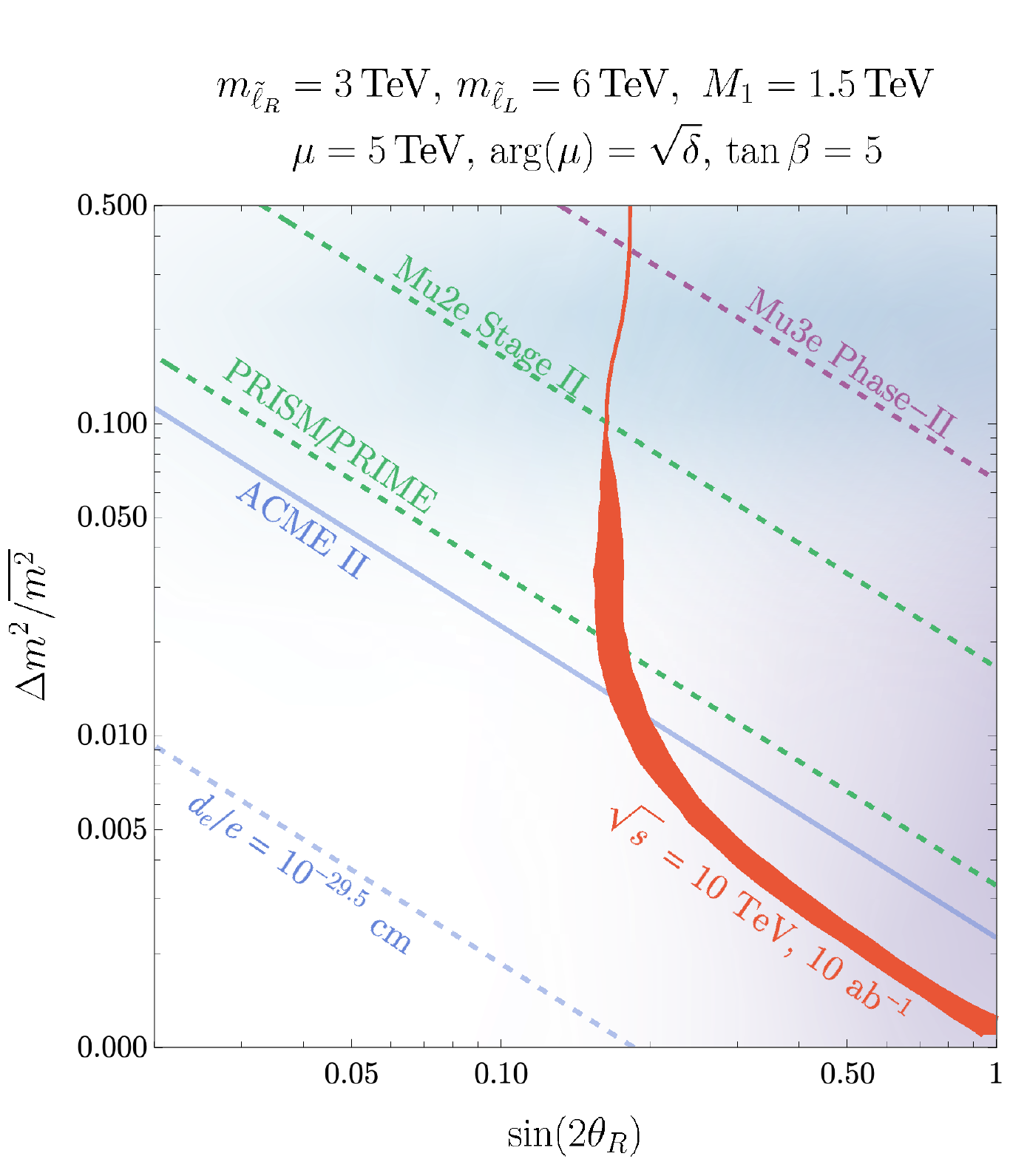}
\end{subfigure}
\hfill
\begin{subfigure}[b]{0.49\textwidth}
 \centering
 \includegraphics[width=\textwidth]{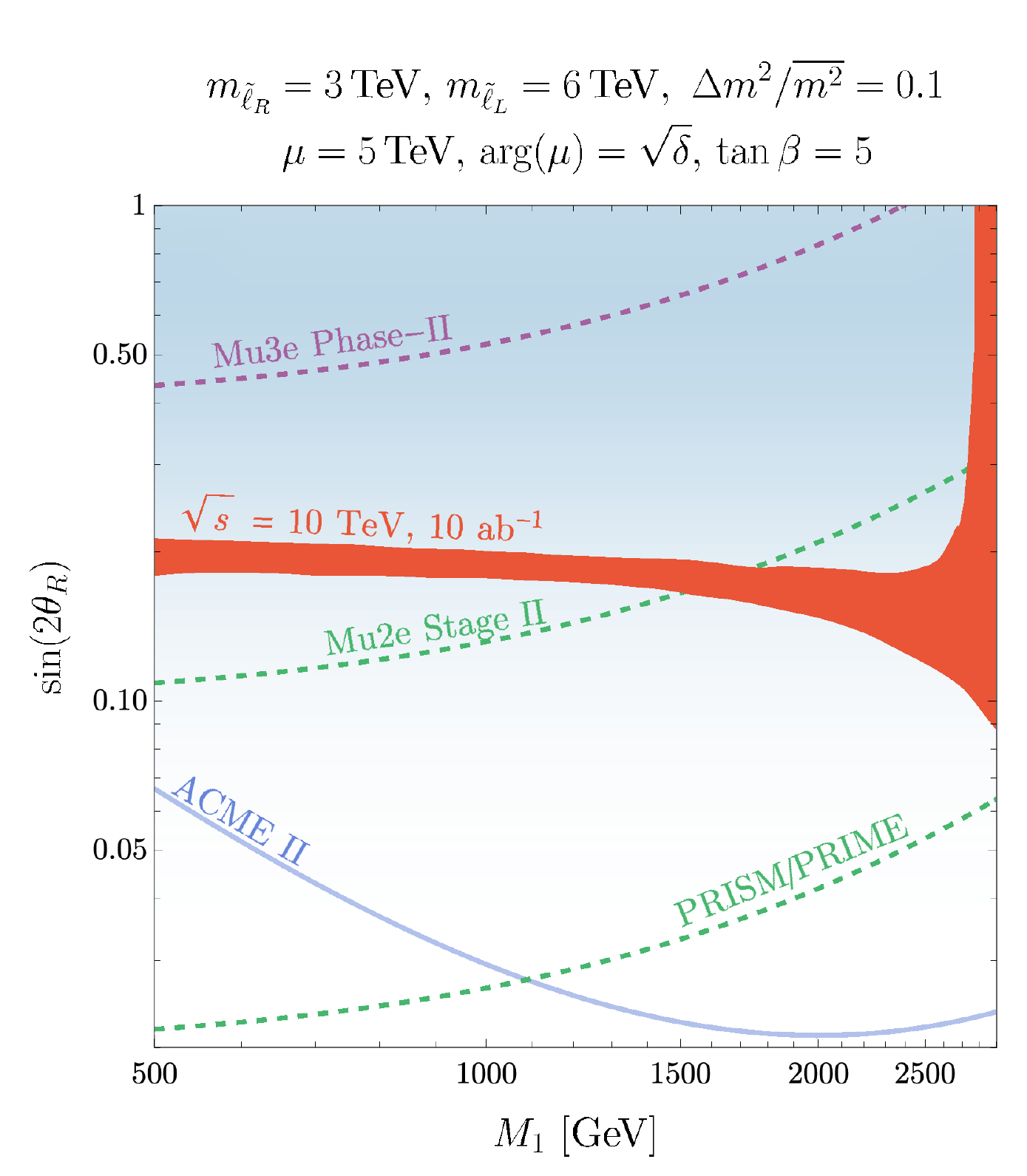}
\end{subfigure}
\hfill
\caption{Complementarity of a 10 TeV muon collider with EDM and $\mu \to e\gamma$ experiments. This figure is similar to the upper row of Fig.~\ref{fig:collider_higherTeV}, but also shows the electron EDM constraint from the contributions shown in Fig.~\ref{fig:eEDMdiagrams}. The phase  in the $\mu$ term is assumed to originate from a subdominant gravity mediated contribution  of order $m_{3/2} \sim \sqrt{\delta} \mu$. The solid blue curve displays the constraint from the current ACME II experiment~\cite{ACME:2018yjb}. The dashed blue curve shows a hypothetical (near-)future constraint.}
\label{fig:nonminimalselectron}
\end{figure}

To compare the reach of electron EDM experiments to LFV and muon collider experiments, we must make additional assumptions about the SUSY spectrum. If the CP-violating phase $\arg(M_1 \mu)$ is large, then the EDM experiment is much more stringent. However, it is plausible that this phase vanishes at leading order and obtains only subdominant contributions. If the off-diagonal mass matrix terms are of order $m_{3/2}^2$ and the component of $\mu$ with a CP-violating phase is of order $m_{3/2}$, then we have $\arg(\mu) \sim \sqrt{\delta}$. We follow this assumption in plotting constraints in Fig.~\ref{fig:nonminimalselectron}, taking $\delta$ to be related to the parameters $\sin(2\theta_R)$ and $\Delta m^2/\overline{m^2}$ on the two plot axes via Eq.~\eqref{eq:deltaRReq}. We further assume that the left-handed sleptons and higgsinos are sufficiently massive that they will not be directly produced at a 10 TeV muon collider (6 TeV and 5 TeV, respectively) and will have a subdominant effect on the $\mu \to e\gamma$ rate, so that the other curves in the figure are unchanged from Fig.~\ref{fig:collider_higherTeV}. The result, illustrated in Fig.~\ref{fig:nonminimalselectron}, is that ACME II is already a stringent constraint on the parameter space, but the muon collider can probe regions that are not yet excluded. Furthermore, models in which $\arg(M_1 \mu)$ is smaller than $\sqrt{\delta}$ by an order of magnitude would continue to evade EDM experiments, while remaining accessible to LFV and muon collider experiments.

\subsection{Complementarity with eEDM and $\mu \to e\gamma$ via stau mixing}

\begin{figure}[!h]
\centering
\includegraphics [width = 0.6\textwidth]{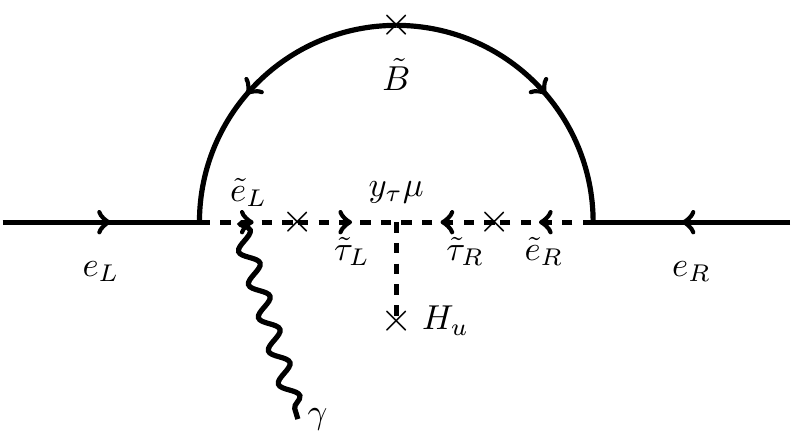}
\caption{
One-loop diagrams contributing to the electron EDM in the presence of flavor violation, proportional to the flavor-violating parameters $\delta^{LL}_{e\tau}$ and $\delta^{RR}_{\tau e}$, which will in general carry $O(1)$ CP-violating phases when arising from gravity mediation. 
}
\label{fig:eEDMstauloop}
\end{figure}

Flavor violation opens up the possibility that the dominant source of the electron EDM will be sensitive to larger couplings, such as  the  tau Yukawa~\cite{Altmannshofer:2009ne,Moroi:2013sfa,McKeen:2013dma, Altmannshofer:2013lfa}. These contributions are enhanced by $y_\tau/y_e \approx 3.5 \times 10^3$, but also suppressed by mixing effects, as illustrated in Fig.~\ref{fig:eEDMstauloop}. In the limit in which the right-handed sleptons are all nearly degenerate with mass $m_{{\tilde \ell}_R}$, and the left-handed sleptons are all  nearly  degenerate with a much larger mass $m_{{\tilde \ell}_L}$, we can estimate the electron EDM induced by loops of the right-handed selectron and stau states and a bino:
\begin{equation}
d_e\approx e  \frac{\alpha_Y}{2\pi} \frac{m_\tau \,\tan\beta}{m_{{\tilde \ell}_L}^2 m_{{\tilde \ell}_R}^2} \mathrm{Im}(\mu M_1\delta^{LL}_{e\tau} \delta^{RR}_{\tau  e})f_{2n}\left(\frac{|M_1|^2}{m_{{\tilde \ell}_R}^2}\right).
  \label{eq:eEDMstaumix1}
\end{equation}
Here (in notation following~\cite{Altmannshofer:2009ne}) the loop function is
\begin{equation}
f_{2n}(r) = \frac{1+(4-5r)r + 2r(2+r)  \log r}{4(1-r)^4}.
\end{equation}
As in the $\mu \to e\gamma$ context discussed in Section~\ref{subsec:precision}, this loop function  is related to the function $B(r)$ appearing in~\eqref{eq:loopfunctionB} by a derivative: $(r B(r))' = 2 f_{2n}(r)$. Also as in that discussion, the mass insertion approximation here is valid even when the states ${\tilde e}_R, {\tilde \tau}_R$ are maximally mixed with each other (and similary ${\tilde e}_L, {\tilde \tau}_L$), provided the $\delta$'s are small. The equation~\eqref{eq:eEDMstaumix1} gets corrections of relative size $m_{{\tilde \ell}_R}^2/m_{{\tilde \ell}_L}^2$. Another approximate expression has been derived in the limit where {\em all} of the sleptons are degenerate at a common scale $m_{\tilde \ell}$~\cite{Altmannshofer:2009ne},
\begin{equation}
d_e \approx  e \frac{\alpha_Y}{4\pi} \frac{m_\tau \,  \tan  \beta}{m_{\tilde \ell}^4} \mathrm{Im}(\mu M_1 \delta^{LL}_{e\tau} \delta^{RR}_{\tau e}) f_{4n}\left(\frac{|M_1|^2}{m_{\tilde \ell}^2}\right)
  \label{eq:eEDMstaumix2}
\end{equation}
Here
\begin{equation}
f_{4n}(r) = \frac{-3-44r +  36r^2  + 12r^3 - r^4 - 12r(3r+2) \log r}{6(1-r)^6}.
\end{equation}
In our numerical results, we use the complete one-loop correction from the set of mixed ${\tilde e}_{L, R}, {\tilde \tau}_{L,R}$ states, rather than either of these approximate formulas.

An analogous contribution to the $\mu \to e\gamma$ decay rate can enhance it beyond the minimal contribution that  we discussed in Sec.~\ref{subsec:precision}. The corresponding Feynman diagram is precisely Fig.~\ref{fig:eEDMstauloop}, but with either $e_L, {\tilde e}_L$ replaced by $\mu_L, {\tilde \mu}_L$ or the analogous change to the right-handed (s)electron. The corresponding amplitude  can be estimated in exactly the same way,  but without taking the imaginary part:
\begin{equation}
{\cal A}^R_{\mu e} \approx  \frac{\alpha_Y}{2\pi} \frac{m_\tau \,\tan\beta}{m_{{\tilde \ell}_L}^2 m_{{\tilde \ell}_R}^2} \mu M_1\delta^{LL}_{\mu\tau} \delta^{RR}_{\tau  e} f_{2n}\left(\frac{|M_1|^2}{m_{{\tilde \ell}_R}^2}\right).
  \label{eq:muegammastaumix1}
\end{equation}
When  the $\delta$ values are relatively large, this effect dominates over the minimal contribution due to the large ratio $m_\tau / m_\mu$.  However, because it is quadratic in the $\delta$'s while the minimal contribution is linear in $\delta^{RR}_{e\mu}$, we expect  that in  most of the parameter space it is subleading.

\begin{figure}[!ht]
\centering
\begin{subfigure}[b]{0.49\textwidth}
 \centering
 \includegraphics[width=\textwidth]{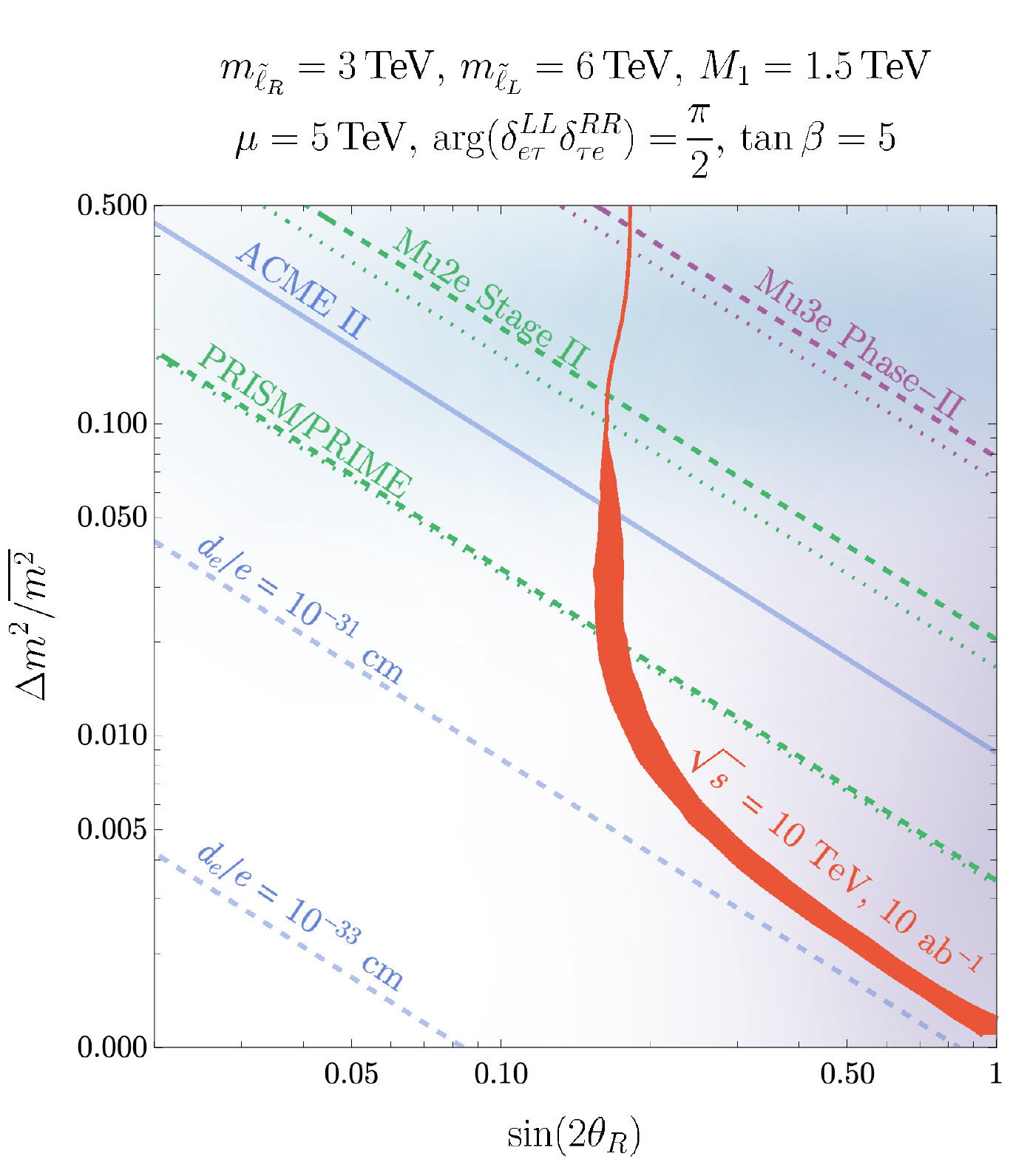}
\end{subfigure}
\hfill
\begin{subfigure}[b]{0.49\textwidth}
 \centering
 \includegraphics[width=\textwidth]{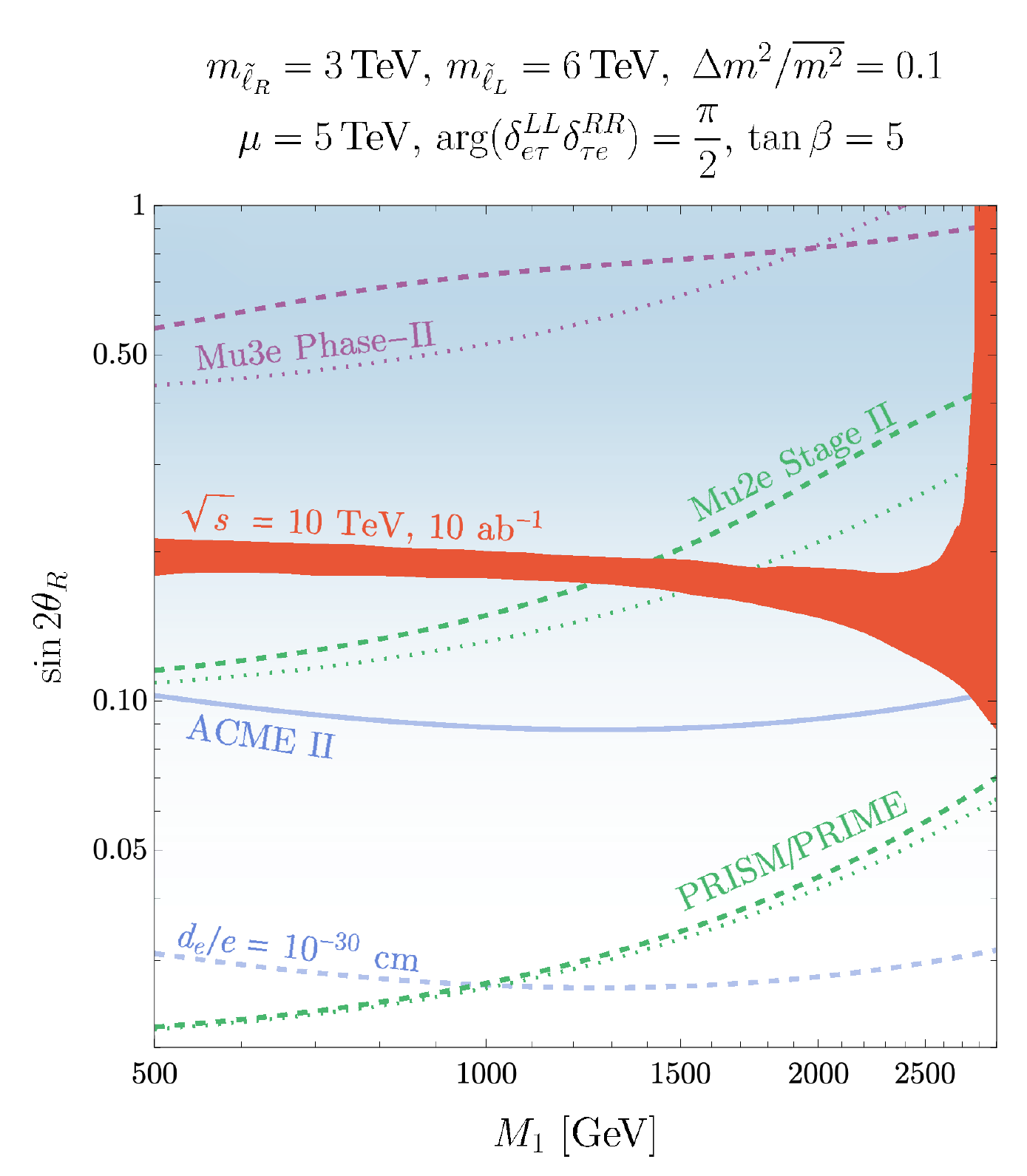}
\end{subfigure}
\hfill
\caption{Complementarity of a 10 TeV muon collider with EDM and $\mu \to e\gamma$ experiments. This figure is similar to Fig.~\ref{fig:collider_higherTeV}, but also shows the electron EDM and $\mu \to e\gamma$ effects from the stau mixing contributions shown in Fig.~\ref{fig:eEDMstauloop}. The solid blue curve displays the constraint from the current ACME II experiment~\cite{ACME:2018yjb}. The dashed blue curves show hypothetical future EDM  constraints. The dotted green and purple curves are the low-energy LFV constraints from Fig.~\ref{fig:collider_higherTeV}, while the dashed green curves add the effects from stau mixing. We see that the stau mixing effect, being quadratic in $\delta$, is only important at larger $\delta$ values.}
\label{fig:nonminimalstau}
\end{figure}

We illustrate the size of these stau-mixing-induced EDM and LFV signals, in comparison to collider reach, in Fig.~\ref{fig:nonminimalstau}. In this case, we do not include the contributions discussed in the preceding subsection (assuming $\arg(M_1 \mu)$ is negligible), but instead assuming a maximal phase in the products of $\delta$ parameters appearing in Eq.~\eqref{eq:eEDMstaumix1} and Eq.~\eqref{eq:muegammastaumix1}. For plotting purposes, we take both $\delta$ parameters to be determined by $\frac{\Delta m^2}{\overline{m^2}}$ and $\sin(2\theta_R)$ as in Eq.~\eqref{eq:deltaRReq}, even though strictly speaking that was an equation only for $\delta^{RR}_{e\mu}$. At least in certain classes of models, all of the $\delta^{LL}$ and $\delta^{RR}$ values could be expected to be parametrically of the same size. For the $\mu \to e\gamma$ contribution, we add the minimal contribution from Eq.~\eqref{eq:dipole_minimal_mia} to that of Eq.~\eqref{eq:muegammastaumix1}, taking each of the $\delta$ parameters to have phase $\pi/4$. (Choosing different phases would have only a mild effect on the low-energy LFV curves in these log-log plots.) The figure shows that ACME II is already a significant constraint on the parameter space. EDM experiments will need to improve by multiple orders of magnitude to fully cover the muon collider's range in this scenario, because the phase scales as $\delta^2$ (though dramatic experimental improvements are expected in the next decade, so this is possible).  Furthermore, we see that at larger values of $\delta$, the new stau-mediated contribution to $\mu \to e\gamma$ has a significant effect on the bounds (dashed green and purple curves) compared to the minimal contribution (dotted green and purple curves). As expected, this effect is negligible for smaller $\delta$, as in the PRISM/PRIME curves.

\medskip

In this section we have focused on some particular contributions to the electron EDM and $\mu \to e$ processes, with a small number of additional particles involved beyond those contributing to our collider observables. In general, the full spectrum of superpartners could participate in loops contributing to these processes. For example, $\mu \to e$ conversion in nuclei can proceed via box diagrams involving squarks, sleptons, and neutralinos or charginos~\cite{Hisano:1995cp, Arganda:2007jw}. However, provided that we remain in the limit where the SUSY spectrum is not split, and we consider the squarks to be modestly heavier than the sleptons, it is reasonable to expect that the contributions we have considered here provide the correct order of magnitude for the observables. More complete discussions of the relative size of various contributions, especially in the split SUSY limit where the variation is more dramatic, may be found in, e.g.,~\cite{Ellis:2016yje}.

\section{Conclusions}
\label{sec:conclusion}

A high-energy muon collider is an exciting possibility for advancing beyond the Standard Model, offering a unique combination of a clean lepton-collider environment for precision measurements, the high energy reach conventionally thought of as the domain of hadron colliders, and both muon fusion processes taking advantage of the full beam energy and vector boson fusion processes sampling a range of lower energies. By colliding muons, new opportunities open up for flavor physics previously inaccessible at colliders based on first-generation fermions. At the same time, new precision experiments are improving constraints on CP-violating observables like the electron EDM and lepton flavor violating observables like $\mu \to e\gamma$ by orders of magnitude. By the time any future collider is built, we may already have dramatic discoveries or stringent constraints from such experiments. Thus, it is important to explore the interplay between them.

In this paper, we have shown that there is a powerful complementarity between these different experiments. A muon collider offers the opportunity to discover new electromagnetically-interacting particles, like sleptons, and to measure their mass and the mass of their decay products with high precision. Such precision measurements further facilitate clean measurements of the rate of flavor-violating processes. We have illustrated this with the case of mixing between right-handed selectrons and smuons. In regions of parameter space with sizable mixing, as motivated by gauge-mediated supersymmetry breaking with subleading (e.g., gravity-mediated) flavor violation, flavor-violating signals could be measured even for fractional mass splittings at the sub-percent level.

The minimal model responsible for the collider signal that we considered also gives rise to $\mu \to e$ decay and conversion processes in the right range to be probed by next-generation experiments. These are complementary, in the sense that each of these experiments can probe parameter space that is inaccessible to the other. Furthermore, they would complement each other well in their region of overlap. A discovery at an LFV experiment could occur first, but this would only tell us a single branching ratio. We would know that physics beyond the Standard Model exists, but we would lack the detailed information needed to build a new, more complete Standard Model. The only way to unravel the physics ultimately responsible for the flavor-violating decay would be to reach high energies and directly produce the particles responsible. We have shown, in an example, that a muon collider could do this.

Models responsible for these signals could also, generically, produce a detectable electron EDM. However, the EDM signal requires CP violation, whereas the LFV and collider signals do not. EDM experiments, then, could measure the complex phase associated with the couplings that might be discovered by the other experiments. A discovery of new physics with an associated small CP phase would (much like the Strong CP problem of the Standard Model) be an important hint about underlying dynamics. It could point, for instance, to the possibility that CP is a spontaneously broken gauge symmetry in our universe, broken dynamically in a way that generates a large CKM phase but suppresses other phases.

In the end, we will need a large suite of experiments to uniquely fingerprint the source of physics beyond the Standard Model. LFV and EDM experiments are making major strides, and a decisive discovery of new physics could come in the near future. In such a case, the crucial next step is to turn to an experiment that can fully characterize the new physics. Only a collider has the power to produce multiple new particles and measure their masses and couplings, unveiling the origin of higher-dimension operators. A muon collider is an ideal choice for such a task.

\subsection*{Acknowledgments}

The work of SH, QL and MR is supported in part by the DOE grant DE-SC0013607 and the by the Alfred P. Sloan Foundation Grant No. G-2019-12504. QL and MR are also supported by the NASA Grant 80NSSC20K0506.

\appendix
\addcontentsline{toc}{section}{Appendix}

\section{Flavor and Supersymmetry Breaking}
\label{app:flavor}

\subsection{Gauge mediation with subdominant gravity mediation}

In our phenomenological studies, we have simply assumed some structure of flavor-conserving and flavor-violating soft SUSY-breaking terms. In this appendix, we provide some brief comments on possible underlying models of SUSY breaking responsible for such structures. Model building is not our primary goal in this paper, so we will keep our comments brief, but having an approximate underlying model in mind will allow us to provide estimates not only of the relative reach of collider observables and precision CLFV observables, but also of electric dipole moment (EDM) observables.

Gauge-mediated supersymmetry breaking (GMSB, reviewed in~\cite{Giudice:1998bp, Kitano:2010fa}) naturally provides flavor-universal supersymmetric soft masses, because the different generations of matter fields have identical gauge interactions. GMSB also produces CP-preserving supersymmetric soft masses. A weakness of GMSB models is that they do not automatically explain the $\mu$ and $b_\mu$ terms, which are needed to give mass to higgsinos and to explain electroweak symmetry breaking. This requires extensions of the underlying model; examples include~\cite{Dvali:1996cu, Giudice:2007ca, Csaki:2008sr}. Some of these extensions predict no nonzero physical CP phase in the gaugino and higgsino sector. GMSB models predict small $A$-terms, generated radiatively from the gaugino masses and $\mu$. Thus, flavor- and CP-violating observables can be highly suppressed in GMSB models.

The soft terms generated by GMSB from messenger fields (which carry SM gauge charges) at a mass scale $M_\mathrm{mess}$ with SUSY-breaking F-term VEV $F$ are of order
\begin{equation}
m_\mathrm{soft} \sim \frac{c \alpha_r}{4\pi} \frac{F}{M_\mathrm{mess}},
\end{equation}
where $\alpha_r$ is determined by the most significant gauge coupling (e.g., $\alpha_s$ for squarks and gluinos, or $\alpha_W$ for left-handed sleptons or winos) and $c$ is a factor that we will take to be order-one.\footnote{In many models, the gauginos are in fact parametrically lighter than the scalars due to a ``gaugino screening'' phenomenon~\cite{Arkani-Hamed:1998mzz,Komargodski:2009jf,Dumitrescu:2010ha,Cohen:2011aa}, leading to split SUSY phenomenology. We have chosen not to focus on split spectra with very light gauginos in this work. Some conditions under which this assumption holds are explained in~\cite{Dumitrescu:2010ha}.} We will assume that gaugino screening is not significant, so that slepton and neutralino masses are the same order of magnitude. One expects in the most minimal variants of GMSB that there are light right-handed sleptons and binos, with somewhat heavier left-handed sleptons and winos, and squarks and gluinos an order of magnitude heavier. However, variants of GMSB can allow a much wider range of spectra~\cite{Cheung:2007es, Meade:2008wd}. The gravitino mass is 
\begin{equation}
m_{3/2} = \frac{F_0}{\sqrt{3} M_\mathrm{Pl}},
\end{equation}
where $F_0 \equiv \mathrm{e}^{\langle K \rangle/2} \langle W \rangle/M_{\rm Pl}$ is expected to be of the same order as the F-term VEV $F$ appearing in the MSSM soft terms. In our studies, we have focused on decays ${\tilde \ell} \to \ell {\tilde \chi}^0$, and omitted subsequent decays to the gravitino, like ${\tilde \chi}^0 \to \gamma {\tilde G}$. The lifetime for such decays is
\begin{equation}
c\tau \sim \frac{16\pi F_0^2}{m_{{\tilde \chi}^0}^5} = \frac{48 \pi m_{3/2}^2 M_\mathrm{Pl}^2}{m_{{\tilde \chi}^0}^5} = 1.7\,\mathrm{km} \left(\frac{m_{3/2}}{1\,\mathrm{GeV}}\right)^2 \left(\frac{10\,\mathrm{TeV}}{m_{{\tilde \chi}^0}}\right)^5.
\end{equation}
Thus, the decays happen well outside the collider and our approach is consistent, provided that $m_{3/2} \gtrsim 100\,\mathrm{MeV}$. This threshold is several orders of magnitude below $m_{{\tilde \chi}^0}$, meaning that it is consistent to assume that GMSB dominates over gravity mediation.

Gravity mediation is the effect of generic Planck-suppressed operators, which contribute to soft SUSY-breaking terms in addition to GMSB. Such contributions are generally assumed to be flavor-violating, that is, we expect to have terms like 
\begin{equation}
\int \mathrm{d}^4\theta\,\frac{X^\dagger X}{M_{\rm Pl}^2} c_{ij} L_i^\dagger L_j \sim c_{ij} m_{3/2}^2 {\tilde \ell}_i^\dagger {\tilde\ell}_j,
\end{equation}
where $c_{ij}$ is a matrix in flavor space with order-one entries. (This assumption is not so easily justified, despite being a common one; we discuss this further below in \S\ref{subsec:why}.) In this way, GMSB with subdominant gravity mediation can explain a nearly flavor-symmetric SUSY spectrum with subleading flavor-violating effects, as we have assumed. We then expect the off-diagonal entries in $\delta^{LL}$ and $\delta^{RR}$ to be of order
\begin{equation}
\delta^{LL}_\text{off-diag} \sim \delta^{RR}_\text{off-diag} \sim \left(\frac{m_{3/2}}{m_{{\tilde \ell}_{L,R}}}\right)^2.
\end{equation}
One can also consider gravity-mediated contributions to the left-right slepton mixing parameters $\delta^{LR}$. These could arise from $A$-terms of order $m_{3/2}$. However, one could consider a scenario in which the fields $X$ acquiring F-term VEVs also carry charges under other symmetries, which would suppress the gravity-mediated contributions to $A$-terms (as well as to gaugino masses). Thus, we will assume that the flavor-violating slepton mixings are localized in $\delta^{LL}$ and $\delta^{RR}$.

Focusing on the right-handed selectron and smuon studied in the main text, in terms of the parametrization Eq.~\eqref{eq:slepton_mass_matrix}, gauge mediation motivates the choice
\begin{equation}
\Delta^{RR}_{ij} = c_{ij} m_{3/2}^2,
\end{equation}
with the $c_{ij}$ order-one numbers. The universal contribution $m_R^2$ is a sum of a gauge-mediated contribution and a D-term contribution:
\begin{equation}
m_R^2 = \left({\tilde m}_{\ell_R}^2\right)_\mathrm{GMSB} + m_Z^2 \cos(2\beta) \sin^2\theta_W + \cdots,
\end{equation}
where $\cdots$ stands in for the universal part of loop corrections. We can absorb various other non-universal, subleading contributions into the definition of the $\Delta_{ij}$ terms, including the Yukawa contributions of $m_e^2$ and $m_\mu^2$ on the diagonal, and shifts due to mixing with the left-handed sleptons, which for the smuon is of order $m_\mu^2 \mu^2 \tan^2\beta / {\tilde m}_{\ell_L}^2$ and is subleading provided that $m_{3/2} \gg m_\mu \tan \beta$.  From Eq.~\eqref{eq:mixing_defs}, we see that for generic, order-one choices of the $c_{ij}$ coefficients, the mixing $\sin(2 \theta_R)$ is $O(1)$.

As with the slepton soft masses, we could also imagine that the higgsino mass parameter $\mu$ is a sum of a dominant term of order $m_\mathrm{soft}$ and a subleading, gravity-mediated term of order $m_{3/2}$. To avoid stringent EDM constraints, the leading contribution should not carry a CP-violating phase, but the subleading contribution could. This may be rather model-dependent, as any structure or symmetries invoked to solve the $\mu$ problem in GMSB could also affect whether the Giudice-Masiero contribution to the $\mu$ and $b_\mu$ terms~\cite{Giudice:1988yz} is allowed at order $m_{3/2}$ or is further suppressed. 

\subsection{A variant: GMSB with gauged $L_\mu - L_\tau$}

The Standard Model (without right-handed neutrinos) can be extended by gauging an anomaly-free $U(1)$ symmetry which is the difference of lepton numbers in two generations~\cite{Foot:1990mn,He:1990pn}. The most interesting case is $L_\mu - L_\tau$, as its gauging can explain why the muon neutrino and tau neutrino are nearly maximally mixed~\cite{Ma:2001md}. This gauge symmetry must be higgsed, in order for neutrino masses to be generated. 

The GMSB formalism can be extended to include contributions from messengers charged under higgsed gauge fields~\cite{Gorbatov:2008qa, Craig:2012yd}. If messengers of SUSY breaking carry charge under $L_\mu - L_\tau$, then the soft terms for the muon and tau lepton fields will have additional contributions. Unlike conventional GMSB, then, this theory would predict a spectrum with nearly-degenerate smuons and staus (and the corresponding sneutrinos) but with selectrons having a different mass. For our purposes, a detailed discussion of the resulting spectrum is not necessary; we simply take this as one possible justification for a spectrum in which the right-handed selectron is lighter, by an $O(1)$ factor, than the right-handed smuon and stau, leading to a larger value of $\Delta \overline{m^2}$. Again, subdominant effects of gravity mediation can contribute flavor-violating soft terms.

\subsection{Should gravity mediation violate flavor maximally?}
\label{subsec:why}

The claim that Planck-suppressed operators should have completely generic flavor structure arises from the expectation that theories of quantum gravity have no approximate global symmetries at the Planck scale. However, there is some tension between this point of view and the expectation that the flavor structure of the Standard Model---the pattern of hierarchical masses and mixings---is explained in a natural way by some physics beyond the Standard Model at high energies. This point seems to be under-appreciated, so we will make a few remarks on how the tension might be resolved. One way to explain small Yukawa couplings and mixings is by invoking horizontal symmetries, such as a $U(1)$ symmetry under which different generations carry different charges~\cite{Froggatt:1978nt, Leurer:1992wg, Nir:1993mx}. In a gravitational theory, one should demand that such a symmetry is either gauged or is an approximate (accidental) symmetry enforced by a gauge symmetry (perhaps a discrete one). In the case that it is a gauge symmetry, it may be anomalous, and hence the corresponding gauge boson could have a large Stueckelberg (4d Green-Schwarz) mass. In any event, such a gauge symmetry would forbid a term of the form $\int \mathrm{d}^4\theta\,X^\dagger X L_i^\dagger L_j$ where $L_i$ and $L_j$ carry different flavor gauge charges. Left-handed lepton superfields could have equal horizontal charges and completely anarchic couplings, as suggested by neutrino mixing~\cite{Hall:1999sn}, but this would not be expected for right-handed leptons or quarks. In such models a phenomenon of {\em alignment}, in which the quark mass matrices and squark mass-squared matrices are approximately diagonal in the same basis, can suppress flavor-changing neutral currents~\cite{Nir:1993mx}. It is possible that CP is a fundamental symmetry spontaneously broken only by fields with horizontal charge, which can suppress EDMs~\cite{Nir:1996am}; this was recently discussed in the lepton sector in~\cite{Aloni:2021wzk,Nakai:2021mha}. 

Although models with such flavor textures are interesting in their own right, an extension of the models can allow the generic Planck-suppressed flavor violation that is conventionally assumed. If there are flavor-charged fields that acquire F-term VEVs at a common scale, then the operators take the form $\int \mathrm{d}^4\theta\,X_k^\dagger X_l L_i^\dagger L_j$ where the horizontal charge of $X_k^\dagger X_l$ is nonzero and compensates that of $L_i^\dagger L_j$. Of course, this constitutes an additional source of flavor-symmetry breaking. Nonetheless, it could be extremely subdominant to the leading flavor breaking, which is often assumed to arise from SUSY-preserving VEVs, $\langle Z \rangle/\Lambda \sim \lambda \sim 0.2$, generated at a much higher scale. In other words, flavor symmetries are broken twice: once at a high scale, generating the correct textures of Yukawa terms, but without breaking supersymmetry; and a second time at low scale, breaking supersymmetry in a substantially flavor-violating manner, but producing only small corrections to the Yukawa textures.

One could consider other explanations of flavor structure and the associated gravity-mediated soft terms. For example, Yukawa couplings are sometimes suppressed not by a symmetry but by physical separation (small wavefunction overlap) of fields in extra dimensions. The three-field overlap integral involved in generating a coupling like $H_d L_i {\bar E}_j$ is quite different from the four-field overlap integral generating a coupling to $X^\dagger X L_i^\dagger L_j$, and (if the wavefunction of $X$ is relatively flat) the latter may be much less suppressed than the former.

Producing complete models along these lines is beyond the scope of this work, where our primary aim is collider phenomenology and an assessment of the complementarity between different experiments. Nonetheless, we believe that if new sources of flavor violation are observed in experiment it will be important to think clearly about how they relate to the flavor problem of explaining the SM Yukawa textures, and considerations along the lines that we have sketched here may be relevant.

\section{Details of the Mass Measurements}
\label{app:mass_meas}

In this Appendix we provide more details on the measurement of the slepton and neutralino masses obtained by fitting to the charged lepton energy spectrum, as described in Eqs.~\eqref{eq:fc_processes} and \eqref{eq:mass_endpoint_relations}.
In Appendix~\ref{subsec:mass_meas_background}, we discuss the backgrounds, details of their generation, and the selection cuts used to mitigate them.
In Appendix~\ref{subsec:likelihood}, we provide more details on the likelihood procedure used to fit the slepton and neutralino masses, and how the resulting precision was determined.
More discussion of the results, including results for other center of mass energies not presented in Fig.~\ref{fig:mass_meas_summary} is provided in Appendix~\ref{subsec:mass_meas_discussion}.

\subsection{Background Mitigation}
\label{subsec:mass_meas_background}

The leading backgrounds to the processes in Eq.~\eqref{eq:fc_processes} are $\ell^+\ell^- \nu \bar{\nu}$ production (via intermediate $W$ bosons) and lepton pair production via neutral vector boson fusion.
The importance of the latter is particular to high-energy muon colliders, where the forward remnants may not be tagged as a result of the shielding for the beam-induced background, thus mimicking the missing momentum of the neutralinos in the signal process.\footnote{\label{tablefoot}The referee brought to our attention the possibility that a forward muon can potentially be detected even in the presence of a BIB-shielding nozzle, since a muon is sufficiently heavy to penetrate the nozzle and reach a forward detector. If the muon detection coverage is increased from $\eta = 2.5$ to $\eta = 4$, then the neutral-VBF background is completely eliminated after applying all the kinematic cuts described in the text. There is potentially room for a better cut strategy. For example, a less aggressive cut on the transverse component of $p_{\ell^+} + p_{\ell^-}$ might be sufficient to eliminate all of the neutral-VBF background, but we leave such detailed cut optimization for future work.}
There are also backgrounds from $\tau$ pair production and charged vector boson fusion (which is kinematically similar to the neutral VBF background, but suppressed by the $W$ parton luminosity), as well as other reducible backgrounds from detector effects, but these are subdominant after imposing simple cuts and we therefore neglect them in this estimate.

To isolate the signal process we impose the following cuts, based in part on those of ref.~\cite{Freitas:2011ti}.
In all cases, we require exactly one $\ell^+$ and one $\ell^-$, each with $|\eta_{\ell}| < 2.5$ and $p_{T,\ell} > 25\,\textrm{GeV}$.
We demand that $|p_{\ell^+} + p_{\ell^-}| > 250, 750$, or $2500\,\textrm{GeV}$,
and that $p_{\textrm{miss}} > 0.5, 1.5$ or $5\,\textrm{TeV}$ for 
$\sqrt{s} = 3, 10$ and $30\,\textrm{TeV}$ respectively.
We further demand that $|\cos(\phi_{\ell^+} - \phi_{\ell^-})| < 0.95$ to reduce back-to-back topologies.
Finally, we require that the transverse component of $p_{\ell^+} + p_{\ell^-}$ be greater than $250, 750$ and $2500\,\textrm{GeV}$ at $3, 10$, and $30\,\textrm{TeV}$.
This final cut is added to substantially limit the VBF background: in the limit where the remnant muons continue exactly in the forward direction (i.e., in the limit used for the PDF approximation to the background), the underlying $\gamma \gamma \to \ell^+ \ell^-$ process has exactly zero transverse momentum. In the full $2 \to 4$ process, a small amount of transverse momentum is imparted to the remnants, and the rejection is no longer perfect, but still removes a great deal of the events.

The backgrounds are generated with \textsc{MadGraph5}~\cite{Alwall:2014hca} and analyzed at parton level.
To generate VBF background events, in order to avoid collinear singularities, we impose a generator-level cut on the maximum allowed pseudorapidity of the muons, $|\eta| < \eta_{\textrm{max}}$. 
Since in the $|\eta| \to \infty$ limit the cut on the transverse momentum of the central di-lepton system removes all the events, we need only take $\eta_{\textrm{max}}$ large enough such that the cross section after applying the selection cuts is constant.
We generated events with $\eta_{\textrm{max}}$ ranging from $4.0$ to $6.0$, finding stable results, and took $|\eta| < 5.0$ as the generator-level cut for the final analysis.
The final, post-cut cross sections of the different background processes at the various center of mass energies are shown in Table~\ref{tab:final_bkg_csxs}.

\begin{table}[h]
\centering
\begin{tabular}{c|ccc}
$\sqrt{s}$ 					& ~~~$3\,\textrm{TeV}$~~~ 	& $10\,\textrm{TeV}$		& $30\,\textrm{TeV}$ \\ \hline
$W^{+}W^-$ 				& $0.616$ 							& $0.609$						& $3.03\times 10^{-4}$ \\
VBF ($e^+e^-$)\textsuperscript{\ref{tablefoot}} 		& $0.0256$  							& $7.41\times 10^{-3}$	& $2.68\times 10^{-4}$ \\
VBF ($\mu^+\mu^-$)\textsuperscript{\ref{tablefoot}}	& $0.0314$  							& $0.0969$ 					& $4.16\times 10^{-4}$ \\ \hline
\end{tabular}
\caption{The cross sections, in fb, for the background processes at various center of mass energies after all the selection cuts described in the text.}
\label{tab:final_bkg_csxs}
\end{table}

To mimic the imperfect resolution of the detectors, the energies of the events were manually smeared by a Gaussian, with the width taken as a function of the energy and pseudorapidity of the particle that mimics the E-Cal resolution of the Muon Collider Collaboration Delphes Card~\cite{deFavereau:2013fsa}.

\subsection{Unbinned Likelihood Procedure}
\label{subsec:likelihood}

Given a finite set of data, $\{ x_i \}$ distributed according to a known distribution $f(x, \theta_a)$ containing unknown parameters, $\theta_a$, one can estimate the parameters of the underlying distribution by maximizing the (extended)  likelihood function of the data, $L(\{x_i\}; \theta_a)$.
In the present situation, the expected functional form of the signal distribution whose parameters we wish to endpoint is simple, so an unbinned likelihood fit to the data is an ideal way to estimate their values. For more details on maximum likelihood methods, see refs.~\cite{Cowan:1998ji, ParticleDataGroup:2020ssz}.

As discussed below Eq.~\eqref{eq:lepton_energy}, the distribution of the lepton energies in slepton pair production is expected to be uniform between $\Emin$ and $\Emax$. We can therefore take the (normalized) signal probability density function (p.d.f.) for a lepton of energy $E$ as
\begin{equation}
\label{eq:sig_pdf}
f_{\textrm{sig}}(E; \Eminhat, \Emaxhat, \sigma_1, \sigma_2) = 
\frac{1}{\Emaxhat - \Eminhat}
\begin{cases}
\frac{1}{2}\bigg[1 + \textrm{erf} \Big( \frac{E - \Eminhat}{\sqrt{2} \sigma_1}\Big) \bigg], & E < \frac{1}{2}(\Emaxhat - \Eminhat), \\[12pt]
\frac{1}{2}\,\textrm{erfc} \Big( \frac{E - \Emaxhat}{\sqrt{2} \sigma_2}\Big)\phantom{\bigg|}, & E > \frac{1}{2}(\Emaxhat - \Eminhat), \\[12pt] 
\end{cases}
\end{equation}
where $\Eminhat$ and $\Emaxhat$ are the estimators for the lower- and upper-endpoints, respectively. The parameters $\sigma_{1,2}$ describe the width of the error functions at the endpoints, included to account for the smearing of the edges of the distribution due to the detector energy resolution.

We must also introduce template p.d.f.s for the backgrounds. These can be chosen to follow simple functional forms based on the energy distributions obtained from simulation.
The $WW$ background leads to events with lepton energies ranging up to $\sqrt{s}/2$, while the leptons from the VBF background have energies peaked at low values.
After the selection cuts, we find reasonable descriptions of the background distributions using a uniform distribution for the $WW$ background and a shape similar to a Gamma distribution for the VBF events:
\begin{align}
\label{eq:vbf_bkg_pdf}
& f_{\textrm{bkg.,VBF}}(E; a_1, a_2) \propto
\bigg(\frac{E}{a_1}\bigg)^{a_2} \exp(-E / a_1) \\
\label{eq:ww_bkg_pdf}
& f_{\textrm{bkg.,}WW}(E) \propto \frac{1}{E_{\textrm{beam}}}
\end{align}
Here, the $a_i$ are nuisance parameters that are fit to the background shapes based on simulation.
The proportionality factors are taken (for given values of the nuisance parameters) such that the integral over energies up to $E_{\textrm{beam}} = \sqrt{s}/2$ is unity.

\begin{figure}[h]
\centering
\vskip 0.5cm
\includegraphics[width=10cm]{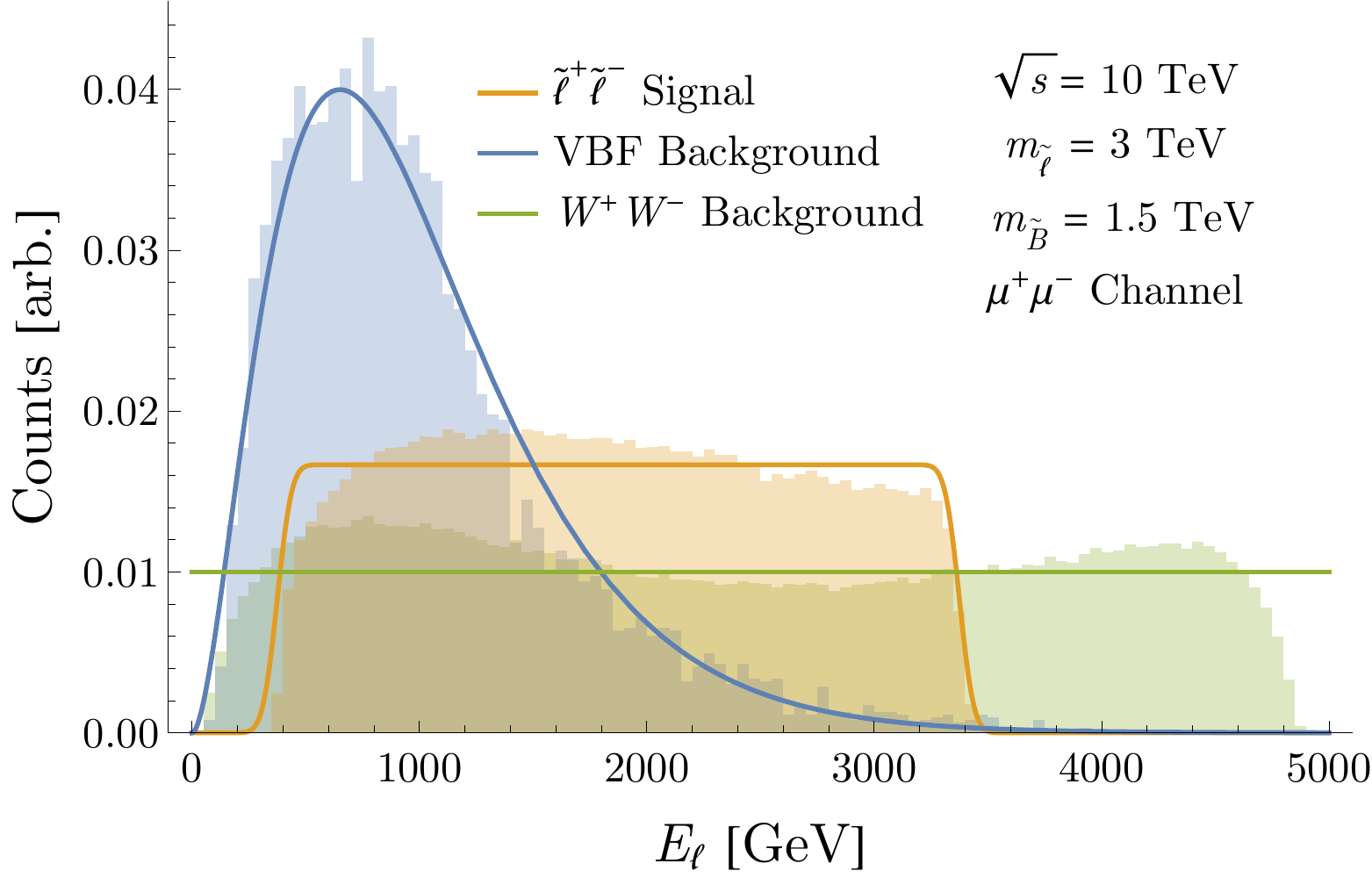}
\caption{Normalized histograms of the slepton-pair signal and the $WW$ and VBF backgrounds, together with plots of the normalized p.d.f.s, with best fit values of the nuisance parameters.}
\label{fig:pdf_examples}
\end{figure}

With the normalized signal and background p.d.f.s in hand, we can compute the negative extended log-likelihood, $L$, of a set of $n$ observed events with energies $\{E_i\}$:
\begin{multline}\label{eq:ext_nll}
- \log L(n, \{E_i\}; \mu_{\textrm{sig.}}, \Eminhat, \Emaxhat, \theta_m)
 = 
\mu_{\textrm{tot.}} - \sum_{i=1}^n\, \log\Big( \mu_{\textrm{tot.}}\, p(E_i; \mu_{\textrm{sig.}}, \Eminhat, \Emaxhat, \theta_m) \Big) .
\end{multline}
Here, the $\mu_a$ indicate the mean number of expected events for the signal and background processes after the selection cuts detailed in the previous section, assuming a given luminosity, and $\mu_{\textrm{tot.}} = \mu_{\textrm{sig.}} + \mu_{\textrm{bkg.,VBF}} + \mu_{\textrm{bkg.,}WW}$.
The $\theta_m$ is short for the set of nuisance parameters for the signal and background p.d.f.s, $\theta_m = \{ \sigma_1, \sigma_2, a_1, a_2 \}$ described in Eqs.~\eqref{eq:sig_pdf} and \eqref{eq:vbf_bkg_pdf}.
The likelihood for an individual event, $p(E_i; \dots)$ is given by
\begin{multline}\label{eq:p_event}
p(E; \mu_{\textrm{sig.}}, \Eminhat, \Emaxhat, \theta_m)
= \\[6pt]
\frac{\mu_{\textrm{sig.}}}{\mu_{\textrm{tot.}}} f_{\textrm{sig}}(E; \Eminhat, \Emaxhat, \sigma_1, \sigma_2)
+ \frac{\mu_{\textrm{bkg.,VBF}}}{\mu_{\textrm{tot.}}} f_{\textrm{bkg.,VBF}}(E; a_1, a_2)
+ \frac{\mu_{\textrm{bkg.,}WW}}{\mu_{\textrm{tot.}}} f_{\textrm{bkg.,}WW}(E)
\end{multline}
We take $\mu_{\textrm{bkg.,VBF}}$ and $\mu_{\textrm{bkg.,}WW}$ as known from simulation and evaluate $\mu_{\textrm{sig.}}$, $\Eminhat$, and $\Emaxhat$ as functions of the estimated slepton and neutralino masses, $\hat{m}_{\tilde{\ell}_R}$, $\hat{m}_{\tilde{\chi}_1^0}$. The signal rate, $\mu_{\textrm{sig}}$ is also a function of the mixing angle and mass splitting, $\sin2\theta_R$ and $\Delta m^2 / \overline{m^2}$.

Given the expected number of events for the signal and background, we create toy datasets by randomly selecting
$n_{\textrm{sig.}}$, $n_{\textrm{bkg.,VBF}}$, and $n_{\textrm{bkg.,}WW}$ events from the corresponding generated samples, with $n_a$ drawn from a Poisson distribution with the corresponding mean, $\mu_a$. 
We can then evaluate the extended log-likelihood (or more practically, the negative log of the likelihood or NLL) on the toy datasets. For large enough sample sizes, Wilks' theorem~\cite{Wilks:1938xx, Cowan:2010js} implies that $2 \Delta \textrm{NLL}$ follows a $\chi^2$ distribution, which we can use to construct confidence intervals for the slepton and neutralino masses.
This process can be repeated for a number of toy datasets with the mean values of the NLL allowing us to estimate the expected precision on the masses. 

Because the cross sections for the $\mu^+\mu^- \to \mu^+\mu^- \tilde{B} \tilde{B}$ process are roughly an order of magnitude larger than the same process with $e^+e^-$ in the final state, the measurement of the smuon mass is generally more precise than the measurement of the selectron mass. This can be significantly ameliorated, however, by using the fact that the neutralino mass measured in both processes is the same, and is driven mostly by the $\mu^+\mu^-$ measurement. 
To improve the precision on the selectron mass, we therefore add the likelihood as a function of the neutralino mass (marginalized over the smuon mass) to the likelihood obtained in the $e^+e^-$ measurement. 

\begin{figure}[h]
\vskip 0.25cm
\centering
\includegraphics[width=8cm]{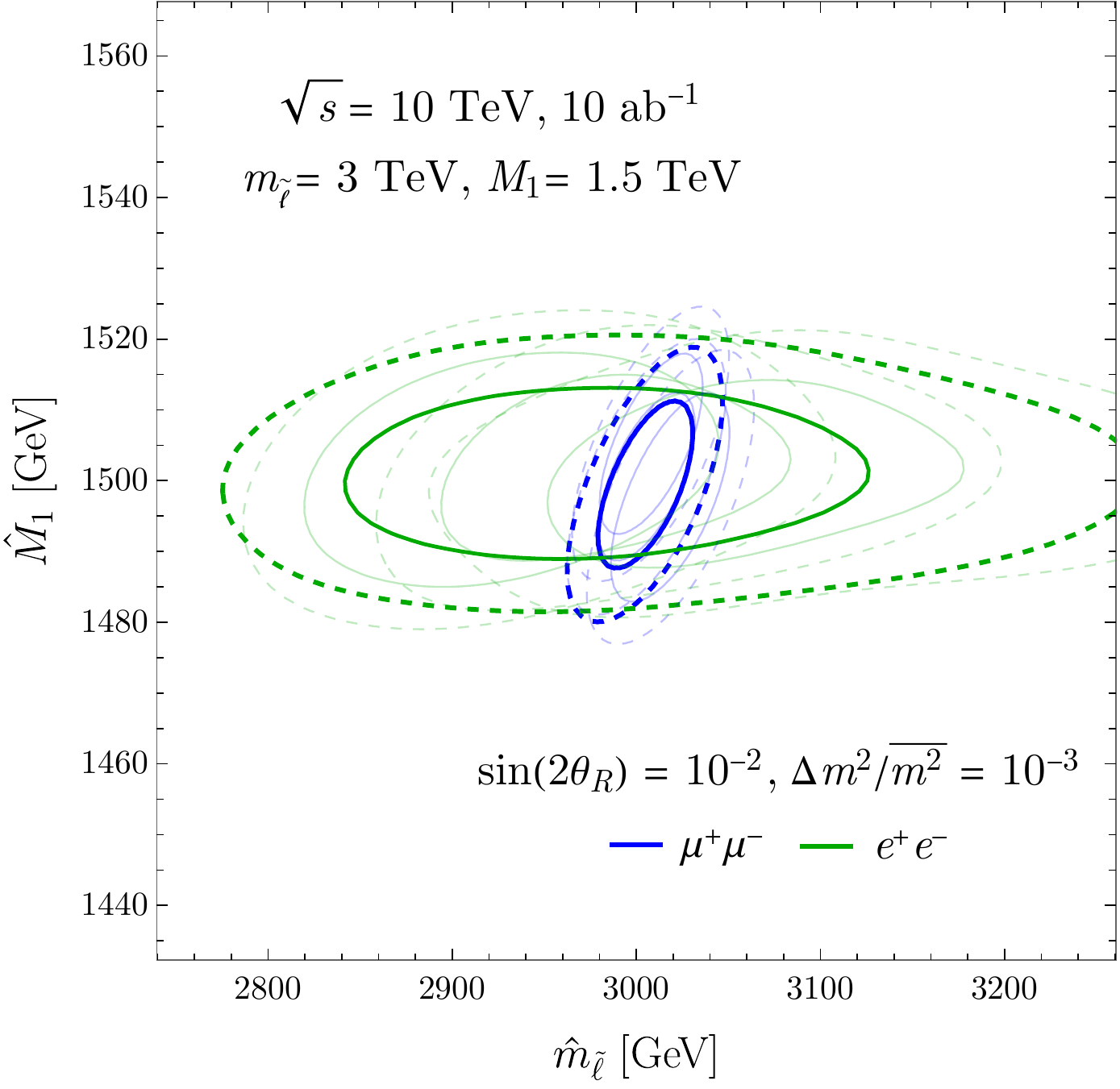}
\caption{
Contours of $\Delta\textrm{NLL}$ evaluated for the various toy datasets, as well as the average $\Delta\textrm{NLL}$ for ten such toys, both for the $\mu^+\mu^-$ (blue) and $e^+e^-$ (green) channels, at 68\% (solid) and 95\% (dashed) C.L.
}
\label{fig:likelihood_demonstration}
\end{figure}

The resulting $1$ and $2\sigma$ contours for several toy datasets, as well as the mean, are illustrated for $\sqrt{s} = 10\,\textrm{TeV}$ with both $\mu^+\mu^-$ (blue) and $e^+e^-$ (green) final states in Fig.~\ref{fig:likelihood_demonstration}. In practice, since the number of signal events tends to be relatively large at all the energies considered, we find a stable mean with 10 toy datasets.
This process can be repeated for different values of the true masses and mixing angles, and the resulting likelihoods are marginalized over to obtain one-dimensional likelihoods for each mass that can be converted to an expected precision.

\subsection{Additional Results and Discussion}
\label{subsec:mass_meas_discussion}

The results summarized here, and in Section~\ref{subsec:mass_measurement}, indicate that, despite the additional challenges present at a muon collider, slepton and neutralino masses can be measured with $\sim$ few $\%$ precision.
Our analysis neglects several important components that would be part of a more rigorous study: we have made no effort to include systematics associated with our parameterization of the background shapes, and taken all the nuisance parameters as fixed. These require a more detailed handling of the detector effects, but we expect they can be handled in a way that does not significantly affect the achievable precision.

Furthermore, as noted below Eq.~\eqref{eq:fc_processes}, we have assumed that the eigenstates we are measuring are flavor eigenstates. Explicitly, while the cross sections for $\mu^+\mu^- \to \mu^+\mu^- \tilde{B}\tilde{B}$ and $\mu^+\mu^- \to e^+e^- \tilde{B}\tilde{B}$ are taken to vary as a function of the mixing angle, we are assuming throughout that the masses measured in $\mu^+\mu^-$ and $e^+e^-$ final states are respectively the ``smuon'' and ``selectron'' masses, $m_{\tilde{\mu}}$ and $m_{\tilde{e}}$. This is certainly the case in the limit $\sin2\theta_R \to 0$, and is also a safe assumption when $\Delta m^2 \to 0$, since then the two mass eigenstates are degenerate. 
The only potential difficulty arises for large $\sin2\theta_R$ {\em and} large $\Delta m^2/\overline{m^2}$, in which case there are two distinct masses (and therefore two distinct endpoints) that would be measured in both flavor final states.
This is a problem, for instance, where $\Delta m^2 / \overline{m^2} = 0.1$, in which case the slepton mass splitting for $m_{\tilde{\ell}} = 3\,\textrm{TeV}$ is $\Delta m = 150\,\textrm{GeV}$, which is greater than the expected precision shown Fig.~\ref{fig:mass_meas_summary}.
However, the large $\sin2\theta_R$ and large $\Delta m^2 / \overline{m^2}$ regions are also where the projected low-energy constraints are the strongest, and in principle it is straightforward to generalize our analysis to measure two endpoints in the same signal spectrum.
Given that the muon collider reach extends to relatively small mixing angles, we do not expect these issues to significantly affect our results, and we leave a detailed study of this scenario to future work.

\section{Slepton Pair Reconstruction Analysis}
\label{app:reconstruction}

In this section we describe in detail how the two neutralino momenta in the process
\be
\mu^+\mu^- \rightarrow \tilde{\ell}^{\pm}_{1,2} \tilde{\ell}^{\mp}_{1,2}\rightarrow e^{\pm}\mu^{\mp}\tilde{B}\,\tilde{B}
\ee
can be partially reconstructed from the single missing momentum measured in the detector, given knowledge of the slepton and neutralino masses. To avoid confusion with the $\{1,2\}$ label for the slepton mass eigenstates, we label the two lepton and neutralinos in the event final state by $\{a, b\}$.

Assume that we have measured two lepton three-momenta ${\vec p}_{\ell a}$, ${\vec p}_{\ell b}$, with magnitude $E_{\ell i} = |{\vec p}_{\ell i}|$. The invisible momentum is taken to be ${\vec p}_\text{inv} = - {\vec p}_{\ell a} - {\vec p}_{\ell b}$. The total center-of-mass energy of the collision is $E_{\rm CM}$. We assume that $\ell_a$ and (invisible) $\tilde{B}_a$ reconstruct ${\tilde \ell}_a$ and that $\ell_b$ and (invisible) $\tilde{B}_b$ reconstruct ${\tilde \ell}_b$, where $\tilde{\ell}_{a, b}$ can be any of $\tilde{\ell}_{1,2}$.

We label the neutralino momenta as ${\vec p}_{\tilde{B} a}$ and ${\vec p}_{\tilde{B} b}$. These momenta can be parameterized in terms of one unknown 3-momentum $\vec k$, where
\begin{equation}
{\vec p}_{\tilde{B} a} = \frac{1}{2} {\vec p}_\text{inv} + {\vec k}, \qquad
{\vec p}_{\tilde{B} b} = \frac{1}{2} {\vec p}_\text{inv} - {\vec k}.
  \label{eq:pchiparam}
\end{equation}
We can rewrite the neutralino momenta in terms of the unknown 3-momentum $\vec{k}$. Writing this in terms of the lepton energies, we find:
\begin{equation}
E_{\tilde{B} a}^2 - E_{\tilde{B} b}^2 = (E_{\tilde{B} a} + E_{\tilde{B} b})(E_{\tilde{B} a} - E_{\tilde{B} b})= 2 {\vec p}_\text{inv} \cdot {\vec k}.
\end{equation}
Using the total energy conservation we have
\begin{equation}
E_{\tilde{B} a} - E_{\tilde{B} b} = \frac{2 {\vec p}_\text{inv} \cdot {\vec k}}{E_{\rm CM} - E_{\ell a} - E_{\ell b}}.
  \label{eq:Ediff}
\end{equation}

\noindent
The sleptons also need to satisfy the mass-shell conditions, $(E_{\tilde{B} i} + E_{\ell i})^2 - ({\vec p}_{\tilde{B} i} + {\vec p}_{\ell i})^2 = m_{{\tilde \ell} i}^2$.
Using that the lepton and neutralino are on-shell, we find that
\begin{equation}
E_{\tilde{B} i} = \frac{{\vec p}_{\tilde{B} i} \cdot {\vec p}_{\ell i} }{E_{\ell i}} + \frac{1}{2 E_{\ell i}} \left(m_{{\tilde \ell}_i}^2 - M_1^2\right).
\end{equation}
Plugging these formulas into \eqref{eq:Ediff}, using the parametrization \eqref{eq:pchiparam}, and writing $\vec{p}_{\rm inv} = -\vec{p}_{\ell a} - \vec{p}_{\ell b}$, we get two linear equations involving the two unknown quantities ${\vec k} \cdot {\vec p}_{\ell a}$, ${\vec k} \cdot {\vec p}_{\ell b}$, in terms of particle masses and visible momenta and energies only. The solution to these two linear equations is
\begin{align}
{\vec k} \cdot {\vec p}_{\ell a} &= +\left[\frac{E_{\rm CM} E_{\ell a}}{2} + \frac{E_{\ell a}}{2E_{\rm CM}} \left(m_{{\tilde \ell} a}^2 - m_{{\tilde \ell} b}^2\right) - \frac{1}{2} \left(m_{{\tilde \ell} a}^2 - M_1^2 + E_{\ell a}^2 - {\vec p}_{\ell a} \cdot {\vec p}_{\ell b}\right)\right], \nonumber \\
{\vec k} \cdot {\vec p}_{\ell b} &= -\left[\frac{E_{\rm CM} E_{\ell b}}{2} + \frac{E_{\ell b}}{2E_{\rm CM}} \left(m_{{\tilde \ell} b}^2 - m_{{\tilde \ell} a}^2\right) - \frac{1}{2} \left(m_{{\tilde \ell} b}^2 - M_1^2 + E_{\ell b}^2 - {\vec p}_{\ell a} \cdot {\vec p}_{\ell b}\right)\right].
   \label{eq:kcomponents}
\end{align}
Thus, we have determined two of the three components of the unknown vector. From this, we can calculate
\begin{align}
{\vec p}_{\tilde{B} a} \cdot {\vec p}_\text{inv} &= \frac{1}{2} E_{\rm CM} (E_{\ell b} - E_{\ell a}) - \frac{E_{\ell a} + E_{\ell b}}{2 E_{\rm CM}} \left(m_{{\tilde \ell} a}^2 - m_{{\tilde \ell} b}^2\right) + \frac{1}{2} \left(2 E_{\ell a}^2 + m_{{\tilde \ell} a}^2 - m_{{\tilde \ell} b}^2 + 2 {\vec p}_{\ell a} \cdot {\vec p}_{\ell b}\right), \nonumber \\
{\vec p}_{\tilde{B} b} \cdot {\vec p}_\text{inv} &= \frac{1}{2} E_{\rm CM} (E_{\ell a} - E_{\ell b}) - \frac{E_{\ell 1} + E_{\ell b}}{2 E_{\rm CM}} \left(m_{{\tilde \ell} b}^2 - m_{{\tilde \ell} a}^2\right) + \frac{1}{2} \left(2 E_{\ell b}^2 + m_{{\tilde \ell} b}^2 - m_{{\tilde \ell} a}^2 + 2 {\vec p}_{\ell a} \cdot {\vec p}_{\ell b}\right).
\end{align}
Note that these equations are exchanged by $a\leftrightarrow b$ and sum to $|{\vec p}_\text{inv}|^2 = E_{\ell a}^2 + E_{\ell b}^2 + 2 {\vec p}_{\ell a} \cdot {\vec p}_{\ell b}$, as they should.

The third component of the unknown momentum should be fixed by conservation of energy. Therefore, if we find that the components we already found lead to more energy than $E_{\rm CM}$, we conclude the event is not reconstructible. At this point, it is useful to introduce a coordinate system. We take the invisible 3-momentum to point along the $z$-axis, and the lepton 3-momenta  to  lie in the $(x,z)$ plane. We can translate our constraints on ${\vec k} \cdot {\vec p}_{\ell a}$ and ${\vec k} \cdot {\vec p}_{\ell b}$ into the $x$ and $z$ components of $k$, with the $y$-component unknown (and only increasing the energy).

We take the invisible 3-momentum and the lepton 3-momenta to be:
\begin{equation}
{\vec p}_\text{inv} = (0, 0,   p_\text{inv}), \qquad
{\vec p}_{\ell a} = (p_{\ell x},  0,  p_{\ell z}), \qquad
{\vec p}_{\ell b} = (-p_{\ell x}, 0,  - p_\text{inv} - p_{\ell z}).
\end{equation}
Note that these quantities can be easily calculated starting from a general frame: $p_\text{inv} = |{\vec p}_{\ell a} + {\vec p}_{\ell b}|$, $p_{\ell z} = ({\vec p}_{\ell a} \cdot {\vec p}_\text{inv})/p_\text{inv}$, and $p_{\ell x} = \sqrt{|{\vec p}_{\ell a}|^2 - p_{\ell z}^2}$.
In this frame, we have
\begin{equation}
{\vec p}_{\tilde{B} a} = (p_{\tilde{B}a, x}, p_{\tilde{B}a, y}, p_{\tilde{B}a, z}), \qquad
{\vec p}_{\tilde{B} b} = (-p_{\tilde{B}a, x}, -p_{\tilde{B}a, y}, p_\text{inv} - p_{\tilde{B}a, z}).
\end{equation}
Using \Eq{eq:kcomponents}, we can solve for the unknown components $p_{\tilde{B}a, x}$ and $p_{\tilde{B}a,z}$:
\begin{align}
p_{\tilde{B}a, x} &= \frac{1}{2 E_{\rm CM} p_\text{inv} p_{\ell x}} \Big[E_{\rm CM}^2 (-E_{\ell b} p_{\ell z} + E_{\ell a} (p_\text{inv} + p_{\ell z})) + (m_{{\tilde \ell} a}^2 - m_{{\tilde \ell} b}^2) (E_{\ell b} p_{\ell z} +  E_{\ell a} (p_\text{inv} + p_{\ell z})) ~+  \nonumber \\
&  \qquad   ~E_{\rm CM}
 \big( p_\text{inv} ( M_1^2 - m_{{\tilde \ell} a}^2+ {\vec p}_{\ell a} \cdot {\vec p}_{\ell b} - E_{\ell a}^2) + p_{\ell z}(E_{\ell b}^2 - m_{{\tilde \ell} a}^2 + m_{{\tilde \ell} 2}^2 - E_{\ell a}^2) \big)\Big], \nonumber \\[0.5em]
p_{\tilde{B}a, z} &= \frac{1}{2 p_\text{inv}} \Big[E_{\rm CM} (E_{\ell b} - E_{\ell a}) - \frac{E_{\ell a} + E_{\ell b}}{E_{\rm CM}} \big(m_{{\tilde \ell} a}^2 - m_{{\tilde \ell} b}^2\big) + \big(E_{\ell a}^2-E_{\ell b}^2 + m_{{\tilde \ell} a}^2 - m_{{\tilde \ell} b}^2 + 2 {\vec p}_{\ell a} \cdot {\vec p}_{\ell b}\big)\Big].
\end{align}
Now, the total energy of the two neutralinos is bounded below:
\begin{equation}
E_{\tilde{B} a} + E_{\tilde{B} b} \geq \mathcal{E}_{\tilde{B}a} + \mathcal{E}_{\tilde{B}b} \equiv 
\sqrt{p_{\tilde{B}a, x}^2 + p_{\tilde{B}a, z}^2 + M_1^2} + \sqrt{p_{\tilde{B}a, x}^2 +  (p_\text{inv} - p_{\tilde{B}a, z})^2 + M_1^2}.
\end{equation}
If $\mathcal{E}_{\tilde{B}a} + \mathcal{E}_{\tilde{B}b}$ is larger than $E_{\rm CM} - E_{\ell a} - E_{\ell b}$, then no real value of $p_{\tilde{B}a, y}$ will lead to a valid solution of the system.

\end{spacing}

\addcontentsline{toc}{section}{References} 
{\small
\bibliographystyle{utphys}
\bibliography{mssm_lfv}

\providecommand{\href}[2]{#2}\begingroup\raggedright\begin{thebibliography}{100}

\bibitem{Isidori:2010kg}
G.~Isidori, Y.~Nir, and G.~Perez, ``{Flavor Physics Constraints for Physics
  Beyond the Standard Model},''
  \href{http://dx.doi.org/10.1146/annurev.nucl.012809.104534}{{\em Ann. Rev.
  Nucl. Part. Sci.} {\bf 60} (2010)  355},
  \href{http://arxiv.org/abs/1002.0900}{{\tt arXiv:1002.0900 [hep-ph]}}.

\bibitem{Delahaye:2019omf}
J.~P. Delahaye, M.~Diemoz, K.~Long, B.~Mansouli\'e, N.~Pastrone, L.~Rivkin,
  D.~Schulte, A.~Skrinsky, and A.~Wulzer, ``{Muon Colliders},''
  \href{http://arxiv.org/abs/1901.06150}{{\tt arXiv:1901.06150
  [physics.acc-ph]}}.

\bibitem{AlAli:2021let}
H.~Al~Ali {\em et al.}, ``{The Muon Smasher's Guide},''
  \href{http://arxiv.org/abs/2103.14043}{{\tt arXiv:2103.14043 [hep-ph]}}.

\bibitem{Aime:2022flm}
C.~Aim\`e {\em et al.}, ``{Muon Collider Physics Summary},''
\newblock 2022.
\newblock \href{http://arxiv.org/abs/2203.07256}{{\tt arXiv:2203.07256
  [hep-ph]}}.

\bibitem{DeBlas:2022wxr}
J.~de~Blas {\em et al.}, ``{The physics case of a 3 TeV muon collider stage},''
\newblock 2022.
\newblock \href{http://arxiv.org/abs/2203.07261}{{\tt arXiv:2203.07261
  [hep-ph]}}.

\bibitem{Baldini:2018uhj}
A.~Baldini {\em et al.}, ``{A submission to the 2020 update of the European
  Strategy for Particle Physics on behalf of the COMET, MEG, Mu2e and Mu3e
  collaborations},'' \href{http://arxiv.org/abs/1812.06540}{{\tt
  arXiv:1812.06540 [hep-ex]}}.

\bibitem{Middleton:2022fvu}
S.~Middleton, M.~Lee, and Y.~Seiya, ``{Experimental Searches for Muon to
  Electron Conversion in a Nucleus: COMET, DeeMe, and Mu2e. A Contributed paper
  for Snowmass 21},''
\newblock 2022.
\newblock \href{http://arxiv.org/abs/2203.07089}{{\tt arXiv:2203.07089
  [hep-ex]}}.

\bibitem{Agashe:2022uih}
K.~Agashe, M.~Ekhterachian, Z.~Liu, and R.~Sundrum, ``{Sleptonic SUSY: From UV
  Framework to IR Phenomenology},'' \href{http://arxiv.org/abs/2203.01796}{{\tt
  arXiv:2203.01796 [hep-ph]}}.

\bibitem{Calibbi:2015kja}
L.~Calibbi, I.~Galon, A.~Masiero, P.~Paradisi, and Y.~Shadmi, ``{Charged
  Slepton Flavor post the 8 TeV LHC: A Simplified Model Analysis of Low-Energy
  Constraints and LHC SUSY Searches},''
  \href{http://dx.doi.org/10.1007/JHEP10(2015)043}{{\em JHEP} {\bf 10} (2015)
  043}, \href{http://arxiv.org/abs/1502.07753}{{\tt arXiv:1502.07753
  [hep-ph]}}.

\bibitem{Gabbiani:1996hi}
F.~Gabbiani, E.~Gabrielli, A.~Masiero, and L.~Silvestrini, ``{A Complete
  analysis of FCNC and CP constraints in general SUSY extensions of the
  standard model},'' \href{http://dx.doi.org/10.1016/0550-3213(96)00390-2}{{\em
  Nucl. Phys. B} {\bf 477} (1996)  321--352},
  \href{http://arxiv.org/abs/hep-ph/9604387}{{\tt arXiv:hep-ph/9604387}}.

\bibitem{Arkani-Hamed:1996bxi}
N.~Arkani-Hamed, H.-C. Cheng, J.~L. Feng, and L.~J. Hall, ``{Probing lepton
  flavor violation at future colliders},''
  \href{http://dx.doi.org/10.1103/PhysRevLett.77.1937}{{\em Phys. Rev. Lett.}
  {\bf 77} (1996)  1937--1940}, \href{http://arxiv.org/abs/hep-ph/9603431}{{\tt
  arXiv:hep-ph/9603431}}.

\bibitem{Krasnikov:1995qq}
N.~Krasnikov, ``{Search for flavor lepton number violation in slepton decays at
  LEP-2 and NLC},'' \href{http://dx.doi.org/10.1016/S0370-2693(96)01213-0}{{\em
  Phys. Lett. B} {\bf 388} (1996)  783--787},
  \href{http://arxiv.org/abs/hep-ph/9511464}{{\tt arXiv:hep-ph/9511464}}.

\bibitem{Hisano:1995cp}
J.~Hisano, T.~Moroi, K.~Tobe, and M.~Yamaguchi, ``{Lepton flavor violation via
  right-handed neutrino Yukawa couplings in supersymmetric standard model},''
  \href{http://dx.doi.org/10.1103/PhysRevD.53.2442}{{\em Phys. Rev. D} {\bf 53}
  (1996)  2442--2459}, \href{http://arxiv.org/abs/hep-ph/9510309}{{\tt
  arXiv:hep-ph/9510309}}.

\bibitem{Kitano:2002mt}
R.~Kitano, M.~Koike, and Y.~Okada, ``{Detailed calculation of lepton flavor
  violating muon electron conversion rate for various nuclei},''
  \href{http://dx.doi.org/10.1103/PhysRevD.76.059902}{{\em Phys. Rev. D} {\bf
  66} (2002)  096002}, \href{http://arxiv.org/abs/hep-ph/0203110}{{\tt
  arXiv:hep-ph/0203110}}. [Erratum: Phys.Rev.D 76, 059902 (2007)].

\bibitem{Masina:2002mv}
I.~Masina and C.~A. Savoy, ``{Sleptonarium: Constraints on the CP and flavor
  pattern of scalar lepton masses},''
  \href{http://dx.doi.org/10.1016/S0550-3213(03)00294-3}{{\em Nucl. Phys. B}
  {\bf 661} (2003)  365--393}, \href{http://arxiv.org/abs/hep-ph/0211283}{{\tt
  arXiv:hep-ph/0211283}}.

\bibitem{Arganda:2005ji}
E.~Arganda and M.~J. Herrero, ``{Testing supersymmetry with lepton flavor
  violating tau and mu decays},''
  \href{http://dx.doi.org/10.1103/PhysRevD.73.055003}{{\em Phys. Rev. D} {\bf
  73} (2006)  055003}, \href{http://arxiv.org/abs/hep-ph/0510405}{{\tt
  arXiv:hep-ph/0510405}}.

\bibitem{Arganda:2007jw}
E.~Arganda, M.~J. Herrero, and A.~M. Teixeira, ``{mu-e conversion in nuclei
  within the CMSSM seesaw: Universality versus non-universality},''
  \href{http://dx.doi.org/10.1088/1126-6708/2007/10/104}{{\em JHEP} {\bf 10}
  (2007)  104}, \href{http://arxiv.org/abs/0707.2955}{{\tt arXiv:0707.2955
  [hep-ph]}}.

\bibitem{Altmannshofer:2013lfa}
W.~Altmannshofer, R.~Harnik, and J.~Zupan, ``{Low Energy Probes of PeV Scale
  Sfermions},'' \href{http://dx.doi.org/10.1007/JHEP11(2013)202}{{\em JHEP}
  {\bf 11} (2013)  202}, \href{http://arxiv.org/abs/1308.3653}{{\tt
  arXiv:1308.3653 [hep-ph]}}.

\bibitem{Moroi:2013sfa}
T.~Moroi and M.~Nagai, ``{Probing Supersymmetric Model with Heavy Sfermions
  Using Leptonic Flavor and CP Violations},''
  \href{http://dx.doi.org/10.1016/j.physletb.2013.04.049}{{\em Phys. Lett. B}
  {\bf 723} (2013)  107--112}, \href{http://arxiv.org/abs/1303.0668}{{\tt
  arXiv:1303.0668 [hep-ph]}}.

\bibitem{Ellis:2016yje}
S.~A.~R. Ellis and A.~Pierce, ``{Impact of Future Lepton Flavor Violation
  Measurements in the Minimal Supersymmetric Standard Model},''
  \href{http://dx.doi.org/10.1103/PhysRevD.94.015014}{{\em Phys. Rev. D} {\bf
  94} (2016) no.~1, 015014}, \href{http://arxiv.org/abs/1604.01419}{{\tt
  arXiv:1604.01419 [hep-ph]}}.

\bibitem{Calibbi:2017uvl}
L.~Calibbi and G.~Signorelli, ``{Charged Lepton Flavour Violation: An
  Experimental and Theoretical Introduction},''
  \href{http://dx.doi.org/10.1393/ncr/i2018-10144-0}{{\em Riv. Nuovo Cim.} {\bf
  41} (2018) no.~2, 71--174}, \href{http://arxiv.org/abs/1709.00294}{{\tt
  arXiv:1709.00294 [hep-ph]}}.

\bibitem{Crivellin:2018mqz}
A.~Crivellin, Z.~Fabisiewicz, W.~Materkowska, U.~Nierste, S.~Pokorski, and
  J.~Rosiek, ``{Lepton flavour violation in the MSSM: exact diagonalization vs
  mass expansion},'' \href{http://dx.doi.org/10.1007/JHEP06(2018)003}{{\em
  JHEP} {\bf 06} (2018)  003}, \href{http://arxiv.org/abs/1802.06803}{{\tt
  arXiv:1802.06803 [hep-ph]}}.

\bibitem{Aloni:2021wzk}
D.~Aloni, P.~Asadi, Y.~Nakai, M.~Reece, and M.~Suzuki, ``{Spontaneous CP
  violation and horizontal symmetry in the MSSM: toward lepton flavor
  naturalness},'' \href{http://dx.doi.org/10.1007/JHEP09(2021)031}{{\em JHEP}
  {\bf 09} (2021)  031}, \href{http://arxiv.org/abs/2104.02679}{{\tt
  arXiv:2104.02679 [hep-ph]}}.

\bibitem{Crivellin:2018qmi}
A.~Crivellin, M.~Hoferichter, and P.~Schmidt-Wellenburg, ``{Combined
  explanations of $(g-2)_{\mu,e}$ and implications for a large muon EDM},''
  \href{http://dx.doi.org/10.1103/PhysRevD.98.113002}{{\em Phys. Rev. D} {\bf
  98} (2018) no.~11, 113002}, \href{http://arxiv.org/abs/1807.11484}{{\tt
  arXiv:1807.11484 [hep-ph]}}.

\bibitem{TheMEG:2016wtm}
{\bf MEG} Collaboration, A.~Baldini {\em et al.}, ``{Search for the lepton
  flavour violating decay $\mu ^+ \rightarrow \mathrm {e}^+ \gamma $ with the
  full dataset of the MEG experiment},''
  \href{http://dx.doi.org/10.1140/epjc/s10052-016-4271-x}{{\em Eur. Phys. J. C}
  {\bf 76} (2016) no.~8, 434}, \href{http://arxiv.org/abs/1605.05081}{{\tt
  arXiv:1605.05081 [hep-ex]}}.

\bibitem{Baldini:2018nnn}
{\bf MEG II} Collaboration, A.~Baldini {\em et al.}, ``{The design of the MEG
  II experiment},''
  \href{http://dx.doi.org/10.1140/epjc/s10052-018-5845-6}{{\em Eur. Phys. J. C}
  {\bf 78} (2018) no.~5, 380}, \href{http://arxiv.org/abs/1801.04688}{{\tt
  arXiv:1801.04688 [physics.ins-det]}}.

\bibitem{Bellgardt:1987du}
{\bf SINDRUM} Collaboration, U.~Bellgardt {\em et al.}, ``{Search for the Decay
  $\mu^+ \to e^+ e^+ e^-$},''
  \href{http://dx.doi.org/10.1016/0550-3213(88)90462-2}{{\em Nucl. Phys. B}
  {\bf 299} (1988)  1--6}.

\bibitem{Mu3e:2020gyw}
{\bf Mu3e} Collaboration, K.~Arndt {\em et al.}, ``{Technical design of the
  phase I Mu3e experiment},''
  \href{http://dx.doi.org/10.1016/j.nima.2021.165679}{{\em Nucl. Instrum. Meth.
  A} {\bf 1014} (2021)  165679}, \href{http://arxiv.org/abs/2009.11690}{{\tt
  arXiv:2009.11690 [physics.ins-det]}}.

\bibitem{Bertl:2006up}
{\bf SINDRUM II} Collaboration, W.~H. Bertl {\em et al.}, ``{A Search for muon
  to electron conversion in muonic gold},''
  \href{http://dx.doi.org/10.1140/epjc/s2006-02582-x}{{\em Eur. Phys. J. C}
  {\bf 47} (2006)  337--346}.

\bibitem{Adamov:2018vin}
{\bf COMET} Collaboration, R.~Abramishvili {\em et al.}, ``{COMET Phase-I
  Technical Design Report},'' \href{http://dx.doi.org/10.1093/ptep/ptz125}{{\em
  PTEP} {\bf 2020} (2020) no.~3, 033C01},
  \href{http://arxiv.org/abs/1812.09018}{{\tt arXiv:1812.09018
  [physics.ins-det]}}.

\bibitem{Angelique:2018svf}
{\bf COMET} Collaboration, J.~Ang\'elique {\em et al.}, ``{COMET - A submission
  to the 2020 update of the European Strategy for Particle Physics on behalf of
  the COMET collaboration},'' \href{http://arxiv.org/abs/1812.07824}{{\tt
  arXiv:1812.07824 [hep-ex]}}.

\bibitem{Bartoszek:2014mya}
{\bf Mu2e} Collaboration, L.~Bartoszek {\em et al.}, ``{Mu2e Technical Design
  Report},'' \href{http://arxiv.org/abs/1501.05241}{{\tt arXiv:1501.05241
  [physics.ins-det]}}.

\bibitem{Abusalma:2018xem}
{\bf Mu2e} Collaboration, F.~Abusalma {\em et al.}, ``{Expression of Interest
  for Evolution of the Mu2e Experiment},''
  \href{http://arxiv.org/abs/1802.02599}{{\tt arXiv:1802.02599
  [physics.ins-det]}}.

\bibitem{Kuno:2005mm}
Y.~Kuno, ``{PRISM/PRIME},''
  \href{http://dx.doi.org/10.1016/j.nuclphysbps.2005.05.073}{{\em Nucl. Phys. B
  Proc. Suppl.} {\bf 149} (2005)  376--378}.

\bibitem{Pasternak:2010zz}
J.~Pasternak {\em et al.}, ``{Accelerator and Particle Physics Research for the
  Next Generation Muon to Electron Conversion Experiment - the PRISM Task
  Force},'' {\em Conf. Proc. C} {\bf 100523} (2010)  WEPE056.

\bibitem{Altmannshofer:2009ne}
W.~Altmannshofer, A.~J. Buras, S.~Gori, P.~Paradisi, and D.~M. Straub,
  ``{Anatomy and Phenomenology of FCNC and CPV Effects in SUSY Theories},''
  \href{http://dx.doi.org/10.1016/j.nuclphysb.2009.12.019}{{\em Nucl. Phys. B}
  {\bf 830} (2010)  17--94}, \href{http://arxiv.org/abs/0909.1333}{{\tt
  arXiv:0909.1333 [hep-ph]}}.

\bibitem{McKeen:2013dma}
D.~McKeen, M.~Pospelov, and A.~Ritz, ``{Electric dipole moment signatures of
  PeV-scale superpartners},''
  \href{http://dx.doi.org/10.1103/PhysRevD.87.113002}{{\em Phys. Rev. D} {\bf
  87} (2013) no.~11, 113002}, \href{http://arxiv.org/abs/1303.1172}{{\tt
  arXiv:1303.1172 [hep-ph]}}.

\bibitem{Capdevilla:2020qel}
R.~Capdevilla, D.~Curtin, Y.~Kahn, and G.~Krnjaic, ``{Discovering the physics
  of $(g-2)_\mu$ at future muon colliders},''
  \href{http://dx.doi.org/10.1103/PhysRevD.103.075028}{{\em Phys. Rev. D} {\bf
  103} (2021) no.~7, 075028}, \href{http://arxiv.org/abs/2006.16277}{{\tt
  arXiv:2006.16277 [hep-ph]}}.

\bibitem{Buttazzo:2020ibd}
D.~Buttazzo and P.~Paradisi, ``{Probing the muon $g-2$ anomaly with the Higgs
  boson at a muon collider},''
  \href{http://dx.doi.org/10.1103/PhysRevD.104.075021}{{\em Phys. Rev. D} {\bf
  104} (2021) no.~7, 075021}, \href{http://arxiv.org/abs/2012.02769}{{\tt
  arXiv:2012.02769 [hep-ph]}}.

\bibitem{Yin:2020afe}
W.~Yin and M.~Yamaguchi, ``{Muon $g-2$ at multi-TeV muon collider},''
  \href{http://arxiv.org/abs/2012.03928}{{\tt arXiv:2012.03928 [hep-ph]}}.

\bibitem{Capdevilla:2021rwo}
R.~Capdevilla, D.~Curtin, Y.~Kahn, and G.~Krnjaic, ``{No-lose theorem for
  discovering the new physics of (g-2)\ensuremath{\mu} at muon colliders},''
  \href{http://dx.doi.org/10.1103/PhysRevD.105.015028}{{\em Phys. Rev. D} {\bf
  105} (2022) no.~1, 015028}, \href{http://arxiv.org/abs/2101.10334}{{\tt
  arXiv:2101.10334 [hep-ph]}}.

\bibitem{Chen:2021rnl}
N.~Chen, B.~Wang, and C.-Y. Yao, ``{The collider tests of a leptophilic scalar
  for the anomalous magnetic moments},''
  \href{http://arxiv.org/abs/2102.05619}{{\tt arXiv:2102.05619 [hep-ph]}}.

\bibitem{Huang:2021biu}
G.-y. Huang, S.~Jana, F.~S. Queiroz, and W.~Rodejohann, ``{Probing the RK(*)
  anomaly at a muon collider},''
  \href{http://dx.doi.org/10.1103/PhysRevD.105.015013}{{\em Phys. Rev. D} {\bf
  105} (2022) no.~1, 015013}, \href{http://arxiv.org/abs/2103.01617}{{\tt
  arXiv:2103.01617 [hep-ph]}}.

\bibitem{Li:2021lnz}
T.~Li, M.~A. Schmidt, C.-Y. Yao, and M.~Yuan, ``{Charged lepton flavor
  violation in light of the muon magnetic moment anomaly and colliders},''
  \href{http://dx.doi.org/10.1140/epjc/s10052-021-09569-9}{{\em Eur. Phys. J.
  C} {\bf 81} (2021) no.~09, 811}, \href{http://arxiv.org/abs/2104.04494}{{\tt
  arXiv:2104.04494 [hep-ph]}}.

\bibitem{Asadi:2021gah}
P.~Asadi, R.~Capdevilla, C.~Cesarotti, and S.~Homiller, ``{Searching for
  leptoquarks at future muon colliders},''
  \href{http://dx.doi.org/10.1007/JHEP10(2021)182}{{\em JHEP} {\bf 10} (2021)
  182}, \href{http://arxiv.org/abs/2104.05720}{{\tt arXiv:2104.05720
  [hep-ph]}}.

\bibitem{Haghighat:2021djz}
G.~Haghighat and M.~Mohammadi~Najafabadi, ``{Search for lepton-flavor-violating
  ALPs at a future muon collider and utilization of polarization-induced
  effects},'' \href{http://arxiv.org/abs/2106.00505}{{\tt arXiv:2106.00505
  [hep-ph]}}.

\bibitem{Bandyopadhyay:2021pld}
P.~Bandyopadhyay, A.~Karan, and R.~Mandal, ``{Distinguishing signatures of
  scalar leptoquarks at hadron and muon colliders},''
  \href{http://arxiv.org/abs/2108.06506}{{\tt arXiv:2108.06506 [hep-ph]}}.

\bibitem{Dermisek:2021mhi}
R.~Dermisek, K.~Hermanek, and N.~McGinnis, ``{Di-Higgs and tri-Higgs boson
  signals of muon g-2 at a muon collider},''
  \href{http://dx.doi.org/10.1103/PhysRevD.104.L091301}{{\em Phys. Rev. D} {\bf
  104} (2021) no.~9, L091301}, \href{http://arxiv.org/abs/2108.10950}{{\tt
  arXiv:2108.10950 [hep-ph]}}.

\bibitem{Capdevilla:2021kcf}
R.~Capdevilla, D.~Curtin, Y.~Kahn, and G.~Krnjaic, ``{Systematically Testing
  Singlet Models for $(g-2)_\mu$},''
  \href{http://arxiv.org/abs/2112.08377}{{\tt arXiv:2112.08377 [hep-ph]}}.

\bibitem{Medina:2021ram}
A.~D. Medina, N.~I. Mileo, A.~Szynkman, and S.~A. Tanco, ``{The Elusive Muonic
  WIMP},'' \href{http://arxiv.org/abs/2112.09103}{{\tt arXiv:2112.09103
  [hep-ph]}}.

\bibitem{Chen:2016wkt}
J.~Chen, T.~Han, and B.~Tweedie, ``{Electroweak Splitting Functions and High
  Energy Showering},'' \href{http://dx.doi.org/10.1007/JHEP11(2017)093}{{\em
  JHEP} {\bf 11} (2017)  093}, \href{http://arxiv.org/abs/1611.00788}{{\tt
  arXiv:1611.00788 [hep-ph]}}.

\bibitem{Han:2020uid}
T.~Han, Y.~Ma, and K.~Xie, ``{High energy leptonic collisions and electroweak
  parton distribution functions},''
  \href{http://dx.doi.org/10.1103/PhysRevD.103.L031301}{{\em Phys. Rev. D} {\bf
  103} (2021) no.~3, L031301}, \href{http://arxiv.org/abs/2007.14300}{{\tt
  arXiv:2007.14300 [hep-ph]}}.

\bibitem{Han:2021kes}
T.~Han, Y.~Ma, and K.~Xie, ``{Quark and gluon contents of a lepton at high
  energies},'' \href{http://dx.doi.org/10.1007/JHEP02(2022)154}{{\em JHEP} {\bf
  02} (2022)  154}, \href{http://arxiv.org/abs/2103.09844}{{\tt
  arXiv:2103.09844 [hep-ph]}}.

\bibitem{Ruiz:2021tdt}
R.~Ruiz, A.~Costantini, F.~Maltoni, and O.~Mattelaer, ``{The Effective Vector
  Boson Approximation in High-Energy Muon Collisions},''
  \href{http://arxiv.org/abs/2111.02442}{{\tt arXiv:2111.02442 [hep-ph]}}.

\bibitem{Eichten:2013ckl}
E.~Eichten and A.~Martin, ``{The Muon Collider as a $H/A$ Factory},''
  \href{http://dx.doi.org/10.1016/j.physletb.2013.11.035}{{\em Phys. Lett. B}
  {\bf 728} (2014)  125--130}, \href{http://arxiv.org/abs/1306.2609}{{\tt
  arXiv:1306.2609 [hep-ph]}}.

\bibitem{Buttazzo:2018qqp}
D.~Buttazzo, D.~Redigolo, F.~Sala, and A.~Tesi, ``{Fusing Vectors into Scalars
  at High Energy Lepton Colliders},''
  \href{http://dx.doi.org/10.1007/JHEP11(2018)144}{{\em JHEP} {\bf 11} (2018)
  144}, \href{http://arxiv.org/abs/1807.04743}{{\tt arXiv:1807.04743
  [hep-ph]}}.

\bibitem{Chakrabarty:2014pja}
N.~Chakrabarty, T.~Han, Z.~Liu, and B.~Mukhopadhyaya, ``{Radiative Return for
  Heavy Higgs Boson at a Muon Collider},''
  \href{http://dx.doi.org/10.1103/PhysRevD.91.015008}{{\em Phys. Rev. D} {\bf
  91} (2015) no.~1, 015008}, \href{http://arxiv.org/abs/1408.5912}{{\tt
  arXiv:1408.5912 [hep-ph]}}.

\bibitem{Bandyopadhyay:2020otm}
P.~Bandyopadhyay and A.~Costantini, ``{Obscure Higgs boson at Colliders},''
  \href{http://dx.doi.org/10.1103/PhysRevD.103.015025}{{\em Phys. Rev. D} {\bf
  103} (2021) no.~1, 015025}, \href{http://arxiv.org/abs/2010.02597}{{\tt
  arXiv:2010.02597 [hep-ph]}}.

\bibitem{Liu:2021jyc}
W.~Liu and K.-P. Xie, ``{Probing electroweak phase transition with multi-TeV
  muon colliders and gravitational waves},''
  \href{http://dx.doi.org/10.1007/JHEP04(2021)015}{{\em JHEP} {\bf 04} (2021)
  015}, \href{http://arxiv.org/abs/2101.10469}{{\tt arXiv:2101.10469
  [hep-ph]}}.

\bibitem{Han:2021udl}
T.~Han, S.~Li, S.~Su, W.~Su, and Y.~Wu, ``{Heavy Higgs bosons in 2HDM at a muon
  collider},'' \href{http://dx.doi.org/10.1103/PhysRevD.104.055029}{{\em Phys.
  Rev. D} {\bf 104} (2021) no.~5, 055029},
  \href{http://arxiv.org/abs/2102.08386}{{\tt arXiv:2102.08386 [hep-ph]}}.

\bibitem{Liu:2021akf}
W.~Liu, K.-P. Xie, and Z.~Yi, ``{Testing leptogenesis at the LHC and future
  muon colliders: a $Z'$ scenario},''
  \href{http://arxiv.org/abs/2109.15087}{{\tt arXiv:2109.15087 [hep-ph]}}.

\bibitem{Chen:2022msz}
S.~Chen, A.~Glioti, R.~Rattazzi, L.~Ricci, and A.~Wulzer, ``{Learning from
  Radiation at a Very High Energy Lepton Collider},''
  \href{http://arxiv.org/abs/2202.10509}{{\tt arXiv:2202.10509 [hep-ph]}}.

\bibitem{Bao:2022onq}
Y.~Bao, J.~Fan, and L.~Li, ``{Electroweak ALP Searches at a Muon Collider},''
  \href{http://arxiv.org/abs/2203.04328}{{\tt arXiv:2203.04328 [hep-ph]}}.

\bibitem{DiLuzio:2018jwd}
L.~Di~Luzio, R.~Gr\"ober, and G.~Panico, ``{Probing new electroweak states via
  precision measurements at the LHC and future colliders},''
  \href{http://dx.doi.org/10.1007/JHEP01(2019)011}{{\em JHEP} {\bf 01} (2019)
  011}, \href{http://arxiv.org/abs/1810.10993}{{\tt arXiv:1810.10993
  [hep-ph]}}.

\bibitem{Buttazzo:2020uzc}
D.~Buttazzo, R.~Franceschini, and A.~Wulzer, ``{Two Paths Towards Precision at
  a Very High Energy Lepton Collider},''
  \href{http://dx.doi.org/10.1007/JHEP05(2021)219}{{\em JHEP} {\bf 05} (2021)
  219}, \href{http://arxiv.org/abs/2012.11555}{{\tt arXiv:2012.11555
  [hep-ph]}}.

\bibitem{Chiesa:2020awd}
M.~Chiesa, F.~Maltoni, L.~Mantani, B.~Mele, F.~Piccinini, and X.~Zhao,
  ``{Measuring the quartic Higgs self-coupling at a multi-TeV muon collider},''
  \href{http://dx.doi.org/10.1007/JHEP09(2020)098}{{\em JHEP} {\bf 09} (2020)
  098}, \href{http://arxiv.org/abs/2003.13628}{{\tt arXiv:2003.13628
  [hep-ph]}}.

\bibitem{Costantini:2020stv}
A.~Costantini, F.~De~Lillo, F.~Maltoni, L.~Mantani, O.~Mattelaer, R.~Ruiz, and
  X.~Zhao, ``{Vector boson fusion at multi-TeV muon colliders},''
  \href{http://dx.doi.org/10.1007/JHEP09(2020)080}{{\em JHEP} {\bf 09} (2020)
  080}, \href{http://arxiv.org/abs/2005.10289}{{\tt arXiv:2005.10289
  [hep-ph]}}.

\bibitem{Han:2020pif}
T.~Han, D.~Liu, I.~Low, and X.~Wang, ``{Electroweak couplings of the Higgs
  boson at a multi-TeV muon collider},''
  \href{http://dx.doi.org/10.1103/PhysRevD.103.013002}{{\em Phys. Rev. D} {\bf
  103} (2021) no.~1, 013002}, \href{http://arxiv.org/abs/2008.12204}{{\tt
  arXiv:2008.12204 [hep-ph]}}.

\bibitem{Chiesa:2021qpr}
M.~Chiesa, B.~Mele, and F.~Piccinini, ``{Multi Higgs production via photon
  fusion at future multi-TeV muon colliders},''
  \href{http://arxiv.org/abs/2109.10109}{{\tt arXiv:2109.10109 [hep-ph]}}.

\bibitem{Cepeda:2021rql}
M.~Cepeda, S.~Gori, V.~M. Outschoorn, and J.~Shelton, ``{Exotic Higgs
  Decays},'' \href{http://arxiv.org/abs/2111.12751}{{\tt arXiv:2111.12751
  [hep-ph]}}.

\bibitem{Chen:2021pqi}
J.~Chen, T.~Li, C.-T. Lu, Y.~Wu, and C.-Y. Yao, ``{The measurement of Higgs
  self-couplings through $2\rightarrow 3$ VBS in future muon colliders},''
  \href{http://arxiv.org/abs/2112.12507}{{\tt arXiv:2112.12507 [hep-ph]}}.

\bibitem{Spor:2022mxl}
S.~Spor and M.~K\"oksal, ``{Investigation of anomalous triple gauge couplings
  in $\mu\gamma$ collision at multi-TeV muon colliders},''
  \href{http://arxiv.org/abs/2201.00787}{{\tt arXiv:2201.00787 [hep-ph]}}.

\bibitem{Buonincontri:2022ylv}
{\bf Muon Collider Physics and Detector Working Group} Collaboration,
  L.~Buonincontri, P.~Andreetto, N.~Bartosik, M.~Casarsa, A.~Gianelle,
  D.~Lucchesi, and L.~Sestini, ``{Higgs boson couplings at muon collider},''
  \href{http://dx.doi.org/10.22323/1.398.0619}{{\em PoS} {\bf EPS-HEP2021}
  (2022)  619}.

\bibitem{Neuffer:1983jr}
D.~Neuffer, ``{Principles and Applications of Muon Cooling},''
  \href{http://dx.doi.org/10.2172/1156195}{{\em Part. Accel.} {\bf 14} (1983)
  75--90}.

\bibitem{Ally:2022rgk}
D.~Ally, L.~Carpenter, T.~Holmes, L.~Lee, and P.~Wagenknecht, ``{Strategies for
  Beam-Induced Background Reduction at Muon Colliders},''
  \href{http://arxiv.org/abs/2203.06773}{{\tt arXiv:2203.06773
  [physics.ins-det]}}.

\bibitem{Antonelli:2015nla}
M.~Antonelli, M.~Boscolo, R.~Di~Nardo, and P.~Raimondi, ``{Novel proposal for a
  low emittance muon beam using positron beam on target},''
  \href{http://dx.doi.org/10.1016/j.nima.2015.10.097}{{\em Nucl. Instrum. Meth.
  A} {\bf 807} (2016)  101--107}, \href{http://arxiv.org/abs/1509.04454}{{\tt
  arXiv:1509.04454 [physics.acc-ph]}}.

\bibitem{MICE:2019jkl}
{\bf MICE} Collaboration, M.~Bogomilov {\em et al.}, ``{Demonstration of
  cooling by the Muon Ionization Cooling Experiment},''
  \href{http://dx.doi.org/10.1038/s41586-020-1958-9}{{\em Nature} {\bf 578}
  (2020) no.~7793, 53--59}, \href{http://arxiv.org/abs/1907.08562}{{\tt
  arXiv:1907.08562 [physics.acc-ph]}}.

\bibitem{Jindariani:2022gxj}
S.~Jindariani {\em et al.}, ``{Promising Technologies and R\&D Directions for
  the Future Muon Collider Detectors},''
\newblock 2022.
\newblock \href{http://arxiv.org/abs/2203.07224}{{\tt arXiv:2203.07224
  [physics.ins-det]}}.

\bibitem{Bartosik:2022ctn}
N.~Bartosik {\em et al.}, ``{Simulated Detector Performance at the Muon
  Collider},''
\newblock 2022.
\newblock \href{http://arxiv.org/abs/2203.07964}{{\tt arXiv:2203.07964
  [hep-ex]}}.

\bibitem{Long:2020wfp}
K.~Long, D.~Lucchesi, M.~Palmer, N.~Pastrone, D.~Schulte, and V.~Shiltsev,
  ``{Muon colliders to expand frontiers of particle physics},''
  \href{http://dx.doi.org/10.1038/s41567-020-01130-x}{{\em Nature Phys.} {\bf
  17} (2021) no.~3, 289--292}, \href{http://arxiv.org/abs/2007.15684}{{\tt
  arXiv:2007.15684 [physics.acc-ph]}}.

\bibitem{Franceschini:2021aqd}
R.~Franceschini and M.~Greco, ``{Higgs and BSM Physics at the Future Muon
  Collider},'' \href{http://dx.doi.org/10.3390/sym13050851}{{\em Symmetry} {\bf
  13} (2021) no.~5, 851}, \href{http://arxiv.org/abs/2104.05770}{{\tt
  arXiv:2104.05770 [hep-ph]}}.

\bibitem{Cesarotti:2022ttv}
C.~Cesarotti, S.~Homiller, R.~K. Mishra, and M.~Reece, ``{Probing New Gauge
  Forces with a High-Energy Muon Beam Dump},''
  \href{http://arxiv.org/abs/2202.12302}{{\tt arXiv:2202.12302 [hep-ph]}}.

\bibitem{Han:2020uak}
T.~Han, Z.~Liu, L.-T. Wang, and X.~Wang, ``{WIMPs at High Energy Muon
  Colliders},'' \href{http://dx.doi.org/10.1103/PhysRevD.103.075004}{{\em Phys.
  Rev. D} {\bf 103} (2021) no.~7, 075004},
  \href{http://arxiv.org/abs/2009.11287}{{\tt arXiv:2009.11287 [hep-ph]}}.

\bibitem{Capdevilla:2021fmj}
R.~Capdevilla, F.~Meloni, R.~Simoniello, and J.~Zurita, ``{Hunting wino and
  higgsino dark matter at the muon collider with disappearing tracks},''
  \href{http://dx.doi.org/10.1007/JHEP06(2021)133}{{\em JHEP} {\bf 06} (2021)
  133}, \href{http://arxiv.org/abs/2102.11292}{{\tt arXiv:2102.11292
  [hep-ph]}}.

\bibitem{Bottaro:2021snn}
S.~Bottaro, D.~Buttazzo, M.~Costa, R.~Franceschini, P.~Panci, D.~Redigolo, and
  L.~Vittorio, ``{Closing the window on WIMP Dark Matter},''
  \href{http://dx.doi.org/10.1140/epjc/s10052-021-09917-9}{{\em Eur. Phys. J.
  C} {\bf 82} (2022) no.~1, 31}, \href{http://arxiv.org/abs/2107.09688}{{\tt
  arXiv:2107.09688 [hep-ph]}}.

\bibitem{AguilarSaavedra:2001rg}
{\bf ECFA/DESY LC Physics Working Group} Collaboration, J.~Aguilar-Saavedra
  {\em et al.}, ``{TESLA: The Superconducting electron positron linear collider
  with an integrated x-ray laser laboratory. Technical design report. Part 3.
  Physics at an e+ e- linear collider},''
  \href{http://arxiv.org/abs/hep-ph/0106315}{{\tt arXiv:hep-ph/0106315}}.

\bibitem{Freitas:2004re}
A.~Freitas, H.-U. Martyn, U.~Nauenberg, and P.~Zerwas, ``{Sleptons: Masses,
  mixings, couplings},'' in {\em {International Conference on Linear Colliders
  (LCWS 04)}}, pp.~939--946.
\newblock 9, 2004.
\newblock \href{http://arxiv.org/abs/hep-ph/0409129}{{\tt
  arXiv:hep-ph/0409129}}.

\bibitem{Conley:2010jk}
J.~Conley, H.~Dreiner, and P.~Wienemann, ``{Measuring a Light Neutralino Mass
  at the ILC: Testing the MSSM Neutralino Cold Dark Matter Model},''
  \href{http://dx.doi.org/10.1103/PhysRevD.83.055018}{{\em Phys. Rev. D} {\bf
  83} (2011)  055018}, \href{http://arxiv.org/abs/1012.1035}{{\tt
  arXiv:1012.1035 [hep-ph]}}.

\bibitem{Blaising:2012vd}
J.-J. Blaising, M.~Battaglia, J.~Marshall, J.~Nardulli, M.~Thomson, A.~Sailer,
  and E.~van~der Kraaij, ``{Physics performances for Scalar Electrons, Scalar
  Muons and Scalar Neutrinos searches at CLIC},'' in {\em {International
  Workshop on Future Linear Colliders (LCWS11)}}.
\newblock 1, 2012.
\newblock \href{http://arxiv.org/abs/1201.2092}{{\tt arXiv:1201.2092
  [hep-ex]}}.

\bibitem{Battaglia:2013bha}
M.~Battaglia, J.-J. Blaising, J.~S. Marshall, S.~Poss, A.~Sailer, M.~Thomson,
  and E.~van~der Kraaij, ``{Physics performance for scalar electron, scalar
  muon and scalar neutrino searches at $\sqrt{s} =$ 3 TeV and 1.4 TeV at
  CLIC},'' \href{http://dx.doi.org/10.1007/JHEP09(2013)001}{{\em JHEP} {\bf 09}
  (2013)  001}, \href{http://arxiv.org/abs/1304.2825}{{\tt arXiv:1304.2825
  [hep-ex]}}.

\bibitem{Neuffer:2018yof}
D.~Neuffer and V.~Shiltsev, ``{On the feasibility of a pulsed 14 TeV c.m.e.
  muon collider in the LHC tunnel},''
  \href{http://dx.doi.org/10.1088/1748-0221/13/10/T10003}{{\em JINST} {\bf 13}
  (2018) no.~10, T10003}, \href{http://arxiv.org/abs/1811.10694}{{\tt
  arXiv:1811.10694 [physics.acc-ph]}}.

\bibitem{Zimmermann:2018wfu}
F.~Zimmermann, ``{LHC/FCC-based muon colliders},''
  \href{http://dx.doi.org/10.1088/1742-6596/1067/2/022017}{{\em Journal of
  Physics: Conference Series} {\bf 1067} (2018)  022017}.

\bibitem{Freitas:2011ti}
A.~Freitas, ``{Feasibility of slepton precision measurements at a muon
  collider},'' \href{http://arxiv.org/abs/1107.3853}{{\tt arXiv:1107.3853
  [hep-ph]}}.

\bibitem{Alwall:2014hca}
J.~Alwall, R.~Frederix, S.~Frixione, V.~Hirschi, F.~Maltoni, O.~Mattelaer,
  H.~S. Shao, T.~Stelzer, P.~Torrielli, and M.~Zaro, ``{The automated
  computation of tree-level and next-to-leading order differential cross
  sections, and their matching to parton shower simulations},''
  \href{http://dx.doi.org/10.1007/JHEP07(2014)079}{{\em JHEP} {\bf 07} (2014)
  079}, \href{http://arxiv.org/abs/1405.0301}{{\tt arXiv:1405.0301 [hep-ph]}}.

\bibitem{Duhr:2011se}
C.~Duhr and B.~Fuks, ``{A superspace module for the FeynRules package},''
  \href{http://dx.doi.org/10.1016/j.cpc.2011.06.009}{{\em Comput. Phys.
  Commun.} {\bf 182} (2011)  2404--2426},
  \href{http://arxiv.org/abs/1102.4191}{{\tt arXiv:1102.4191 [hep-ph]}}.

\bibitem{ACME:2018yjb}
{\bf ACME} Collaboration, V.~Andreev {\em et al.}, ``{Improved limit on the
  electric dipole moment of the electron},''
  \href{http://dx.doi.org/10.1038/s41586-018-0599-8}{{\em Nature} {\bf 562}
  (2018) no.~7727, 355--360}.

\bibitem{Ibrahim:2007fb}
T.~Ibrahim and P.~Nath, ``{CP Violation From Standard Model to Strings},''
  \href{http://dx.doi.org/10.1103/RevModPhys.80.577}{{\em Rev. Mod. Phys.} {\bf
  80} (2008)  577--631}, \href{http://arxiv.org/abs/0705.2008}{{\tt
  arXiv:0705.2008 [hep-ph]}}.

\bibitem{Cesarotti:2018huy}
C.~Cesarotti, Q.~Lu, Y.~Nakai, A.~Parikh, and M.~Reece, ``{Interpreting the
  Electron EDM Constraint},''
  \href{http://dx.doi.org/10.1007/JHEP05(2019)059}{{\em JHEP} {\bf 05} (2019)
  059}, \href{http://arxiv.org/abs/1810.07736}{{\tt arXiv:1810.07736
  [hep-ph]}}.

\bibitem{Ellis:2008zy}
J.~R. Ellis, J.~S. Lee, and A.~Pilaftsis, ``{Electric Dipole Moments in the
  MSSM Reloaded},'' \href{http://dx.doi.org/10.1088/1126-6708/2008/10/049}{{\em
  JHEP} {\bf 10} (2008)  049}, \href{http://arxiv.org/abs/0808.1819}{{\tt
  arXiv:0808.1819 [hep-ph]}}.

\bibitem{Giudice:1998bp}
G.~F. Giudice and R.~Rattazzi, ``{Theories with gauge mediated supersymmetry
  breaking},'' \href{http://dx.doi.org/10.1016/S0370-1573(99)00042-3}{{\em
  Phys. Rept.} {\bf 322} (1999)  419--499},
  \href{http://arxiv.org/abs/hep-ph/9801271}{{\tt arXiv:hep-ph/9801271}}.

\bibitem{Kitano:2010fa}
R.~Kitano, H.~Ooguri, and Y.~Ookouchi, ``{Supersymmetry Breaking and Gauge
  Mediation},''
  \href{http://dx.doi.org/10.1146/annurev.nucl.012809.104540}{{\em Ann. Rev.
  Nucl. Part. Sci.} {\bf 60} (2010)  491--511},
  \href{http://arxiv.org/abs/1001.4535}{{\tt arXiv:1001.4535 [hep-th]}}.

\bibitem{Dvali:1996cu}
G.~R. Dvali, G.~F. Giudice, and A.~Pomarol, ``{The Mu problem in theories with
  gauge mediated supersymmetry breaking},''
  \href{http://dx.doi.org/10.1016/0550-3213(96)00404-X}{{\em Nucl. Phys. B}
  {\bf 478} (1996)  31--45}, \href{http://arxiv.org/abs/hep-ph/9603238}{{\tt
  arXiv:hep-ph/9603238}}.

\bibitem{Giudice:2007ca}
G.~F. Giudice, H.~D. Kim, and R.~Rattazzi, ``{Natural mu and B mu in gauge
  mediation},'' \href{http://dx.doi.org/10.1016/j.physletb.2008.01.030}{{\em
  Phys. Lett. B} {\bf 660} (2008)  545--549},
  \href{http://arxiv.org/abs/0711.4448}{{\tt arXiv:0711.4448 [hep-ph]}}.

\bibitem{Csaki:2008sr}
C.~Csaki, A.~Falkowski, Y.~Nomura, and T.~Volansky, ``{New Approach to the
  mu-Bmu Problem of Gauge-Mediated Supersymmetry Breaking},''
  \href{http://dx.doi.org/10.1103/PhysRevLett.102.111801}{{\em Phys. Rev.
  Lett.} {\bf 102} (2009)  111801}, \href{http://arxiv.org/abs/0809.4492}{{\tt
  arXiv:0809.4492 [hep-ph]}}.

\bibitem{Arkani-Hamed:1998mzz}
N.~Arkani-Hamed, G.~F. Giudice, M.~A. Luty, and R.~Rattazzi, ``{Supersymmetry
  breaking loops from analytic continuation into superspace},''
  \href{http://dx.doi.org/10.1103/PhysRevD.58.115005}{{\em Phys. Rev. D} {\bf
  58} (1998)  115005}, \href{http://arxiv.org/abs/hep-ph/9803290}{{\tt
  arXiv:hep-ph/9803290}}.

\bibitem{Komargodski:2009jf}
Z.~Komargodski and D.~Shih, ``{Notes on SUSY and R-Symmetry Breaking in
  Wess-Zumino Models},''
  \href{http://dx.doi.org/10.1088/1126-6708/2009/04/093}{{\em JHEP} {\bf 04}
  (2009)  093}, \href{http://arxiv.org/abs/0902.0030}{{\tt arXiv:0902.0030
  [hep-th]}}.

\bibitem{Dumitrescu:2010ha}
T.~T. Dumitrescu, Z.~Komargodski, N.~Seiberg, and D.~Shih, ``{General Messenger
  Gauge Mediation},'' \href{http://dx.doi.org/10.1007/JHEP05(2010)096}{{\em
  JHEP} {\bf 05} (2010)  096}, \href{http://arxiv.org/abs/1003.2661}{{\tt
  arXiv:1003.2661 [hep-ph]}}.

\bibitem{Cohen:2011aa}
T.~Cohen, A.~Hook, and B.~Wecht, ``{Comments on Gaugino Screening},''
  \href{http://dx.doi.org/10.1103/PhysRevD.85.115004}{{\em Phys. Rev. D} {\bf
  85} (2012)  115004}, \href{http://arxiv.org/abs/1112.1699}{{\tt
  arXiv:1112.1699 [hep-ph]}}.

\bibitem{Cheung:2007es}
C.~Cheung, A.~L. Fitzpatrick, and D.~Shih, ``{(Extra)ordinary gauge
  mediation},'' \href{http://dx.doi.org/10.1088/1126-6708/2008/07/054}{{\em
  JHEP} {\bf 07} (2008)  054}, \href{http://arxiv.org/abs/0710.3585}{{\tt
  arXiv:0710.3585 [hep-ph]}}.

\bibitem{Meade:2008wd}
P.~Meade, N.~Seiberg, and D.~Shih, ``{General Gauge Mediation},''
  \href{http://dx.doi.org/10.1143/PTPS.177.143}{{\em Prog. Theor. Phys. Suppl.}
  {\bf 177} (2009)  143--158}, \href{http://arxiv.org/abs/0801.3278}{{\tt
  arXiv:0801.3278 [hep-ph]}}.

\bibitem{Giudice:1988yz}
G.~F. Giudice and A.~Masiero, ``{A Natural Solution to the mu Problem in
  Supergravity Theories},''
  \href{http://dx.doi.org/10.1016/0370-2693(88)91613-9}{{\em Phys. Lett. B}
  {\bf 206} (1988)  480--484}.

\bibitem{Foot:1990mn}
R.~Foot, ``{New Physics From Electric Charge Quantization?},''
  \href{http://dx.doi.org/10.1142/S0217732391000543}{{\em Mod. Phys. Lett. A}
  {\bf 6} (1991)  527--530}.

\bibitem{He:1990pn}
X.~G. He, G.~C. Joshi, H.~Lew, and R.~R. Volkas, ``{New Z-prime
  Phenomenology},'' \href{http://dx.doi.org/10.1103/PhysRevD.43.R22}{{\em Phys.
  Rev. D} {\bf 43} (1991)  22--24}.

\bibitem{Ma:2001md}
E.~Ma, D.~P. Roy, and S.~Roy, ``{Gauged L(mu) - L(tau) with large muon
  anomalous magnetic moment and the bimaximal mixing of neutrinos},''
  \href{http://dx.doi.org/10.1016/S0370-2693(01)01428-9}{{\em Phys. Lett. B}
  {\bf 525} (2002)  101--106}, \href{http://arxiv.org/abs/hep-ph/0110146}{{\tt
  arXiv:hep-ph/0110146}}.

\bibitem{Gorbatov:2008qa}
E.~Gorbatov and M.~Sudano, ``{Sparticle Masses in Higgsed Gauge Mediation},''
  \href{http://dx.doi.org/10.1088/1126-6708/2008/10/066}{{\em JHEP} {\bf 10}
  (2008)  066}, \href{http://arxiv.org/abs/0802.0555}{{\tt arXiv:0802.0555
  [hep-ph]}}.

\bibitem{Craig:2012yd}
N.~Craig, M.~McCullough, and J.~Thaler, ``{The New Flavor of Higgsed Gauge
  Mediation},'' \href{http://dx.doi.org/10.1007/JHEP03(2012)049}{{\em JHEP}
  {\bf 03} (2012)  049}, \href{http://arxiv.org/abs/1201.2179}{{\tt
  arXiv:1201.2179 [hep-ph]}}.

\bibitem{Froggatt:1978nt}
C.~D. Froggatt and H.~B. Nielsen, ``{Hierarchy of Quark Masses, Cabibbo Angles
  and CP Violation},''
  \href{http://dx.doi.org/10.1016/0550-3213(79)90316-X}{{\em Nucl. Phys. B}
  {\bf 147} (1979)  277--298}.

\bibitem{Leurer:1992wg}
M.~Leurer, Y.~Nir, and N.~Seiberg, ``{Mass matrix models},''
  \href{http://dx.doi.org/10.1016/0550-3213(93)90112-3}{{\em Nucl. Phys. B}
  {\bf 398} (1993)  319--342}, \href{http://arxiv.org/abs/hep-ph/9212278}{{\tt
  arXiv:hep-ph/9212278}}.

\bibitem{Nir:1993mx}
Y.~Nir and N.~Seiberg, ``{Should squarks be degenerate?},''
  \href{http://dx.doi.org/10.1016/0370-2693(93)90942-B}{{\em Phys. Lett. B}
  {\bf 309} (1993)  337--343}, \href{http://arxiv.org/abs/hep-ph/9304307}{{\tt
  arXiv:hep-ph/9304307}}.

\bibitem{Hall:1999sn}
L.~J. Hall, H.~Murayama, and N.~Weiner, ``{Neutrino mass anarchy},''
  \href{http://dx.doi.org/10.1103/PhysRevLett.84.2572}{{\em Phys. Rev. Lett.}
  {\bf 84} (2000)  2572--2575}, \href{http://arxiv.org/abs/hep-ph/9911341}{{\tt
  arXiv:hep-ph/9911341}}.

\bibitem{Nir:1996am}
Y.~Nir and R.~Rattazzi, ``{Solving the supersymmetric CP problem with Abelian
  horizontal symmetries},''
  \href{http://dx.doi.org/10.1016/0370-2693(96)00571-0}{{\em Phys. Lett. B}
  {\bf 382} (1996)  363--368}, \href{http://arxiv.org/abs/hep-ph/9603233}{{\tt
  arXiv:hep-ph/9603233}}.

\bibitem{Nakai:2021mha}
Y.~Nakai, M.~Reece, and M.~Suzuki, ``{Supersymmetric alignment models for (g
  \ensuremath{-} 2)$_{\mu}$},''
  \href{http://dx.doi.org/10.1007/JHEP10(2021)068}{{\em JHEP} {\bf 10} (2021)
  068}, \href{http://arxiv.org/abs/2107.10268}{{\tt arXiv:2107.10268
  [hep-ph]}}.

\bibitem{deFavereau:2013fsa}
{\bf DELPHES 3} Collaboration, J.~de~Favereau, C.~Delaere, P.~Demin,
  A.~Giammanco, V.~Lema\^\i{}tre, A.~Mertens, and M.~Selvaggi, ``{DELPHES 3, A
  modular framework for fast simulation of a generic collider experiment},''
  \href{http://dx.doi.org/10.1007/JHEP02(2014)057}{{\em JHEP} {\bf 02} (2014)
  057}, \href{http://arxiv.org/abs/1307.6346}{{\tt arXiv:1307.6346 [hep-ex]}}.

\bibitem{Cowan:1998ji}
G.~Cowan, {\em {Statistical data analysis}}.
\newblock Oxford University Press, 1998.

\bibitem{ParticleDataGroup:2020ssz}
{\bf Particle Data Group} Collaboration, P.~A. Zyla {\em et al.}, ``{Review of
  Particle Physics},'' \href{http://dx.doi.org/10.1093/ptep/ptaa104}{{\em PTEP}
  {\bf 2020} (2020) no.~8, 083C01}.

\bibitem{Wilks:1938xx}
S.~S. Wilks, ``{The Large-Sample Distribution of the Likelihood Ratio for
  Testing Composite Hypotheses},''
  \href{http://dx.doi.org/10.1214/aoms/1177732360}{{\em The Annals of
  Mathematical Statistics} {\bf 9} (1938) no.~1, 60 -- 62}.
  \url{https://doi.org/10.1214/aoms/1177732360}.

\bibitem{Cowan:2010js}
G.~Cowan, K.~Cranmer, E.~Gross, and O.~Vitells, ``{Asymptotic formulae for
  likelihood-based tests of new physics},''
  \href{http://dx.doi.org/10.1140/epjc/s10052-011-1554-0}{{\em Eur. Phys. J. C}
  {\bf 71} (2011)  1554}, \href{http://arxiv.org/abs/1007.1727}{{\tt
  arXiv:1007.1727 [physics.data-an]}}. [Erratum: Eur.Phys.J.C 73, 2501 (2013)].

\end{thebibliography}\endgroup
}

\end{document}